\documentclass[a4paper,structabstract]{aa}

\pdfoutput=1

\usepackage[utf8]{inputenc}
\usepackage{amsmath} 
\usepackage{txfonts} 
\usepackage{bm} 
\usepackage{url}
\usepackage{natbib}
\usepackage{graphicx}
\newcommand{\transposee}[1]{{#1}^\intercal}
\DeclareRobustCommand{\rchi}{{\mathpalette\irchi\relax}}
\newcommand{\irchi}[2]{\raisebox{\depth}{$#1\chi$}} 
\begin{document}

  \title{Probabilistic multi-catalogue positional cross-match}
  \titlerunning{Probabilistic multi-catalogue xmatch}

  \author{F.-X. Pineau\inst{\ref{inst:oas}} \and S. Derriere\inst{\ref{inst:oas}} \and C. Motch\inst{\ref{inst:oas}}
          \and F. J. Carrera\inst{\ref{inst:ifca}} \and F. Genova\inst{\ref{inst:oas}} \and L. Michel\inst{\ref{inst:oas}}
	  \and B. Mingo\inst{\ref{inst:lei}} \and A. Mints\inst{\ref{inst:aip},\ref{inst:maxp}} \and A. Nebot G\'omez-Mor\'an\inst{\ref{inst:oas}}
	  \and S. R. Rosen\inst{\ref{inst:lei}} \and A. Ruiz Camu\~nas\inst{\ref{inst:ifca}}
  }
  \authorrunning{Pineau, Derriere, Motch et al.}

  \institute{Observatoire astronomique de Strasbourg, Université de Strasbourg, CNRS, UMR 7550,
              11 rue de l’Université, F-67000 Strasbourg, France
              \email{francois-xavier.pineau@astro.unistra.fr} \label{inst:oas}
         \and
	     IFCA (CS-IC-UC), Avenida de los Castros, 39005 Santander, Spain
	     \label{inst:ifca}
	 \and
	    Department of Physics \& Astronomy, University of Leicester, Leicester, LEI 7RH, UK
	    \label{inst:lei}
	 \and
	   Leibniz-Institut f\"ur Astrophysik Potsdam (AIP), An der Sternwarte 16, 14482 Potsdam, Germany
	   \label{inst:aip}
	 \and
	   Max Planck Institute for Solar System Research, Justus-von-Liebig-Weg 3, 37077 G\"ottingen, Germany
	   \label{inst:maxp}
  }

  \date{Received June 30, 2016; accepted August 28, 2016}

  \abstract
{
Catalogue cross-correlation is essential to building large sets of
multi-wavelength data, whether it be to study the properties of
populations of astrophysical objects or to build reference
catalogues (or timeseries) from survey 
observations.
Nevertheless, resorting to automated processes with limited
sets of information available on large numbers of sources detected
at different epochs with various filters and instruments inevitably leads to
spurious associations.
We need both statistical criteria to select detections to be
merged as unique sources, and statistical indicators helping
in achieving compromises between completeness and reliability of selected associations.
}
{
We lay the foundations of a statistical framework
for multi-catalogue cross-correlation and cross-identification
based on explicit simplified catalogue models. 
A proper identification process should rely on both 
astrometric and photometric data. 
Under some conditions, the astrometric part and the photometric
part can be processed separately and merged a posteriori
to provide a single global probability of identification.
The present paper addresses almost exclusively the astrometrical part
and specifies the proper probabilities to be merged with photometric likelihoods.
}
{
To select matching candidates in $n$ catalogues, we used the Chi (or, indifferently, the Chi-square) test
with $2(n-1)$ degrees of freedom. We thus call this cross-match a $\rchi$-match.
In order to use Bayes' formula, we considered exhaustive sets of hypotheses
based on combinatorial analysis.
The volume of the $\rchi$-test domain of acceptance
-- a $2(n-1)$-dimensional acceptance ellipsoid -- is used 
to estimate the expected numbers of spurious associations.
We derived priors for those numbers using a frequentist approach
relying on simple geometrical considerations.
Likelihoods are based on standard Rayleigh, $\rchi$ and Poisson
distributions that we normalized over the $\rchi$-test acceptance domain.
We validated our theoretical results by generating and cross-matching
synthetic catalogues.
}
{
The results we obtain do not depend on the order used to cross-correlate the catalogues.
We applied the formalism described in the present paper to build the
multi-wavelength catalogues used for the science cases of the ARCHES (Astronomical Resource Cross-matching for High Energy Studies) project.
Our cross-matching engine is publicly available through a multi-purpose web interface.
In a longer term, we plan to integrate this tool into the CDS XMatch Service.
}
{}


  \maketitle

  \section{Introduction}
The development of new detectors with high throughput over large areas has
revolutionized observational astronomy during recent decades.
These technological advances, aided by a considerable increase of computing
power, have opened the way to outstanding ground-based and space-borne all-sky
or very large area imaging projects (e.g. the 2MASS \citep{Skrutskie2006,Vizier2MASS},
SDSS \citep{SDSS9,VizierSDSS9} and WISE \citep{Wright2010,CatAllWISE} surveys).
These surveys have provided an essential astrometric and photometric reference frame
and the first true digital maps of the entire sky.
\newline
As an illustration of this flood of data, the number of catalogue entries in the VizieR
service at the Centre de Donn\'ees astronomiques de Strasbourg (CDS) which was about
500 million in 1999 has reached almost 18 billion as on February 2016.
At the 2020 horizon, European space missions such as GAIA and EUCLID together
with the Large Synoptic Survey Telescope (LSST) will provide a several-fold
increase in the number of catalogued optical objects while providing measurements of
exquisite astrometric and photometric quality. 
\newline
This exponentially increasing flow of high quality multi-wavelength data has
radically altered the way astronomers now design observing strategies and
tackle scientific issues. The former paradigm, mostly focusing on a single wavelength
range, has in many cases evolved towards a systematic fully multi-wavelength study.
In fact, modelling the spectral energy distributions over the widest range of
frequencies, spanning from radio to the highest energy gamma-rays has been instrumental
in understanding the physics of stars and galaxies.
\newline
Many well designed and useful tools have been developed worldwide concurrently with
the emergence of the virtual observatory. Most if not all of these tools can handle and
process multi-band images and catalogues. When assembling spectral energy distributions
using surveys obtained at very different wavelengths and with discrepant spatial resolution,
one of the most acute problems is to find the correct counterpart across the various bands.
Several tools such as TOPCAT \citep{Taylor2005} or the CDS XMatch Service
\citep{Pineau2011b, Boch2012} offer basic cross-matching facilities.
However, none of the publicly available tools handles the statistics inherent to the
cross-matching process in a fully coherent manner.
A standard method for a dependable and robust association of a physical source to instances
of it in different catalogues (cross-identification) and in diverse spectral ranges is still absent. 
\newline
The pressing need for a multi-catalogue probabilistic cross-matching tool was
one of the strong motivations of the FP7-Space European program ARCHES
\citep{Motch2016}\footnote{http://www.arches-fp7.eu/}.
Designing a cross-matching tool able to process, in a single pass,
a theoretically unlimited number of catalogues, while computing probabilities of associations
for all catalogue configurations, using the background of sources, positional errors and eventually
introducing priors on the expected shape of the spectral energy distribution is one of the most
important outcomes of the project. A preliminary description of this algorithm was presented
in \cite{Pineau2015}.
Although ARCHES was originally focusing on the cross-matching of XMM-Newton sources,
the algorithms developed in this context are clearly applicable to any combination of catalogues
and energy bands \citep[see for example][]{Mingo2016}.

  \section{Going beyond the two-catalogue case}
Computing probabilities of identifications when cross-correlating two
catalogues in a given area can be quite straightforward
(provided the area is small enough so that the density of sources
can be considered more or less constant, but large enough to provide sufficient statistics).
For each possible pair of sources (one from each catalogue),
we compute the distance normalized by positional errors (called
normalized distance, $\sigma$-distance, $\rchi$-distance or more generally in
this paper Mahalanobis distance $D_M$).
Then we build the histogram of the number of associations per bin of $D_M$.
This histogram is the sum of two components (see Fig. \ref{fig:likelihoods_n2}): the ``real'' or ``true''
associations ($T$) for which the distribution $p(D_M|T)$
follows a Rayleigh distribution; the spurious or ``false'' associations,
for which the distribution $p(D_M|F)$ follows a linear (Poisson) distribution.
Knowing these two distributions and the total number of associations
($n_{T+F}$), we may fit the histogram with the function
\begin{equation}
  f(D_M) = (n_{T+F} - n_F) p(D_M|T)
          + n_F p(D_M|F)
\end{equation}
to estimate the number of spurious associations ($n_F$)
and thus the number of good matches ($n_T = n_{T+F} - n_F$).
Hence, we are able to attribute to an association with a given normalized distance
the probability of being a good match:
\begin{equation}
  p(T|D_M) = \frac{(n_{T+F} - n_F)p(D_M|T)}
                  {f(D_M)}.
\end{equation}
Dividing both the numerator and the denominator by $n_{T+F}$ we recognize
Bayes' formula considering $(n_{T+F} - n_F) / n_{T+F}$ as the prior $p(T)$
and considering either $n_F/n_{T+F}$ as the prior $p(F)$ or $f(D_M)=n_{T+F}p(D_M)$.

The present paper basically extends this simple approach to more
than two catalogues. Instead of fitting histograms to find the number of spurious
associations, we directly compute them from the input catalogues data and
from geometrical considerations.

Previously, \cite{Budavari2008} developed a multi-catalogue cross-match.
For a given set of $n$ sources from $n$ distinct catalogues,
they compute a ``Bayes' factor'' based on both astrometric and photometric data.
The ``Bayes' factor'' is then used as a score: a pre-defined threshold on its value
is applied to select or reject the given set of $n$ sources.
We discuss the astrometric part of \cite{Budavari2008} ``Bayes' factor''
and compare it to our selection criterion in \S \ref{sec:rmkbudav} and \ref{sec:generalities}.

Throughout the present paper we consider a set of $n$ catalogues.
We use a Chi-square criterion based on individual elliptical
positional errors to select, in these catalogues, sets of associations
containing at most one source per catalogue.
We call this selection a $\rchi$-match.
We then compute probabilities for each set of associations.
To compute probabilities, we consider only the result set in which each set
of associations contains exactly $n$ sources (one per catalogue, see below for partial matches).
For people familiar with databases, it can be seen as the result of
inner joins, joining successively each catalogue using a Chi-square criteria.
The probabilities we then compute are only based on positional coincidences.
Although we show how it is possible to add likelihoods based on photometric considerations,
the computation of such photometric likelihoods is beyond the scope of this paper.

As the result of a $\rchi$-match, two distinct sets of associations may have sources in common:
a source having a large positional error in one catalogue may for example
be associated to several sources with smaller errors in another catalogue.
We do not take into account in our probabilities the ``one-to-several'' and
the ``one-to-one'' associations paradigms defined in \cite{Fioc2014}:
it becomes far too complex when dealing with a generic number of catalogues and it is not that simple
when a source may be blended, etc. 
We use a several-to-several-(to-several-...) paradigm.
In other words, we compute probabilities for a set of associations regardless of the fact that
a source in the set can be in other sets of associations.
So a same detection in one catalogue may have very high probabilities of associations with several
(sets of) candidates in the other catalogues.
We think it is the responsibility of the photometric part to disentangle such cases.

Requiring one candidate per catalogue for each set of associations (i.e. each tuple) is somewhat restrictive.
But, if one or several catalogues do not contain any candidates for a tuple,
then we compute the probabilities from the cross-match of the subset of catalogues providing
one candidate to that tuple. Those probabilities are computed independently of the ``full' $n$
catalogues probabilities.
For example, if we cross-match three catalogues 
and if a set of associations (a tuple) contains one source per catalogues (A, B and C), then 
we will compute five probabilities: one for each
possible configuration ($ABC$, $AB\_C$, $A\_BC$, $AC\_B$ and $A\_B\_C$ in which the underscore '\_' separates the 
catalogue entries associated to different actual sources, see \S \ref{sec:3cathyp}).
Now, if one source from A has a candidate in B and no candidate in C, we will compute
only two probabilities ($AB$ and $A\_B$, see \S \ref{sec:2cathyp})
considering only the result of the cross-match
of A with B.
Likewise for A and C only and for B and C only.
These four cross-matches will yield eleven distinct probabilities.
It is possible to deal with ``missing'' detections when computing photometrically based likelihoods
(taking into account limit fluxes, ...) but it is not the case in the astrometric part of this work.

When $\rchi$-matching $n$ catalogues, the number of hypotheses to be tested,
and thus the number of probabilities to be computed for a given set of associations,
increases dramatically with $n$.
This number is 203 for 6 catalogues and reaches 877 for seven catalogues (see Table \ref{tab:bell} in \S \ref{sec:bellnumber}).
To be able to compute probabilities when $\rchi$-matching
more than seven catalogues we may start by merging catalogues for which the probability of making spurious associations is very low
(e.g. catalogues of similar wavelength and similar astrometric accuracy), and handle the merged catalogue as a single input catalogue.\\

In section \S \ref{sec:assumptions} we lay down the assumptions we use to work
on a simplified problem. We then (\S \ref{sec:notations}) define the notations
and the standards used throughout the paper and link them to the standards 
adopted in a few catalogues.
We then describe in detail the candidate selection criterion
(\S \ref{sec:candselect.all}) 
before providing (\S \ref{sec:hypotheses}) an exhaustive list of all
hypotheses we have to account for to apply Bayes' formula.
In \S \ref{sec:candselect.all} we also show how the ``Bayesian cross-match'' of \cite{Budavari2008}
may be interpreted as an inhomogeneous $\rchi$-match.
Then (\S \ref{sec:spurest}) we show how it is possible to estimate the rates of spurious associations
and hence ``priors''.
In \S \ref{sec:integrate} we compute an integral which is related to the probability
the selection criterion has to select a set of $n$ sources for a given hypothesis.
This integral is crucial to compute likelihoods defined in
\S \ref{sec:1model} and to normalize likelihoods in \S \ref{sec:2model}.
Finally, after showing how to introduce the photometric data into the probabilities
(\S \ref{sec:phot}), and
before concluding (\S \ref{sec:conclu}), we explain the tests we carried out
on synthetic catalogues in \S \ref{sec:tests}.
Since this paper is long and technical, we put a summary of the steps to follow
to perform a probabilistic $\rchi$-match in \S \ref{sec:recipe}.

\section{Simplifying assumptions \label{sec:assumptions}}
Cross-correlating catalogues taking into account an accurate model of the sky
on one hand, and the effects and biases due to the catalogue building process
on the other hand is a daunting task.
To make progress towards this objective, we have to start by making
simplifying assumptions.

First of all, we assume that there are no systematic offsets
between the positions of each possible pair of catalogues.
It means that the positions are accurate (no bias).
We also assume that positional errors provided
in catalogues are trustworthy. It means that they are neither overestimated
nor underestimated: for instance, no systematic have to be quadratically added or removed.
The first point supposes an accurate astrometric calibration of all catalogues.
This is somewhat the ``dog chasing its tail'' problem since a proper astrometric
calibration should be based on secure identifications, themselves based on...
cross-identification!
Ideally the astrometric calibration and the cross-identification should be
performed simultaneously in an iterative process.
It will not be developed here but we point out that the present work can be used
to calibrate astrometrically $n$ catalogues at the same time from one reference catalogue,
taking into account all possible associations in all possible catalogue sub-sets. 
However, carrying out careful identification of primary or secondary
astrometric standards is only important when the density of bright astrometric
references is very low, typically in deep small field exposures.
Reliable cross-identification is also crucial when the wavelength band of the
image to calibrate differs widely from that of the astrometric reference image.
In most large scale surveys such as 2MASS \citep{Skrutskie2006} or SDSS
\citep{Pier2003} the density of bright Tycho-2 \citep{Hog2000} or
UCAC \citep{Zacharias2004} astrometric reference stars is high enough
to ensure an excellent overall calibration without any
ambiguity in the associations.

Although the idealized vision of an immutable and static sky is long gone,
we ignore proper motions in this analysis.
There are at least two ways of taking them into account: either we may force
associations to include at least one source from a catalogue containing
measured proper motions; or we may try to fit proper motions during the
cross-match process. 
In this last case, if a set of $n$ sources detected at different
epochs in $n$ distinct catalogues does not satisfy the candidate selection
defined in \S \ref{sec:candselect}, we may make the hypothesis that they
nonetheless are from a same source but having a proper motion.
We can then estimate the proper motion and the associated error based
on positions, (Gaussian) positional errors and epochs (see appendix \ref{sec:pmfit}).
From the $n$ observed positions and associated errors and from
the $n$ theoretical estimated positions and associated errors we can
compute a Mahalanobis distance which follows a $\rchi$ distribution with
$2(n-2)$ degrees of freedom.
Similarly to the candidate selection criterion in \S \ref{sec:candselect} 
we can then reject the hypothesis ``same source with proper motion''
if the Mahalanobis distance is larger than a given threshold.

We neglect clustering effects. We suppose that in a given area $\Omega$,
source properties are homogeneous. This implies that the local density of sources,
the positional error distributions and the associations priors (probabilities of true
associations that in principle depend on the astrophysical nature of the sources
and on the limiting flux) are uniform over the sky area considered.
As usual we have to face the following dilemma: on the one hand,
the larger the area $\Omega$, the better the statistic;
on the other hand, the larger the area $\Omega$, the less probable
the uniform density, errors distributions and priors hypothesis.
In the ARCHES project, for instance, we grouped the individual XMM-Newton EPIC
fields of view of $\approx$\,0.126\,deg$^{2}$ each into installments of homogeneous exposure
times and galactic latitude so as to ensure as much uniformity as possible.
Each installment contained on the order of several hundred sources. 

Finally, we neglect blending.
If two sources are separated in one catalogue and blended in the other one,
the position of the blended source will be something like the photocentre
of the two sources.
Either the blended source will not match any of the two distinct sources,
or only one of the two distinct sources will match, the match
likely being 
in the tail of the Rayleigh distribution, possibly leading to
a low probability of identification.
It will then not be problematic to consider the match as spurious since
the observed flux is contaminated by the flux of the nearby source.
Finally, if the positional accuracy of the blended source is well below that of the distinct
sources, both distinct sources will match the blended source,
leading to a non-unique association requiring further investigations to be disentangled.

\section{Notations and links with catalogues \label{sec:notations}}
\subsection{Notations}
This article uses almost exclusively the notations defined in the
ISO 80000-2:2009(E) international standard.
Exceptionally we waive the notation $\det \bm{A}$ for determinant and replace it by the equivalent
but more compact notation $|\bm{A}|$.

We consider $n$ catalogues defined on a common surface of area $\Omega$.
We assume that each catalogue source has individual elliptical positional errors
defined by a bivariate normal (or binormal) distribution.
For this, we assimilate locally the surface of the sphere to its zenithal (or azimuthal)
equidistant projection \citep[see ARC projection in][]{Calabretta2002},
that is to its local Euclidean tangent plane.
In this frame, the position of a point at distance $d$ arcsec from the
origin $O$ (the tangent point) and having a position angle $\varphi$ (East of North)
is simply
\begin{eqnarray}
  x & = & d\sin\varphi, \\
  y & = & d\cos\varphi.
\end{eqnarray}
This approximation is acceptable since typical positional errors,
distances and surfaces locally considered are small.

We note $\mathcal{N}$ the binormal probability density function (p.d.f)
representing the position of a source $S$ and its associated uncertainty:
\begin{equation}
  \mathcal{N}_{\vec{\mu},\bm{V}}(\vec{p})
    = \frac{1}{2\pi\sqrt{\det \bm{V}}} \exp\{-\frac{1}{2}Q(\vec{p})\} \mathrm{d}\vec{p},
\end{equation}
with
  \begin{itemize}
    \item $\vec{\mu} = \transposee{(\mu_x, \mu_y)}$ the position of the source $S$ provided in a catalogue, that is the mean of the binormal distribution;
    \item $\bm{V}$ the provided variance-covariance -- also simply called covariance -- matrix which defines the error on the source position;
    \item $\vec{p} = \transposee{(x, y)}$ any given two-dimensional position;
    \item $Q(\vec{p})$ the quadratic form $Q(\vec{p}) = \transposee{(\vec{p}-\vec{\mu})}\bm{V}^{-1}(\vec{p}-\vec{\mu})$,
          that is the square of the weighted distance between a given position \vec{p} and the position of the source $S$
    \begin{eqnarray}
      \bm{V} & = &
      \begin{pmatrix}
          \sigma_x^2           & \rho\sigma_x\sigma_y \\
          \rho\sigma_x\sigma_y & \sigma_y^2 
      \end{pmatrix},\\
      \bm{V}^{-1} & = &  
      \frac{1}{\sigma_x^2\sigma_y^2(1-\rho^2)}
      \begin{pmatrix}
          \sigma_y^2            & -\rho\sigma_x\sigma_y \\
          -\rho\sigma_x\sigma_y & \sigma_x^2 
      \end{pmatrix},
    \end{eqnarray}
    where
    \begin{itemize}
      \item $\sigma_x$ is the standard deviation along the $x$-axis (i.e. the East axis),
      \item $\sigma_y$ is the standard deviation along the $y$-axis (i.e. the North axis),
      \item $\rho$ the correlation factor between $\sigma_x$ and $\sigma_y$;
    \end{itemize}
    \item $\det \bm{V} = \sigma_x^2\sigma_y^2(1-\rho^2)$ the determinant of $\bm{V}$;
    \item $\mathrm{d}\vec{p}=\mathrm{d}x\mathrm{d}y$.
  \end{itemize}
A covariance matrix $\bm{V}$ represents a $1\sigma$ ellipse.
The ``real'' position of the source $S$ has $\approx 39$\% chances to be located inside
this $1\sigma$-ellipse.
It must not be confused with the $1$-dimensional $1\sigma$-segment
which contains a real ``value'' with a probability of $\approx 68$\%.

\subsection{Classical positional errors in catalogues}

In astronomical catalogues like the 2MASS All-Sky Catalog of Point Sources
\citep[2MASS-PSC,][]{Vizier2MASS} positional errors are
described by three parameters defining the $1\sigma$ positional uncertainty
ellipse
\footnote{from the 2MASS online user's guide
\url{http://www.ipac.caltech.edu/2mass/releases/allsky/doc/sec2_2a.html}}: 
$err\_maj$ or $a$ the semi-major axis,
$err\_min$ or $b$ the semi-minor axis,
and $err\_ang$ or $\psi$ the positional angle (East of North) of the semi-major axis.
We give the formula to transform the ellipse into a covariance matrix
(see appendix A.2 of \cite{Pineau2011} and footnote 11 in \cite{Fioc2014}):
\begin{eqnarray}
  \sigma_x & = & \sqrt{a^2\sin^2\psi+b^2\cos^2\psi}, \\
  \sigma_y & = & \sqrt{a^2\cos^2\psi+b^2\sin^2\psi}, \\
  \rho\sigma_x\sigma_y & = & \cos\psi\sin\psi(a^2-b^2). \label{eq:rhosigxsigy}
\end{eqnarray}
In the AllWISE catalogue \citep{CatAllWISE}, the coefficients of the covariance
matrix are (almost) directly available.  
Instead of providing the unitless correlation factor $\rho$ or the covariance
$\rho\sigma_x\sigma_y$ (in arcsec$^2$), the authors chose to provide
the co-sigma ($\sigma_{\alpha\delta}$) because, as they state\footnote{\url{http://wise2.ipac.caltech.edu/docs/release/allwise/expsup/sec2_1a.html}},
the latter is in the same units as the other uncertainties. We thus have
\begin{eqnarray}
  \sigma_x & = & \sigma_\alpha = sigra, \\
  \sigma_y & = & \sigma_\delta = sigdec, \\
  \rho\sigma_x\sigma_y & = & \sigma_{\alpha\delta}\times|\sigma_{\alpha\delta}| = sigradec\times |sigradec|.
\end{eqnarray}
In catalogues like the Sloan Digital Sky Survey since its eighth data release
\citep[SDSS-DR8,][]{SDSSDR8} positional errors contain two terms:
the error on RA ($raErr$) and the error on Dec ($decErr$).
In this case the parameters of the covariance matrix are simply
\begin{eqnarray}
  \sigma_x & = & raErr, \\
  \sigma_y & = & decErr, \\
  \rho\sigma_x\sigma_y & = & 0.
\end{eqnarray} 
In catalogues like the XMM catalogues \citep[e.g. the 3XMM-DR5,][]{Rosen2016}
a single error is provided.
Ideally, one would like to have access to the two one-dimensional errors,
even if their respective values are often very close.
The column named $radecErr$ is the total error, so
the quadratic sum of the two computed (but not provided) $1$-dimensional errors,
one computed on RA and one computed on Dec.
If one uses $\sigma_x=radecErr$ and $\sigma_y=radecErr$, the total error
will be $\sigma=\sqrt{\sigma_x^2+\sigma_y^2}=\sqrt{2}radecErr$ instead of $radecErr$.
In output of the astrometric calibration process, the XMM pipeline provides a systematic error $sysErrCC$
which is quadratically added to $radecErr$ to compute the ``total radial position uncertainty''\footnote{\url{http://xmmssc.irap.omp.eu/Catalogue/3XMM-DR6/Coordinates.html}} $posErr$.
As for $radecErr$, we must divide $posErr$ by $\sqrt{2}$ to obtain the $1$-dimensional error.
The appropriate errors to be used (including a systematic) are then
\begin{eqnarray}
  \sigma_x & = & \sigma_y = posErr/\sqrt{2}, \\
  \rho\sigma_x\sigma_y & = & 0.
\end{eqnarray}
The factor $\sqrt{2}$ has not been taken into account in \cite{Pineau2011}.
It partly explains why the fit of the curve in the right panel of Fig. 3
mentioned in \S 5 of this paper does not lead to a Rayleigh scale parameter
equal to 1.\\
Similarly to the XMM case, the error $posErr$ provided
in the GALEX All-Sky Survey Source Catalog (GASC)
catalogue\footnote{\url{http://www.galex.caltech.edu/wiki/GCAT_Manual#Catalog_Column_Description}}
(which also includes the systematic) is a ``total radial error''.
It is thus the Rayleigh parameter $\sigma$ which is the quadratic sum of two one-dimensional errors.
As for XMM, the appropriate errors to be used are
\begin{eqnarray}
  \sigma_x & = & \sigma_y = posErr/\sqrt{2}, \\
  \rho\sigma_x\sigma_y & = & 0.
\end{eqnarray} \\
In catalogues like the ROSAT All-Sky Bright Source Catalogue
\citep[1RXS, ][]{Rosat1999} the error provided is the radius of the cone
containing the real position of a source with a probability of $\approx68.269$\%
(the 1 dimensional $1\sigma$). 
Authors like \cite{Rutledge2003} \citep[given the details provided in][\S 3.3.3]{Rosat1999}
call this radius the $1\sigma$-radius.
We note it $r_{68\%}$.
But, in the Rayleigh distribution, the scale parameter $\sigma$
is defined such that the cone of radius $r=\sigma$
contains the real position with a probability $100\times (1-\exp(-1/2)) \approx 39.347$\%.
Adjusting such that $1-\exp(-1/2\times r_{68\%}^2/\sigma^2)=0.6827$ leads to
\begin{equation}
  \sigma_x = \sigma_y = \sigma = \frac{r_{68\%}}{\sqrt{-2\ln(1-0.6827)}} = \frac{r_{68\%}}{\sqrt{2\ln(3.1515)}} \approx \frac{r_{68\%}}{1.51517}.
\end{equation} 
Similarly if the provided error is the radius of the cone
containing the real position with a probability of 90\%
\cite[e.g in the WGACAT,][]{VizierWGACAT}
\begin{equation}
  \sigma_x = \sigma_y = \sigma = \frac{r_{90\%}}{\sqrt{-2\ln(1-0.90)}} = \frac{r_{90\%}}{\sqrt{2\ln(10)}} \approx \frac{r_{90\%}}{2.14597}.
\end{equation} 
The description \citep{White1997} and the on-line documentation\footnote{\url{http://sundog.stsci.edu/first/catalogs/readme.html}}
of the FIRST catalogue \citep{Helfand2015,Helfand2015Viz} provide an ``empirical expression''
to compute the semi-major and semi-minor axis of the 90\% positional accuracy associated to each source:
\begin{eqnarray}
  a_{90\%} & = & \text{fMaj} (\frac{\text{RMS}}{(\text{Fpeak} - 0.25)} + \frac{1}{20}), \\
  b_{90\%} & = & \text{fMin} (\frac{\text{RMS}}{(\text{Fpeak} - 0.25)} + \frac{1}{20}), 
\end{eqnarray}
in which $\text{fMaj}$ ($\text{fMin}$) is the major (minor) axis of the fitted FWHM,
$\text{RMS}$ ``is a local noise estimate at the source position'' and
$\text{Fpeak}$ is the peak flux density.
The position angle $\psi$ of the accuracy equals the fitted FWHM angle $\text{fPA}$.
We first obtain the $1\sigma$ accuracy ellipse by resizing the $90\%$ ellipse axes
divinding them by the same factor as for the WGACAT (i.e. $\sqrt{2\ln(10)}$)
\begin{eqnarray}
  a & = & \frac{a_{90\%}}{\sqrt{2\ln(10)}}, \label{eq:first.a}\\
  b & = & \frac{b_{90\%}}{\sqrt{2\ln(10)}}. \label{eq:first.b}
\end{eqnarray}
After possibly adding systematics, the variance-covariance matrix is obtained applying the equations
used for the 2MASS catalogue.\\
Errors in catalogues like the Guide Star Catalog Version 2.3.2\footnote{\url{http://vizier.u-strasbg.fr/viz-bin/VizieR?-source=I/305}} 
\citep[GSC2.3]{Lasker2007}
should not be used in the framework of this paper.
As stated in Table 3 of \cite{GSC2008}:
These astrometric and photometric errors are not formal statistical
uncertainties but a raw and conservative estimate to be used for telescope
operations.

Table \ref{tab:catserr} summarizes the transformation of catalogues positional errors
into the coefficients of covariance matrices $\bm{V}$.
\begin{table*}
  \caption{Summary of the transformations of positional errors provided in various astronomical catalogues
    into the coefficients of error covariance matrices (before adding quadratically possible systematics).}
  \label{tab:catserr}      
  \centering          
  \begin{tabular}{l|cccccc} 
     \hline\hline  
                           & 2MASS / FIRST$^1$              & AllWISE          & SDSS   & XMM /GASC                    & 1RXS & WGACAT \\ \hline
    $\sigma_x$             & $\sqrt{a^2\sin^2\psi+b^2\cos^2\psi}$ & $\sigma_\alpha$  & raErr  & $\frac{posErr}{\sqrt{2}}$ & $\frac{r_{68\%}}{\sqrt{2\ln(3.1515)}}$ & $\frac{r_{90\%}}{\sqrt{2\ln(10)}}$ \\
    $\sigma_y$             & $\sqrt{a^2\cos^2\psi+b^2\sin^2\psi}$ & $\sigma_\delta$  & decErr & $\frac{posErr}{\sqrt{2}}$ & $\frac{r_{68\%}}{\sqrt{2\ln(3.1515)}}$ & $\frac{r_{90\%}}{\sqrt{2\ln(10)}}$ \\
    $\rho\sigma_x\sigma_y$ & $\cos\psi\sin\psi(a^2-b^2)$      & $\sigma_{\alpha\delta}\times|\sigma_{\alpha\delta}|$    & 0  & 0                           & 0                                       & 0 \\
    \hline                  
  \end{tabular}\\
  \footnotesize{$^1$ In FIRST, the $90\%$ confidence ellipse semi-axes must be first divided by $\sqrt{2\ln(10)}$ to obtain the $39.347\%$ confidence ellipse.}
\end{table*}

\section{Candidates selection: the $\rchi$-match \label{sec:candselect.all}}
We make the hypothesis that $n$ sources from $n$ distinct catalogues are
$n$ independent detections of a same real source.
With $\vec{p}$ the unknown position of the real source and $\vec{\mu_i}$
the observed position of detection $i$, 
the probability for the $n$ detections to be located at the observed positions
is expressed by the joint density function:
\begin{eqnarray}
  f_p(\vec{\mu_1},\vec{\mu_2},...,\vec{\mu_n}|\vec{p})
    & = & \prod\limits_{i=1}^n\mathcal{N}_{\vec{\mu_i},\bm{V_i}}(\vec{p})
          \nonumber, \\
    & = & \frac{\exp\left\{-\frac{1}{2} \sum\limits_{i=1}^nQ_i(\vec{p})\right\}}
               {(2\pi)^n\prod\limits_{i=1}^n\sqrt{\det\bm{V_i}}}
	  \mathrm{d}\vec{p}.
  \label{eq:likelihood}
\end{eqnarray}

\subsection{Estimation of the real position given $n$ observations}

We introduce the notations $\vec{\mu_\Sigma}$ and $\bm{V_\Sigma}$
for the weighted mean position of the $n$ sources and its associated error respectively.
The inverse of the covariance matrix $\bm{V_\Sigma}$ is
\begin{equation}
  \bm{V_\Sigma^{-1}} = \sum\limits_{i=1}^n\bm{V_i}^{-1},
  \label{eq:VSigma}
\end{equation}
leading to (see demonstration in \S \ref{sec:vsigma})
\begin{equation}
  \bm{V_\Sigma} = \frac{1}{\det\bm{V_\Sigma^{-1}}}
                  \sum\limits_{i=1}^n\frac{\bm{V_i}}{\det\bm{V_i}} 
  \label{eq:wposerr}
\end{equation}
which is used in the weighted mean position expression
\begin{equation}
  \vec{\mu_\Sigma} = \bm{V_\Sigma} \sum\limits_{i=1}^n\bm{V_i}^{-1}\vec{\mu_i}.
  \label{eq:wmeanpos}
\end{equation}
Using both the weighted mean position and its error,
the sum of quadratics in Eq. (\ref{eq:likelihood}) can be divided into two parts and written as
(see demonstration \S\ref{sec:quad})
\begin{eqnarray}
    \sum\limits_{i=1}^n Q_i(\vec{p})
      & = & \sum\limits_{i=1}^n \transposee{(\vec{p}-\vec{\mu_i})}\bm{V_i}^{-1}
                                            (\vec{p}-\vec{\mu_i}),
	    \label{eq:sumqi} \\
      & = & Q_p(\vec{p};\vec{\mu_1},\vec{\mu_2},...,\vec{\mu_n})
          + Q_{\rchi^2}(\vec{\mu_1},\vec{\mu_2},...,\vec{\mu_n}),
\end{eqnarray}
with
\begin{eqnarray}
  Q_p(\vec{p};\vec{\mu_1},\vec{\mu_2},...,\vec{\mu_n})
    & = & \transposee{(\vec{p}-\vec{\mu_\Sigma})}\bm{V_\Sigma}^{-1}(\vec{p}-\vec{\mu_\Sigma}), \\
  Q_{\rchi^2}(\vec{\mu_1},\vec{\mu_2},...,\vec{\mu_n})
    & = & \sum\limits_{i=1}^n\transposee{(\vec{\mu_i} - \vec{\mu_\Sigma})}\bm{V_i}^{-1}
                                         (\vec{\mu_i} - \vec{\mu_\Sigma}).
    \label{eq:qchi2form1}
\end{eqnarray}
In the case of two catalogues the latter term can be written as in Eq. (\ref{eq:q2cats}).
Moreover, if both covariances are null, it takes the simple and common form
\begin{equation}
  Q_{\rchi^2} = \frac{\Delta \alpha^2}{\sigma_{\alpha_1}^2 + \sigma_{\alpha_2}^2} + \frac{\Delta \delta^2}{\sigma_{\delta_1}^2 + \sigma_{\delta_2}^2}.
\end{equation}
Back to the general case, the term $Q_{\rchi^2}$ can also be put in the more
computationally efficient form (only one loop over $i$)
\begin{eqnarray}
  Q_{\rchi^2}(\vec{\mu_1},\vec{\mu_2},...,\vec{\mu_n})
    & = & \sum\limits_{i=1}^n\transposee{\vec{\mu_i}}\bm{V_i}^{-1}\vec{\mu_i}
	    - \transposee{\vec{\mu_\Sigma}}\bm{V_\Sigma}^{-1}\vec{\mu_\Sigma}.
    \label{eq:qchi2form2}
\end{eqnarray}

From those formulae, it appears that the weighted mean position
(Eq. \ref{eq:wmeanpos}) is 
the maximum likelihood estimator of the ``true'' position of the source:
the second term ($Q_{\rchi^2}$) is constant with respect to $\vec{p}$
so the maximum of the likelihood
function $\mathcal{L}(\vec{p};\vec{\mu_1},\vec{\mu_2},...,\vec{\mu_n})=
f_p(\vec{\mu_1},\vec{\mu_2},...,\vec{\mu_n}|\vec{p})$
is obtained when the first term ($Q_p$) is null, so when
$\vec{p}=\vec{\mu_\Sigma}$.
The error on this estimate is simply $\bm{V_\Sigma}$, the inverse of
the Hessian of the likelihood function.

\subsection{Candidates selection criterion \label{sec:candselect}}

For the candidate selection, we are interested in the probability the 
$n$ sources have to be located at the same position.
Let's first rewrite Eq. (\ref{eq:likelihood}) to exhibit a product of a
binormal distribution by another multi-dimensional normal law:
\begin{eqnarray}
  \prod\limits_{i=1}^n\mathcal{N}_{\vec{\mu_i},\bm{V_i}}(\vec{p}) 
   & = & \frac{1}{2\pi\sqrt{\det\bm{V_\Sigma}}}
         \exp\left\{-\frac{1}{2}Q_p\right\} \times \nonumber\\
   &   & \times \frac{1}{(2\pi)^{n-1}\sqrt{
           \frac{\prod\limits_{i=1}^n\det\bm{V_i}}
                {\det\bm{V_\Sigma}}}
         }
	 \exp\left\{ -\frac{1}{2}Q_{\rchi^2} \right\}.
  \label{eq:likely2components}
\end{eqnarray}
When integrating Eq. (\ref{eq:likely2components}) over all possible positions
(i.e. over $\vec{p}$) the first term integrates to 1, since it is the p.d.f of
a normal law in $\vec{p}$, so we obtain
\begin{equation}
  \int\int
    \prod\limits_{i=1}^n\mathcal{N}_{\vec{\mu_i},\bm{V_i}}(\vec{p})
    \text{d}\vec{p}
  = \sqrt{\frac{\det\bm{V_\Sigma}}{\prod\limits_{i=1}^n\det\bm{V_i}}}
    \frac{\exp\left\{-\frac{1}{2}Q_{\rchi^2}\right\}}{(2\pi)^{n-1}}.
  \label{eq:gausschi2}
\end{equation} 
We are supposed to integrate on the surface of the unit sphere.
But the errors being small, we consider the infinity being at a relatively
close distance, before effects of the sphere curvature become non-negligible. \\
In the previous equation, only the $Q_{\rchi^2}$ term remains. 
It can also be written (see demonstration \S\ref{sec:q2alldist})
\begin{equation}
  Q_{\rchi^2}(\vec{\mu_1},\vec{\mu_2},...,\vec{\mu_n})
    = \sum\limits_{i=1}^n\sum\limits_{j=i+1}^n
      \transposee{(\vec{\mu_i}-\vec{\mu_j})}
      \bm{V_i^{-1}}\bm{V_\Sigma}\bm{V_j^{-1}}
      (\vec{\mu_i}-\vec{\mu_j}).
  \label{eq:qchi2form3}
\end{equation}
Eq. (\ref{eq:gausschi2}) is equivalent to $P(D|H)$ in \cite{Budavari2008}
and Eq. (\ref{eq:qchi2form3}) -- multiplied by the $-\frac{1}{2}$ factor in
the exponential (Eq. \ref{eq:gausschi2}) -- is the generalization for elliptical
errors of  Eq. (B12) in \cite{Budavari2008}.
In practice, we never use Eq. (\ref{eq:gausschi2}) since the number of terms to be computed
increases with  $O(n(n-1)/2)$ while it increases with $O(n)$ in Eq. (\ref{eq:qchi2form2})
or in its iterative form (see Eq. (\ref{eq:qchi2iter}) in \S \ref{sec:iterativeform}).
We use here the big O notation, to be read as ``the order of''.

We can see Eq. (\ref{eq:qchi2form1}) as the result of a $2n$-dimensional weighted
least squares in which the model is the ``real'' position of the source and
the solution is $\vec{\mu_\Sigma}$ (by similarity with Eq. \ref{eq:sumqi}).
Putting all positional errors matrices in a $2n\times 2n$ block diagonal matrix $\bm{M}$, 
$Q_{\rchi^2}$ is the square of the Mahalanobis distance $D_M^2(\vec{\mu})$ defined by
\begin{equation}
  D_M^2(\vec{\mu}) = Q_{\rchi^2}(\vec{\mu})
                   = \transposee{\vec{v}}\bm{M^{-1}}\vec{v},
  \label{eq:dm}
\end{equation}
\begin{equation}
  \vec{v} =
  \begin{pmatrix}
    \vec{\mu_1} - \vec{\mu_\Sigma} \\
    \vec{\mu_2} - \vec{\mu_\Sigma} \\
    \vdots      \\
    \vec{\mu_n} - \vec{\mu_\Sigma} 
  \end{pmatrix}
  \mathrm{, }
  \bm{M^{-1}} = 
  \begin{pmatrix}
    \bm{V_1}^{-1} & 0             & \hdots & 0 \\
                0 & \bm{V_2}^{-1} & \ddots & 0 \\
           \vdots & \ddots        & \ddots & 0 \\
                0 & \hdots        & 0 &\bm{V_n}^{-1}
  \end{pmatrix},
\end{equation}
which follows in our particular case a $\rchi^2$ distribution with
$2(n-1)$ degrees of freedom, or equivalently, $(n-1)$ $\rchi^2$ distributions with
two degrees of freedom.
Eq. (\ref{eq:dm}) is probably the Mahalanobis distance mentioned without giving its expression in \cite{Adorf2006}.\\
If $D_M^2(\vec{\mu})$ follows a $\rchi_{dof=2(n-1)}^2$ distribution, 
then its square root, the distance $D_M(\vec{\mu})$, follows a $\rchi_{dof=2(n-1)}$ distribution.

We perform a statistical hypothesis test on a set of $n$ sources,
defining the null hyothesis $H_0$ as follows:
all sources in the set are detections of the same ``real'' source.
The alternative hypothesis $H_1$ would thus be:
not all sources in the set are detections of the same ``real'' source;
in other words the set of $n$ sources contains at least one spurious source;
or, expressed differently, the $n$ sources are $n$ observations of at least two
distinct real sources.
We adopt Fisher's approach, that is we will reject the null hypothesis if, the null
hypothesis being true, the observed data is significantly unlikely.\\
From now on, we indifferently write $x$ or $D_M$ the Mahalanobis distance.
Assuming the null hypothesis is true, the ``theoretical'' probability we had to
get the actual computed (square of) Mahalanobis distance is given by a Chi(-square)
distribution with $2(n-1)$ degrees of freedom:
\begin{eqnarray}
  p(X=x) & = & \rchi_{dof=2(n-1)}(X=x)\mathrm{d}X, \\
  p(X=x^2) & = & \rchi_{dof=2(n-1)}^2(X=x^2)\mathrm{d}X.
\end{eqnarray}
The probability we had to get an actual computed (square of) Mahalanobis
distance less than or equal to a given threshold (or critical value) $k_\gamma^{(2)}$
is given by the value of the cumulative distribution function of a the Chi(-square)
at the given threshold
\begin{equation}
  \gamma = \int_0^{k_\gamma^2}p(X)
         = \int_0^{k_\gamma^2}\rchi_{2(n-1)}^2(X)\mathrm{d}X
	 = F_{\rchi_{2(n-1)}^2}(k_\gamma^2).
\end{equation}
We can indifferently work on $x$ with the $\rchi$ distribution
or on $x^2$ with the $\rchi^2$ distribution.
The threshold $k_\gamma$ we obtain on $x$ is simply the square root of
the threshold $k_\gamma^2$ we obtain on $x^2$.
Although we find the Chi test more natural in the present case,
most astronomers are familiar with the Chi-square test. \\
In the framework of statistical hypothesis tests, it is the
complementary cumulative distribution (or tail distribution) function
which is usually used by defining the $p$-value
\begin{equation}
  p\mathrm{-value} = \int_{x^2}^{+\infty}\rchi_{k=2(n-1)}^2(X)\mathrm{d}X
          = 1 -  F_{\rchi_{2(n-1)}^2}(x^2),
\end{equation}
and a significance level $\alpha$ defined by
\begin{equation}
  \alpha = \int_{k_\gamma^2}^{+\infty}\rchi_{k=2(n-1)}^2(X)\mathrm{d}X
	 = 1 - F_{\rchi_{2(n-1)}^2}(k_\gamma^2)
	 = 1 - \gamma
  \label{eq:alpha}
\end{equation}
is fixed.
The null hypothesis is then rejected if $p\mathrm{-value}<\alpha$.
In the Neyman-Pearson framework $\alpha$ is the type I error, or
the false positive rate, that is the probability the null hypothesis has to be rejected 
(positive rejection test)  while it is true (wrong/false decision).
In our case we fix $\gamma$ (hereafter called completeness),
the fraction of real associations we ``theoretically''
select over all real associations. 
The candidates selection criterion, or fail of rejection criterion, we
use is then
\begin{equation}
  D_M(\vec{\mu}) \le k_\gamma
  \label{eq:candselectcri}
\end{equation}
in which $k_\gamma^2=F_{\rchi_{2(n-1)}^2}^{-1}(\gamma)$
or, equivalently, $k_\gamma=F_{\rchi_{2(n-1)}}^{-1}(\gamma)$.
This inequality is equivalent to $p\mathrm{-value}<\alpha$.
It is important to write ``fail of rejection'' since nothing proves
that if Eq. (\ref{eq:candselectcri}) is satisfied the null hypothesis is true:
at this point the selected set of sources is nothing else than a set of
candidates.
Nevertheless we do call region of acceptance the set of $D_M(\vec{\mu})$
values satisfying Eq. (\ref{eq:candselectcri}).
This region of acceptance
will be useful to define the domain of integration used to normalize likelihoods
when computing probabilities for each hypothesis from \S \ref{sec:spurest}.
Its volume (see e.g. Eq. \ref{eq:ellvolume}) is the volume of the $2n$-ellipsoid
defined by $M$ (see Eq. \ref{eq:dm}) divided by the error ellipse associated
to the weighted mean position $\mu_\Sigma$ and defined by $\bm{V_\Sigma}$ 
(it thus is a volume in a $2(n-1)$ space).

In practice, the value $k_\gamma^2$ is computed numerically using Newton's method
to solve $F_{\rchi_{2(n-1)}^2}(X)-\gamma=0$.
The initial guess we use is the approximate value returned by Eq. (A.3) of \cite{Inglot2010}.

The value of $\gamma$ we fix is independent of $n$,
the number of candidates.
In practice we often set this input parameter to $\gamma=0.9973$.
In one dimension this value leads to $k_\gamma=3$, that is the famous $3\sigma$ rule.
It means that for 10\,000 real associations in a dataset, we theoretically
miss 27 of them by applying the candidate selection criterion.
From now on we call this cross-correlation a $\rchi_\gamma$-match, or simply a
$\rchi$-match.

In the particular case of two catalogues $D_M(\vec{\mu})$ follows
a $\rchi$ distribution with 2 degrees of freedom -- that is a Rayleigh distribution --
and $k_{\gamma=0.9973}=3.443935$.
This latter value is used in the two-catalogues $\rchi$-match of \cite{Pineau2011}.

\subsection{Iterative form: catalogue by catalogue\label{sec:iterativeform}}

  Somewhat similarly to the Bayes factor in \citet[\S 6]{Budavari2008}
  it is noteworthy that $Q_{\rchi^2}$ can be computed iteratively,
  summing $(n-1)$ successive $\rchi^2$ with two degrees of freedom computed
  from $(n-1)$ successive two-catalogues cross-matches.\\
  After each iteration, the new position to be used for the next
  cross-match is the weighted mean of all already matched positions
  and the new associated error is the error on this weighted mean.
  The strict equality between Eq. (\ref{eq:qchi2iter}) and
  the non iterative form, for example Eq. (\ref{eq:qchi2form1}), proves that the result
  is independent of the successive cross-matches order.\\
  The maximum number of cross-matches to be performed must be known in advance
  in order to put an upper limit on $k_\gamma$ since it depends on the degree
  of freedom of the total $\rchi^2$.
  The iteration formula is simply
  \begin{equation}
    Q_{\rchi^2} = \sum\limits_{i=2}^n
      \transposee{(\vec{\mu_{\Sigma_{i-1}}} - \vec{\mu_i})}
                  (\bm{V_{\Sigma_{i-1}}}+\bm{V}_i)^{-1}
                  (\vec{\mu_{\Sigma_{i-1}}} - \vec{\mu_i})
    \label{eq:qchi2iter}
  \end{equation}
  in which
  \begin{eqnarray}
    \bm{V_{\Sigma_{i-1}}}^{-1} & = & \sum\limits_{k=1}^{i-1}\bm{V_k}^{-1}, \label{eq:iterror} \\
    \vec{\mu_{\Sigma_{i-1}}} & = & \bm{V}_{\Sigma_{i-1}}
                            \sum\limits_{k=1}^{i-1}\bm{V_k}^{-1}\vec{\mu_k}. 
  \end{eqnarray}
  We find it from the 2-catalogues case, for which (see \S \ref{sec:sum2q})
  \begin{equation}
    Q_{\rchi^2} = \transposee{(\vec{\mu_1} - \vec{\mu_2})}
                  (\bm{V}_1+\bm{V}_2)^{-1}
                  (\vec{\mu_1} - \vec{\mu_2}).
    \label{eq:q2cats}
  \end{equation}
  We can demonstrate by direct calculation that
  \begin{equation}
    \det (\bm{V_1} + \bm{V_2}) \det \bm{V_{\Sigma_2}} = \det \bm{V_1} \det \bm{V_2}
    \label{eq:detequals}
  \end{equation}
  and so, iteratively, we find the general expression
  \begin{equation}
    \prod_{i=2}^n \det(\bm{V_{\Sigma_{i-1}}} + \bm{V_i})
    = \prod_{i=2}^n \frac{\det \bm{V_{\Sigma_{i-1}}} \det \bm{V_i}}{ \det \bm{V_{\Sigma_i}} }
    = \frac{\prod\limits_{i=1}^n\det \bm{V_i}}{\det \bm{V_{\Sigma}}}
    \label{eq:itdet}
  \end{equation}
  which is consistent with Eq. (\ref{eq:gausschi2}).
  The volume of the acceptance region of the statistical hypothesis test
  is the volume of a $2(n-1)$ dimensional ellipsoid. More precisely, it
  is the product of the previous equation Eq. (\ref{eq:itdet}) by the volume
  of a $2(n-1)$-sphere of radius $k_\gamma$.
  This will be crucial when computing the rate of spurious associations.

\subsection{Iterative form: by groups of catalogues \label{sec:candselectbygroups}}

  Instead of iterating over catalogues one by one, we can also perform
  $G$ sub-cross-matches, each associating 
  $n_g$ distinct sources such that $\sum\limits_{g=1}^G n_g = n$.
  We note $Q_{\rchi^2,\{g\}}$ the square of the Mahalanobis distance associated
  with the group $g$:
  \begin{equation}
    Q_{\rchi^2,\{g\}} = \sum\limits_{i=2}^{n_g}
                   \transposee{(\vec{\mu_{\Sigma_{i-1}}} - \vec{\mu_i})}
                  (\bm{V}_{\Sigma_{i-1}}+\bm{V}_i)^{-1}
                  (\vec{\mu_{\Sigma_{i-1}}} - \vec{\mu_i}).
  \end{equation}
  We show that we can compute $Q_{\rchi^2}$ iteratively from the
  $G$ weighted mean positions $\vec{\mu_{\Sigma_{\{g\}}}}$
  and their associated errors $\bm{V}_{\Sigma_{\{g\}}}^{-1}$.
  The square of the Mahalanobis distance can be written
  \begin{eqnarray}
    Q_{\rchi^2} & = & \sum_{g=2}^G
      \transposee{(\vec{\mu_{\Sigma_{g-1}}} - \vec{\mu_{\Sigma_{\{g\}}}})}
                  (\bm{V}_{\Sigma_{g-1}}+\bm{V}_{\Sigma_{\{g\}}})^{-1}
		  (\vec{\mu_{\Sigma_{g-1}}} - \vec{\mu_{\Sigma_{\{g\}}}}) \nonumber \\
               &  & + \sum\limits_{g=1}^G
	       Q_{\rchi^2,\{g\}}.
    \label{eq:qchi2iterG}
  \end{eqnarray}
  In other words, the square of the Mahalanobis distance is the sum of the
  square of the intra-group Mahalanobis distances plus the inter-group iterative one.
  With $k$ being an index defined inside each of the $G$ groups $\{g\}$
  \begin{eqnarray}
    \bm{V_{\Sigma_{\{g\}}}}^{-1} & = & \sum\limits_{k=1}^{n_g} \bm{V_k}^{-1}, \\
    \vec{\mu_{\Sigma_{\{g\}}}} & = & \bm{V_{\Sigma_{\{g\}}}}
                            \sum\limits_{k=1}^{n_g}\bm{V_k}^{-1}\vec{\mu_k}, \\
    \bm{V_{\Sigma_{g-1}}}^{-1} & = & \sum\limits_{g'=1}^{g-1}
                                \sum\limits_{k=1}^{n_{g'}}
				 \bm{V_k}^{-1}, \\
    \vec{\mu_{\Sigma_{g-1}}} & = & \bm{V_{\Sigma_{g-1}}} 
                                \sum\limits_{g'=1}^{g-1}
                                \sum\limits_{k=1}^{n_{g'}}
                                \bm{V_k}^{-1} \vec{\mu_k}.
  \end{eqnarray}
  In fact, it is a straightforward generalization of the $G=2$ groups case
  for which
  \begin{eqnarray}
    \bm{V_{\Sigma_{g=1}}}^{-1}
      & = & (\bm{V_{\Sigma_{\{1\}}}} + \bm{V_{\Sigma_{\{2\}}}})^{-1}
        = \sum\limits_{i=1}^n \bm{V_{\Sigma_i}}
        = \bm{V_\Sigma}^{-1}, \\
    \vec{\mu_{\Sigma_{g=1}}} & = & \bm{V_{\Sigma_{g=1}}}( 
            \bm{V_{\Sigma_{\{1\}}}}^{-1} \vec{\mu_{\Sigma_{\{1\}}}}
	  + \bm{V_{\Sigma_{\{2\}}}}^{-1} \vec{\mu_{\Sigma_{\{2\}}}}), \\
      & = & \bm{V_\Sigma} \sum\limits_{i=1}^n \bm{V_i}^{-1} \vec{\mu_i}
        = \vec{\mu_\Sigma}.
  \end{eqnarray}
  Here again,
  \begin{equation}
    \prod_{i=2}^{G} \det(V_{\Sigma_{g-1}} + V_{\Sigma_g})
     = \prod_{g=2}^{G} \frac{ \det V_{\Sigma_{g-1}} \prod\limits_{k=1}^{n_g} \det V_k }{\det V_{\Sigma_{g}}}
     = \frac{\prod\limits_{i=1}^n\det V_i}{\det V_{\Sigma_n}}.
      \label{eq:detprodgroups}
  \end{equation}

  Again, $k_\gamma$
  depends on the number of degrees of freedom of the total
  $\rchi^2$, thus on the total number of cross-correlated tables.
  It means that to be complete, all sub-cross-correlations must use
  the candidate selection threshold $k_\gamma(2(n-1))$ computed from the total number of tables
  instead of $k_\gamma(2(n_g-1))$ computed from the number of tables in a group.

\subsection{Summary and Interpretation}

  Equations (\ref{eq:qchi2form1}), (\ref{eq:qchi2form2}),
  (\ref{eq:qchi2form3}), (\ref{eq:dm}), (\ref{eq:qchi2iter})
  and (\ref{eq:qchi2iterG}) are all equivalent and they lead to the
  same value, that is to the same squared Mahalanobis distance.
  All sources are retained as possible candidates if Eq. (\ref{eq:candselectcri})
  is verified, so if the Mahalanobis distance is smaller or equal to
  $k_\gamma$. This threshold is the inverse of
  the cumulative $\rchi$ distribution function at the chosen completeness $\gamma$,
  for $2(n-1)$ degrees of freedom.

  As this criterion is no other than a $\rchi$-test criterion
  (or $\rchi^2$-test criterion if we work on squared Mahalanobis distances)
  we call the result of such a criterion a $\rchi$-match.
  
  The $\rchi$-match criterion defines a region of acceptance which is
  a $2(n-1)$-ellipsoid of radius $k_\gamma$.
  Its volume is computed from Eq. (\ref{eq:itdet}):
  \begin{equation}
    \mathcal{V}_n(k_\gamma) = \left[\frac{\prod\limits_{i=1}\det\bm{V_i}}{\det\bm{V_\Sigma}} \right]^{1/2}
                  \frac{\pi^{n-1}k_\gamma^{2(n-1)}}{(n-1)!},
    \label{eq:ellvolume}
  \end{equation}
  with $\pi^{n-1} k_\gamma^{2(n-1)} / (n-1)!$ the volume of a $2(n-1)$-sphere of radius $k_\gamma$.
  It will be later used to compute the expected number of spurious associations.

\subsection{Comment on the ``Bayesian cross-match'' of \cite{Budavari2008} }\label{sec:rmkbudav}

  We mention in \S \ref{sec:generalities} what appears to be a conceptual problem in
  calling $B$ (Eq. \ref{eq:bbudav}) a Bayes factor for more than 2 catalogues in the astrometrical part
  of \cite{Budavari2008}.\\
  Performing a cross-match by fixing a lower limit $L$ on the ``Bayes factor'' $B$
  defined in Eq. (18) of \cite{Budavari2008} is no other than performing a
  $\rchi$-match with a significance level which depends
  both on the number of sources $n$ and on the volume of the $2(n-1)$-ellipsoid of radius 1.
  In fact, using the factor $B$ of \cite{Budavari2008} in which $w_i$ is the inverse of
  the cirular error on the position of the source $i$ and $\phi_{ij}$ is the angular
  distance between sources $i$ and $j$, we have the equivalence
  \begin{eqnarray}
     & B=2^{n-1}\frac{\prod w_i}{\sum w_i} exp \left\{ -\frac{\sum_{i<j} w_i w_j \phi_{ij}^2}{2\sum w_i} \right\} \ge L \label{eq:bbudav} \nonumber\\
    \Leftrightarrow & \frac{\sum_{i<j} w_i w_j \phi_{ij}^2}{\sum w_i} \le 2 \ln\left( \frac{2^{n-1}}{L}\frac{\prod w_i}{\sum w_i} \right).
  \end{eqnarray}
  We showed that the quantity on the left side of the inequality is
  equal to Eq. (\ref{eq:qchi2form3}) in the present paper and thus follows a $\rchi^2$ distribution
  for ``real'' associations.
  It means that the ``Bayesian'' candidate selection criterion $B\ge L$ is equivalent to 
  a $\rchi^2$ test having a significance level equal to
  \begin{equation}
    \alpha = \int_{2 \ln\left( \frac{2^{n-1}}{L}\frac{\prod w_i}{\sum w_i} \right)}^{+\infty} \rchi_{2(n-1)}^2(x)\mathrm{d}x.
  \end{equation}
  The larger the volume of the $2(n-1)$-ellipsoid of radius 1 ($\propto \sum w_i/\prod w_i$),
  the more ``real'' associations are missed and the less spurious associations are retrieved.
  We could replace the criterion $B\ge L$ by $x\le 1 - \alpha(n, \prod V_i / V_\Sigma)$.
  This is somewhere between the fixed radius cone search and
  the fixed significance level $\rchi$-match.
  The rate of missed ``real'' associations is not homogeneous but depends on the positional errors.
  Only if positional errors are constant in all catalogues, then the $B\ge L$ constraint becomes
  equal to the $\rchi$-match which is equal to a fixed radius cross-match.

\section{Hypotheses from combinatorial considerations \label{sec:hypotheses}}
A $\rchi$-match output is made of sets of associations, each set of associations
containing one source per catalogue.
For each set of associations we want to compute the probability all sources of the set
have to come from a same actual source.
In this section, especially in \S \ref{sec:handbn} we make explicit the sets
$\{h_i\}$ of hypotheses we have to formulate to compute probabilities of identification
when cross-correlating $n$ catalogues.

\subsection{Generalities \label{sec:generalities}}

Given a set $\{h_k\}$ of pairwise disjoint hypotheses
whose union is the entire set of possibilities, 
the law of total probabilities for an observable $x$ is
\begin{equation}
  p(x) = \sum_{i=1}^k p(x|h_i)p(h_i).
\end{equation}
Leading to Bayes' theorem
\begin{equation}
  p(h_j|x) = \frac{p(x|h_j)p(h_j)}{\sum\limits_{i=1}^k p(x|h_i)p(h_i)}.
\end{equation}
We stress that Bayes' factor (also called likelihood ratio)
is defined only in cases involving two and only two hypotheses
\begin{equation}
 LR = K = \frac{p(x|h_1)}{p(x|h_2)},
\end{equation}
and is used when no trustworthy priors $p(h_1)$ and $p(h_2)$ are available.
We can transform any set of pairwise disjoint hypotheses into
two disjoint hypotheses. In this case, using the negation notation $\neg$
\begin{equation}
  LR = \frac{p(x|h_j)}{p(x|\neg h_j)},
  \label{eq:lrmulti}
\end{equation}
with
\begin{equation}
  p(x|\neg h_j) = \frac{\sum\limits_{i\ne j} p(x|h_i)p(h_i)}{p(\neg h_j)},
\end{equation}
and
\begin{equation}
  p(\neg h_j) = \sum\limits_{i\ne j} p(h_i).
\end{equation}
Such a likelihood ratio (Eq. \ref{eq:lrmulti}) is not interesting since it is not only computed
from likelihoods, but also from priors.\\
The term $B(H|K)$ in \citet[Eq. 8]{Budavari2008} is improperly called Bayes
factor when dealing with more than two catalogues.
As a matter of fact, the union of the two hypotheses -- all sources
are from the same real source and each sources is from a distinct real source --
is only a subset of all possibilities so the law of total probabilities
and hence Bayes' formula are not valid.

In \cite{Pineau2011} the term $LR(r)$ in Eq. (9) is also improperly called
likelihood ratio since a likelihood is a probability density function
and so integrates to 1 over its domain of definition.
It is obviously not the case of $\mathrm{d}p(r|spur)$ in Eq. (8).
The built quantity is related to the ratio between the probability the association
has to be ``real'' over the probability it has to be spurious, but formally
it is not a likelihood ratio.
The very same ``abuse of term'' is made in \cite{Wolstencroft1986} (who, moreover,
adds a prior in the likelihood ratio), in \cite{Rutledge2000}, \cite{Brusa2007}
and probably other publications.

\subsection{Possible combinations and the Bell number \label{sec:handbn}}

  Let's suppose we have
  selected one set of $n$ distinct sources from $n$ different catalogues,
  one source per catalogue.
  Those $n$ sources possibly are $n$ detections of $k$ distinct real sources,
  with $k \in [1, n]$. The case $k=1$ corresponds to the situation where
  all sources are $n$ observations of the same real source and the
  case $k=n$ corresponds to the situation where 
  there are $n$ distinct real sources detected independently, one in each catalogue. 
  
  We call $A$ the source from catalogue number one, $B$ the source from catalogue 
  number two and so on.
  
  \subsubsection{Two-catalogues case: two hypotheses \label{sec:2cathyp}}

    The classical two-catalogues case is trivial.
    We formulate only two hypotheses:
    \begin{itemize}
     \item $AB$, the match is a real match, the two sources are two
           observations of a same real source, that is $k=1$;
     \item $A\_B$, the match is spurious, the two sources are two observations
           of two different real sources, that is $k=2$.
    \end{itemize}

  \subsubsection{Three-catalogues case: five hypotheses \label{sec:3cathyp}}

    For three sources $A$, $B$ and $C$ from three different catalogues,
    we formulate five hypotheses:
    \begin{itemize}
      \item $ABC$, all three sources come from a same real source, that is $k=1$;
      \item $AB\_C$, $A$ and $B$ are from a same real source and $C$ is from
            a different real source, that is $k=2$;
      \item $AC\_B$, $A$ and $C$ are from a same real source and $B$ is from
            a different real source, that is $k=2$;
      \item $A\_BC$, $B$ and $C$ are from a same real source and $A$ is from 
            a different real source, that is $k=2$;
      \item $A\_B\_C$, all three sources are from three different real sources, that is $k=3$.
    \end{itemize}

  \subsubsection{Four-catalogues case: 15 hypotheses \label{sec:4cathyp}}

    For four sources $A$, $B$, $C$ and $D$
    we have to formulate 15 hypotheses:
    \begin{itemize}
      \item $ABCD$, when $k=1$;
      \item $ABC\_D$, $ABD\_C$, $ACD\_B$ and $BCD\_A$, but also
      \item $AB\_CD$, $AC\_BD$ and $AD\_BC$ for $k=2$;
      \item $AB\_C\_D$, $AC\_B\_D$, $AD\_B\_C$, $BC\_A\_D$, $BD\_A\_C$ and
            $DC\_A\_B$ when $k=3$;
      \item $A\_B\_C\_D$ when $k=4$.
    \end{itemize}

  \subsubsection{$n$-catalogues case: Bell number of hypotheses \label{sec:bellnumber}}

\begin{figure}[!h]
     \includegraphics[width=0.5\textwidth]{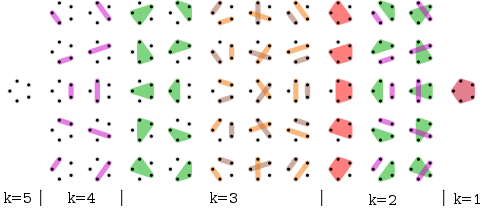}
     \caption{The 52 partitions of a set with n=5 elements. 
     Each partition corresponds to one hypothesis for five distinct sources from five distinct catalogues.
     Left: $k=5$, the five sources are from five distinct real sources.
     Right: $k=1$, the five sources are from a same real source.
     (Tilman Piesk - CC BY 3.0 - modified - link in footnote).}
     \label{fig:bell5}
   \end{figure}

    We now generalize to $n$ catalogues.
    For each possible value of $k$, the number of ways the set of $n$ sources
    can be partitioned into $k$ non-empty subsets -- each subset correspond to 
    a real source -- is given by the Stirling number of the second kind denoted
    ${n \brace k}$.
    The total number of hypotheses to be formulated is equal to the Bell number.
    The Bell number counts the number of partitions of a set and is given by
    \begin{equation}
      B_n = \sum\limits_{k=1}^n{n \brace k}
          = \sum\limits_{k=1}^{n-1} C_{n-1}^kB_k \\
          = \sum\limits_{k=1}^{n-1}\frac{(n-1)!}{(n-1-k)!k!}B_k.
      \label{eq:bellnumber}
    \end{equation}
    Its seven first values are provided in
    Table \ref{tab:bell} and a graphic illustration representing
    all possible partitions for five catalogues is provided in Fig. \ref{fig:bell5}\footnote{Original figure: \url{https://commons.wikimedia.org/wiki/File:Set_partitions_5;_cirlces.svg}}.
    \begin{table}[!h]
      \centering
      \caption{Values of the seven first Bell numbers. They provide the number of hypothesis to be formulated for a set of $n=2$ to $7$ distinct sources from different catalogues.}
      \begin{tabular}{l|rrrrrrr}
          \hline \hline
          $n$   & 1 & 2 & 3 &  4 & 5  & 6   & 7  \\ \hline
          $B_n$ & 1 & 2 & 5 & 15 & 52 & 203 & 877 \\ \hline
      \end{tabular}
      \label{tab:bell}
   \end{table}
   
   We face a combinatorial explosion of the number of hypotheses to be tested
   when increasing the number of catalogues.
   Although the theoretical developments presented here deal with any number
   of catalogues, the exhaustive analysis may be in practice limited to 
   a few catalogues ($n<10$).

   Hereafter we note $h_i$ the hypothesis number $i$, we explicit it with letters
   for example $h_{AB}$, and we note $h_{k=i}$ an hypothesis in which $n$ observed sources
   are associated to $i$ real sources. 

\section{Frequentist estimation of spurious associations rates \& priors \label{sec:spurest}}
We have defined a candidate selection criterion to perform $\rchi$-matches.
We recall that we note $x$ the Mahalanobis distance,
and we note $s$ the ``event'' $x\le k_\gamma$,
that is a given set of sources satisfies the selection criterion.

In a first step we want to estimate the number of ``fully spurious'' associations
we would expect to find in a $\rchi$-match output and derive the prior 
$p(h_{k=n}|s)$ from this estimate.
By ``fully spurious'' we mean that each candidate from each catalogue is
actually associated with a different ``real'' source.
A good such estimate is simply the mean sky area of the test
acceptance region (see Eq. \ref{eq:ellvolume}) over all possible sources of all catalogues, multiplied by
the number of sources in one of the catalogues and by the density of sources in
the other ones.
Written differently for $n$ catalogues of $n_i$ sources each,
on a common surface area $\Omega$, it leads to an estimated number of spurious
associations $\hat{n}_{\Omega_{spur}}$ equals to:
\begin{equation}
  \hat{n}_{\Omega_{spur}} = \frac{\pi^{n-1}k_\gamma^{2(n-1)}}{(n-1)!\Omega^{n-1}}
                       \sum\limits_{i_1=1}^{n_1} \sum\limits_{i_2=1}^{n_2} ...
                       \sum\limits_{i_n=1}^{n_n}
                       \left[\frac{\prod\limits_{j=1}^n \det \bm{V_{i_j}}}{\det \bm{V_\Sigma}}\right]^{1/2}.
  \label{eq:nfullyspur}
\end{equation}
Or, having histograms or more generally discretized positional error distributions:
\begin{equation}
  \hat{n}_{\Omega_{spur}} = \frac{\pi^{n-1}k_\gamma^{2(n-1)}}{(n-1)!\Omega^{n-1}}
                       \sum\limits_{b_1=1}^{N_1} \sum\limits_{b_2=1}^{N_2} ...
                       \sum\limits_{b_n=1}^{N_n}
                       \prod\limits_{k=1}^n c_{b_k}\left[\frac{\prod\limits_{j=1}^n \det \bm{V_{b_j}}}{\det \bm{V_\Sigma}}\right]^{1/2},
    \label{eq:nfullyspurdisc}
\end{equation}
in which $N_k$ are the numbers of bins in histograms
-- or number of points in a discrete distribution-- and
$c_{b_k}$ are number of counts in given bins of a histogram.
The number of counts may be replaced by the value of the discrete distribution
(or weight $w_{b_k}$) times the number of elements: $c_{b_k} = n_kw_{b_k}$.

To perform quick estimations using only a one dimensional error histogram
per catalogue, we approximate elliptical errors by circular errors of same
surface area.

The remainder of this section explains how we can compute priors from
the rate of ``fully spurious'' associations and the number of associations
found in all possible sub-cross-matches.

\subsection{Case of two catalogues}

    Let's suppose that we have two catalogues $A$ and $B$ and each
    catalogue contains only one source in the common surface area $\Omega$.
    We note $\vec{\mu}_{a1}$, $\bm{V_{a1}}$ and $\vec{\mu}_{b1}$, $\bm{V_{b1}}$
    the position of the source and associated covariance matrix in
    $A$ and $B$ respectively.
    If we fix the position $\vec{\mu}_{a1}$ of the first source, the second
    source will be associated with the first one by a $\rchi_\gamma$-match
    if Eq. (\ref{eq:candselectcri}) is satisfied.
    So if the second source is located in an ellipse of surface area
    $\pi\sqrt{\det(\bm{V_{a1}}+\bm{V_{b1}})} k_\gamma^2$ centred around
    the position of the first source.
    We temporarily waive the ISO 80000-2 notation $\det M$
    and replace it by the equivalent and more compact notation $|\bm{M}|$.
    We also replace $\bm{V_{a1}}+\bm{V_{b1}}$ by $\bm{V_{1,1}}$ to rewrite the last term in the
    pithier form $\pi|\bm{V_{1,1}}|^\frac{1}{2}k_\gamma^2$.
    We now suppose that both sources are unrelated and that $\vec{\mu}_{a1}$
    and $\vec{\mu}_{b1}$ are uniformly distributed in $\Omega$.
    Then, neglecting border effects, the probability that the two sources are
    associated by chance when performing a $\rchi_\gamma$-match
    is given by the ratio of the acceptance ellipse to the total surface
    area $\Omega$:
    \begin{equation}
      p = \frac{\int_{x=0}^{x=k_\gamma} \mathrm{d}(\vec{\mu_{a1}}-\vec{\mu_{b1}})}{\Omega}
        = \frac{|V_{1,1}|^\frac{1}{2}\int_{0}^{k_\gamma}\int_{0}^{2\pi}x\mathrm{d}x\mathrm{d}\theta}{\Omega}
        = \frac{\pi|V_{1,1}|^\frac{1}{2}k_\gamma^2}{\Omega}.
    \end{equation}
    We now suppose that the second catalogue contains $n_B$ sources uniformly
    distributed in $\Omega$. And if all of them are unrelated to the source of the
    first catalogue, then the estimated number of spurious associations is simply
    the sum of the previous probability over the $n_B$ sources of the second catalogue
    \begin{equation}
      \hat{n}_{A\_B} = \sum_{j=1}^{n_B} p_{1,j} = \sum_{j=1}^{n_B}
        \frac{\pi|V_{1,j}|^\frac{1}{2}k_\gamma^2}{\Omega}.
    \end{equation}
    We now suppose that the first catalogue contains $n_A$ sources also
    uniformly distributed in $\Omega$, all unrelated to catalogue $B$ sources.
    Still neglecting border effects, the estimated number of spurious associations
    is simply the sum of the previous estimation over all catalogue $A$ sources 
    \begin{equation}
      \hat{n}_{A\_B} = \sum_{i=1}^{n_A}\sum_{j=1}^{n_2} p_{i,j} =
        \sum_{i=1}^{n_1}\sum_{j=1}^{n_B}
          \frac{\pi|V_{i,j}|^\frac{1}{2}k_\gamma^2}{\Omega}.
      \label{eq:noise}
    \end{equation}

    In practice, evaluating this quantity can be time-consuming since
    we have to compute and sum $n_A \times n_B$ terms.
    Fortunately, we can evaluate it exactly for circular errors and
    approximately for elliptical errors computing only $n_A+n_B$ terms.
    In fact
    \begin{eqnarray}
      |V_{i,j}|^\frac{1}{2}
        & = & |V_{ai}+V_{bj}|^\frac{1}{2} \\
        & = & \left(|V_{ai}|^\frac{1}{2} + |V_{bj}|^\frac{1}{2}\right)
              \sqrt{1+\frac{C}{ (|V_{ai}|^\frac{1}{2} + |V_{bj}|^\frac{1}{2})^2}} \\
        & \approx & \left(|V_{ai}|^\frac{1}{2} + |V_{bj}|^\frac{1}{2}\right) \times\\
        & &  \times \left(
                1+\frac{1}{2}\frac{C}{ (|V_{ai}|^\frac{1}{2} + |V_{bj}|^\frac{1}{2})^2}
                -\dots+\dots
              \right)
    \end{eqnarray}
    in which
    \begin{eqnarray}
      C  & = & (\sigma_{x_i}\sigma_{y_j} - \sigma_{y_i}\sigma_{x_j})^2 \nonumber \\
         &   & + 2\sigma_{x_i}\sigma_{y_i}\sigma_{x_j}\sigma_{y_j}
              \left(
                1 + \rho_i\rho_j - \sqrt{(1-\rho_i^2)(1-\rho_j^2)}
              \right),
    \end{eqnarray}
    and thus
    \begin{equation}
      C =
      \begin{cases}
        (\sigma_{x_i}\sigma_{y_j} - \sigma_{y_i}\sigma_{x_j})^2,& \text{if } \rho_i=\rho_j=0; \\
        0,& \text{if errors are circular}.
      \end{cases}
    \end{equation}
    For ordinary ellipses, that is ellipses having a position angle
    different from 0 and $\pi/2$, the approximation is valid if
    $C\ll (|V_{ai}|^\frac{1}{2} + |V_{bj}|^\frac{1}{2})^2$.
    In the particular case of circular errors, Eq. (\ref{eq:noise}) becomes
    \begin{equation}
      \hat{n}_{A\_B} = n_A n_B k_\gamma^2
        \frac{ \overline{\Omega}_{e_A} + \overline{\Omega}_{e_B} }{\Omega},
       \label{eq:estimnspur1}
    \end{equation}
    in which $\overline{\Omega}_{e_A}$ and $\overline{\Omega}_{e_B}$ are the mean surface
    area of all positional error ellipses in catalogues $A$ and $B$ respectively:
    \begin{equation}
      \overline{\Omega}_{e_A} = \frac{1}{n_A}\sum_{i=1}^{n_A}\pi|V_{ai}|^\frac{1}{2}
      \text{ and }
      \overline{\Omega}_{e_B} = \frac{1}{n_B}\sum_{i=1}^{n_B}\pi|V_{bi}|^\frac{1}{2}.
    \end{equation}
    For simple circular errors $\sigma_{ai}$ and $\sigma_{bi}$, this reduces to
    \begin{equation}
      \overline{\Omega}_{e_A} = \frac{1}{n_A}\sum_{i=1}^{n_A}\pi\sigma_{ai}^2
      \text{ and }
      \overline{\Omega}_{e_B} = \frac{1}{n_B}\sum_{i=1}^{n_B}\pi\sigma_{bi}^2.
    \end{equation}
    If errors are constant for all sources in each catalogue, this reduces to
    \begin{equation}
      \overline{\Omega}_{e_A} = \pi\sigma_{a}^2 \text{ and }
      \overline{\Omega}_{e_B} = \pi\sigma_{b}^2.
    \end{equation}
    These estimates based on geometrical considerations have the advantage
    of being very fast to compute.\\

    Theoretically, we should remove from the double summation in Eq. (\ref{eq:noise})
    the pairs $(i, j)$ which are real associations.
    We have no mean to do this since we do not known in advance the result of the cross-identification. 
    Fortunately this effect is negligible in common cases.
    Indeed, if the result of the cross-match of the two catalogues contains $n_{AB}$
    real associations -- that is sources of both catalogues from a same
    real source -- and supposing that the positional error distribution
    of sources having a counterpart is similar to the global error distribution
    we should modify Eq. (\ref{eq:estimnspur1}) by
    \begin{equation}
      \hat{n}_{A\_B} = (n_A n_B - n_{AB}) k_\gamma^2
          \frac{\overline{\Omega}_{e_A}+\overline{\Omega}_{e_B}}{\Omega}.
      \label{eq:estimnspur2}
    \end{equation}
    In practice this estimate will tend to be overestimated since the distribution
    of sources in a catalogue cannot be uniform because of the limited angular resolution
    preventing the detection of very close sources in a same image.
    This effect is usually deemed to be of negligible importance.
    However one can detect its presence in particular circumstances.
    For instance, if the actual counterpart is located in the wings of a much brighter nearby source it may not be detected.
    This effect probably accounts for the presence of a fraction of the stellar identifications in high Galactic latitude X-ray surveys,
    in particular those with a much higher $F_x/F_{opt}$ flux ratios and harder X-ray spectra than normal for active coronae
    in which cases a faint AGN may be the correct identification \citep{Watson2012,Menzel2016}.
    One way to account for this effect and to limit the overestimation
    is to remove from the surface area $\Omega$ small
    areas around each source. The value of those areas depends for example on the 
    source brightness.
    In addition, again because of the angular resolution:
    for real associations in catalogues having similar positional errors,
    the chance a source has to be also associated with a spurious source is low.
    More precisely, the start of the Poisson distribution will be truncated.
    In extreme cases in which the Poisson distribution is truncated for $x<k_\gamma$,
    meaning that sources in a real association cannot be part of a spurious association,
    we should remove those sources from the estimate $\hat{n}_{A\_B}$.
    We thus have to rewrite the previous equation Eq. (\ref{eq:estimnspur2}) as
    \begin{equation}
      \hat{n}_{A\_B} = (n_A - n_{AB}) (n_B - n_{AB}) k_\gamma^2
          \frac{\overline{\Omega}_{e_A}+\overline{\Omega}_{e_B}}{\Omega}.
      \label{eq:estimnspur3}
    \end{equation}
    
    Knowing the total number of associations, $n_T$, resulting from the $\rchi$-match,
    we can estimate from Eq. (\ref{eq:estimnspur2}) the number of spurious associations,
    and thus the number of real associations is estimated by
    \begin{equation}
      \hat{n}_{AB} = \frac{n_T - n_A n_B k_\gamma^2\frac{\overline{\Omega}_{e_A}+\overline{\Omega}_{e_B}}{\Omega}}
                 {1 -       k_\gamma^2\frac{\overline{\Omega}_{e_A}+\overline{\Omega}_{e_B}}{\Omega}}.
    \end{equation}
    If mean error ellipses in both catalogues are very small compared to the
    total surface area -- that is$\overline{\Omega}_{e_A}+\overline{\Omega}_{e_B}<<\Omega$ --
    we can use the approximation
    \begin{equation}
     \hat{n}_{AB} \approx n_T - n_A n_B k_\gamma^2\frac{\overline{\Omega}_{e_A}+\overline{\Omega}_{e_B}}{\Omega},
    \end{equation}
    which is equivalent to using directly equation Eq. (\ref{eq:estimnspur1}),
    that is without taking care of removing real associations.
    $\hat{n}_{AB}$ is but an estimate and nothing prevents it from being
    negative due to count statistics in cross-matches with very few real associations and a lot of
    spurious associations. In practice, we have to define a lower limit
    such as $\hat{n}_{AB}>0$.

    Hence we can estimate the priors in the sample of associations satisfying the 
    selection criterion ($s$)
    \begin{eqnarray}
      p(h_{AB}|s)   & = & \frac{ \hat{n}_{AB} }{ n_T }, \label{eq:estimatedPriorAB} \\
      p(h_{A\_B}|s) & = & 1 - p(h_{AB}|s). \label{eq:estimatedPriorApasB}
    \end{eqnarray}

    After a first two-catalogues cross-match, we may compare the expected
    histogram of $\det\bm{V_{i,j}}$ for spurious associations with
    the same histogram obtained from all associations.
    We may then derive the estimated distribution of this quantity
    ($\det\bm{V_{i,j}}$) for ``real'' associations
    and compute the two likelihoods $p(\det\bm{V_{i,j}}|h_{AB},s)$ and
    $p(\det\bm{V_{i,j}}|h_{A\_B},s)$.

    Similarly we may build the histograms of the quantity
    $\det\bm{V_\Sigma}$ for both the spurious and the ``real'' associations.
    This quantity is the determinant of the covariance matrix
    -- that is the positional error --  associated with the weigthed mean positions.
    We proceeded likewise in \cite{Pineau2011} using the
    ``likelihood ratio'' (see our comment on the abuse of term likelihood ratio)
    quantity instead of positional uncertainties.

\subsection{Case of three catalogues}

  We recall that for $3$ catalogues, the output contains five components (see \S \ref{sec:3cathyp}):
  $ABC$, $AB\_C$, $A\_BC$, $AC\_B$, $A\_B\_C$.
  We would like to estimate the number of spurious associations, that is the
  number of associations in the four components other than $ABC$.
  To do so, we need to perform the three two-catalogue cross-matches
  $A$ with $B$, $A$ with $C$ and $B$ with $C$.
  We are thus able to estimate $n_{AB}$ and $n_{A\_B}$, $n_{AC}$ and $n_{A\_C}$
  and finally $n_{BC}$ and $n_{B\_C}$ respectively.
  To compute $n_{AB\_C}$, we proceed like in the previous section
  considering the two catalogues $AB$ and $C$. $AB$ is the result of the
  $\rchi$-match of $A$ with $B$: the positions in catalogue $AB$ are the
  weighted mean positions ($\vec{\mu_\Sigma}$, Eq. \ref{eq:wmeanpos})
  of associated $A$ and $B$ sources and the associated errors (or covariance matrix)
  are given by $\bm{V_\Sigma}$ (Eq. \ref{eq:wposerr}). 
  The only difference with the two-catalogues case is that for the
  first catalogue ($AB$) we replace the simple mean elliptical error surface
  $\overline{\Omega}_{e_{AB}}$ over the $n_{T_{AB}}$ entries
  by the weighted mean accounting for the probabilities the $AB$ associations have to
  be ``real'' (i.e. not spurious)
  \begin{equation}
    \overline{\Omega}_{e_{AB}} =
      \frac{1}
           {\sum\limits_{i=1}^{n_{T_{AB}}} p(h_{AB}|...)}
      \sum_{i=1}^{n_{T_{AB}}} p(h_{AB}|...) \pi|V_{\Sigma_{AB}i}|^\frac{1}{2},
      \label{eq:omegaAB}
  \end{equation}
  in which $p(h_{AB}|...)$ is the probability the association has to be
  a real association knowing some parameters (``...''), and $V_{\Sigma_{AB}i}$ is the 
  covariance matrix of the error on the weighted mean position $i$
  (see \S \ref{sec:iterativeform}, particulary Eq. \ref{eq:iterror}).
  Such a probability will be computed in the next sections.
  We then compute $\overline{\Omega}_{e_{C}}$ and derive $\hat{n}_{AB\_C}$
  like in the two catalogues case replacing $A$ by $AB$ and $B$ by $C$. 
  Similarly to $\hat{n}_{AB\_C}$, we can estimate $\hat{n}_{AC\_B}$
  and $\hat{n}_{A\_BC}$.\\

  We now want to estimate $n_{A\_B\_C}$, with a result which is independent
  from the cross-correlation order.
  Although we may use Eq. (\ref{eq:nfullyspur}), it is possibly time consuming.
  Another solution is to use its discretized form Eq. (\ref{eq:nfullyspurdisc}).
  To do so quickly at the cost of an approximation we may circularize the errors
  by replacing the coefficients of the covariance matrix $\bm{V}$ by a single error
  equal to $\sqrt{\det \bm{V}}$ and setting the correlation (or covariance) parameter equal to 0.
  It means that the new covariance matrix is diagonal and both diagonal elements
  are equal to $\sqrt{\det \bm{V}}$.
  We choose this value to preserve the surface area of the 2D-error
  since the determinant ($\propto$ area) of the circular error equals the determinant
  ($\propto$ area) of the ellipse.
  This approximation is the same as the one made in the previous section.
  For each catalogue we then make the histogram of $\sqrt{\det \bm{V}}$ values
  using steps of for example 1 mas and we apply Eq. (\ref{eq:nfullyspurdisc}).
  In this case -- circular errors -- we simplify the equation using
  \begin{equation}
    \det \bm{V_\Sigma} = \frac{1}{\left(\sum\limits_{i=1}^n\frac{1}{\sqrt{\det \bm{V_i}}}\right)^2}
    \label{eq:detvsig}
  \end{equation}
  and thus
  \begin{equation}
    \frac{\prod\limits_{i=1}^n \det \bm{V_i}}{\det \bm{V_\Sigma}} = 
      \sum\limits_{i=1}^n\prod_{j=1, j \ne i}^n \sqrt{\det \bm{V_j}}.
    \label{eq:detvsigAnotherform}
  \end{equation}
  We do not use this last form but give it for comparison with the denominator 
  of Eq. (17) in \cite{Budavari2008}.

  Another option is to compute the number of ``fully'' spurious associations
  three times by following what was done in the
  previous section (and in the beginning of this section), but computing
  $\overline{\Omega}_{e_{A\_B}}$ instead of $\overline{\Omega}_{e_{AB}}$.
  Similarly to Eq. (\ref{eq:omegaAB}):
  \begin{equation}
    \overline{\Omega}_{e_{A\_B}} =
      \frac{1}
           {\sum\limits_{i=1}^{n_{T_{AB}}} p(h_{A\_B}|...)}
      \sum_{i=1}^{n_{T_{AB}}} p(h_{A\_B}|...) \pi|V_{\Sigma_{AB}i}|^\frac{1}{2}.
  \end{equation}
  Computing $\overline{\Omega}_{e_{C}}$ we derive $\hat{n}_{A\_B\_C}$.
  Similarly we can compute $\overline{\Omega}_{e_{A\_C}}$ and $\overline{\Omega}_{e_{B\_C}}$
  and estimate the number of fully spurious associations taking the
  mean of $\hat{n}_{A\_B\_C}$, $\hat{n}_{A\_C\_B}$ and $\hat{n}_{B\_C\_A}$.
  
  Having the estimated number of associations being part of the components
  $AB\_C$, $AC\_B$, $A\_BC$ and $A\_B\_C$ plus knowing the total number of
  associations $n_T$, we are able to estimate $\hat{n}_{ABC}$ and to compute the
  priors, for example
  \begin{equation}
    p(h_{ABC}|s) = \frac{\hat{n}_{ABC}}{n_T}.
  \end{equation}

\subsection{Case of $n$ catalogues}
  
  We can easily generalise the previous section using recursion.
  For $n=4$ catalogues, we estimate the number of associations in component $A\_B\_C\_D$
  knowing the number of associations in the result of the four-catalogue $\rchi$-match
  and estimating recursively (from the three-catalogue $\rchi$-matches)
  the number of associations in the 14 other components ($AB\_C\_D$, ...).
  So for $n$ catalogues, the total number of distinct (sub)-cross-matches to be performed
  to compute all priors recursively is
  \begin{equation}
    N_{\rchi\mathrm{-match}} = \sum\limits_{k=2}^{n-1} C_{n-1}^k
  \end{equation}
  in which terms $C_{n-1}^k$ are the binomial coefficients $(n-1)!/(k!(n-1-k)!)$.
  For five catalogues, $N_{\rchi\mathrm{-match}} = 26$ and for 6 catalogues, $N_{\rchi\mathrm{-match}} = 57$.

\section{Probability of being $\rchi$-matched under hypothesis $h_i$ \label{sec:integrate}}
In this section we compute $p(s|h_i)$, the probability that $n$ sources from $n$ distinct catalogues
  have to satisfy the candidate selection criteria under hypothesis $h_i$. 
  We will show in section \S \ref{sec:1model} that the p.d.f of the Mahalanobis
  distance for $\rchi$-match associations under hypothesis $h_i$ is the p.d.f of
  the Mahalanobis distance without applying the candidate selection criteria, normalized by
  the probability $p(s|h_i)$ we compute in this section:
  \begin{equation}
   p(x|h_i,s) = \frac{p(x|h_i)}{p(s|h_i)}.
  \end{equation}
  We show here that $p(s|h_i)$ is proportional to the integral we note
  $I_{h_i,n}(k_\gamma)$ (see Eq. (\ref{eq:ikndef})) which is independent of positional uncertainties and which also plays
  a role in \S \ref{sec:2model}.
  We will see also that $p(s|h_i)$ and $I_{h_i,n}(k_\gamma)$ can be simplified to $p(s|h_k)$
  and $I_{k,n}(k_\gamma)$ respectively, that is the probability
  $n$ sources from $n$ distinct catalogues have to satisfy
  the candidate selection criteria knowing they are actually associated with
  $k$ distinct real sources.\\
  If $k=1$, that is all sources are from a same real source,
  we -- logically -- find $p(s|h_{k=1})=\gamma$,
  the cumulative $\rchi$ distribution function evaluated at the threshold $k_\gamma$.\\
  If $k=n$, all sources are spurious, we -- also logically (see Eq. \ref{eq:nfullyspur}) -- find for
  $I_{k=n,n}(k_\gamma)$ the volume of a $2(n-1)$-dimensional sphere of radius
  $k_\gamma$, and $p(s|h_{k=n})$ equals the volume of the 
  $2(n-1)$-dimensional ellipsoid defined by the test acceptance region divided
  by the common $\rchi$-match surface area raised to the power of the number of
  $\rchi$-matches (i.e. $n-1$).

  We note $x$ the total Mahalanobis distance,
  that is the square root of Eq. (\ref{eq:qchi2form2}).
  The vectorial form $\vec{x}=(x_1, x_2, ...x_{n-1})$ denotes
  the $n-1$ terms, also Mahalanobis distances, which are summed
  in the catalogue by catalogue iterative form Eq. (\ref{eq:qchi2iter}).
  We rewrite this equation with the new notations
  \begin{equation}
    x^2 = x_1^2 + x_2^2 + \dots + x_{n-1}^2.
  \end{equation}
  So $x$ is the radius of an hypersphere in the $n-1$ successive Mahalanobis
  distances space.
  The relation between $x$ and $\vec{x}$ of dimension $n-1$
  is the polar transformation
  $F: \mathbb{R}^{n-1} \to \mathbb{R}^{n-1}$,
  $(x_1,x_2,\dots,x_{n-1}) = F(x,\theta_1,\dots,\theta_{n-2})$,
  \begin{equation}
    F:
    \begin{pmatrix}
      x \\
      \theta_1 \\
      \vdots \\
      \theta_{n-2}
    \end{pmatrix}
    \to
    \begin{pmatrix}
      x_1 = f_1(x, \theta_1, \dots, \theta_{n-2}) \\
      x_2 = f_2(x, \theta_1, \dots, \theta_{n-2}) \\
      \vdots \\
      x_{n-1} = f_{n-1}(x, \theta_1, \dots, \theta_{n-2})
    \end{pmatrix},
  \end{equation}
  with 
  \begin{equation}
    f_j(x, \theta_1, \dots, \theta_{n-2})  =
    \begin{cases}
      x\prod\limits_{i=1}^{n-2}\cos\theta_i,& \text{if } j=1; \\
      x\sin\theta_{n-j}\prod\limits_{i=1}^{n-j-1}\cos\theta_i,& \forall j>1.
    \end{cases}
  \end{equation}
  The associated differential transform is
  \begin{equation}
    \mathrm{d}x_1\mathrm{d}x_2\dots\mathrm{d}x_{n-1}
      = |\det J_F(x,\theta_1,\dots,\theta_{n-2})|
        \mathrm{d}x\mathrm{d}\theta_1\dots\mathrm{d}\theta_{n-2},
  \end{equation}
  with $J_F$ the determinant of the Jacobian of $F$ which is for example computed in
  \citet[chap. II ``Exact sampling distributions'', p. 375]{Kendall1994}:
  \begin{equation}
    \det J_F = x^{n-2}\prod_{i=1}^{n-3}\cos^{n-i-2}\theta_i.
  \end{equation}

  We now define the $I_{k,n}(k_\gamma)$ integral which will be crucial in the
  next sections
  \begin{equation}
    I_{k,n}(k_\gamma) = \int_{x=0}^{x \le k_\gamma} 
                       \prod_{i=1}^{n-k} \rchi_2(x_i)
                       \prod_{i=n-k+1}^{n-1} 2\pi x_i
		       \prod_{i=1}^{n-1}\mathrm{d}x_i
    \label{eq:ikndef}
  \end{equation}
  in which $k$ denotes the hypothetical number of real sources and so ranges from 1 to $n$.
  Written this way, the integral is simpler than the equivalent form:
  \begin{equation}
    I_{k,n}(k_\gamma) = \int_{x=0}^{x \le k_\gamma} 
                       \prod_{i=1}^{n_1-1} \rchi_2(x_i)\mathrm{d}x_i
		       \left[
		         \prod_{g=2}^k\left(
                           2\pi x_{g-1}\mathrm{d}x_{g-1}
			   \prod_{j=1}^{n_g-1} \rchi_2(x_j)\mathrm{d}x_j
			 \right)
		       \right]
    \label{eq:ikndef2}
  \end{equation}
  in which we have $k$ groups containing each $n_g$ sources associated to a same real source
  so that $\sum\limits_{g=1}^k n_g = n$ (see the iterative candidate selection by groups of 
  catalogues in \S \ref{sec:candselectbygroups}).
  Here Eq. (\ref{eq:qchi2iterG}) takes the form
  \begin{equation}
    x^2=\sum\limits_{g=2}^kx_{g-1}^2 + \sum\limits_{g=1}^k\sum\limits_{i=1}^{n_g}x_i^2
  \end{equation}
  where $x_{g-1}$ are inter-group Mahalanobis distances and $x_i$ are intra-group Mahalanobis distances.
  In this version of the formula, we suppose that we iteratively cross-correlate the catalogues
  by groups.
  We suppose that each group corresponds to one real source.
  So inside each group, we multiply Rayleigh distributions and when associating each group, we multiply
  by 2d-Poisson distributions.
  
  We compute $I_{k,n}$ using recursive integration by parts, leading to
  (see \S \ref{sec:app:chicumul} and \ref{sec:app:ikn})
  \begin{equation}
    I_{k, n}(k_\gamma) = 
    \begin{cases}
      1-e^{-\frac{1}{2}k_\gamma^2}\sum\limits_{i=2}^{n}\frac{2^{2-i}}{(i-2)!}k_\gamma^{2(i-2)}& \text{, if } k=1; \\
      I_{k, n - 1}(k_\gamma) - 2\pi I_{k - 1, n - 1}(k_\gamma)& \text{, if } 1<k<n; \\
      \pi^{n-1}k_\gamma^{2(n-1)}/(n-1)! & \text{, if } k=n.
    \end{cases}
    \label{eq:ikn}
  \end{equation}
  We provide the exhaustive list of values of $I_{k, n}(k_\gamma)$ for $n=2,3,4$
  and $5$ in Table \ref{tab:imn}.
  Remark: $k_\gamma$ depends on the selected completeness $\gamma$ and on the
  number of catalogues $n$.
  When we call $I_{k, n-k}(k_\gamma)$, the $k_\gamma$ to be used in the integral
  is always the $k_\gamma$ computed for $n$ catalogues. So $I_{1, n - k}$ will
  no longer be equal to $\gamma$ but to $F_{\rchi_{2(n-k-1)}}(k_\gamma)$.
  For example, we fix $\gamma$ to $0.9973$.
  Then for a $n=2$ catalogues $\rchi$-match, $k_\gamma\approx 3.4$ and $I_{1,2}=\gamma$.
  But for a $n=3$ catalogues $\rchi$-match, $k_\gamma\approx 4.0$,  $I_{1,3}=\gamma$
  and $I_{1, 3-1=2} \neq \gamma$.
   \begin{table}[!h]
    \centering
    \caption{Values of the normalization integrals $I_{k, n-k}(k_\gamma)$ for a number of catalogue ranging from two to five.}
    \begin{tabular}{c|c|l}
      k & n & $I_{k, n}(k_\gamma)$ \\ \hline
      1 & 2 & $\gamma=1-e^{-\frac{1}{2}k_\gamma^2}$ \\
      2 & 2 & $\pi k_\gamma^2$ \\
      1 & 3 & $\gamma=1-e^{-\frac{1}{2}k_\gamma^2}(1+\frac{1}{2}k_\gamma^2)$ \\
      2 & 3 & $\pi\left[k_\gamma^2 - 2(1-e^{-\frac{1}{2}k_\gamma^2})\right]$ \\
      3 & 3 & $\frac{\pi^2}{2} k_\gamma^4$ \\
      1 & 4 & $\gamma=1-e^{-\frac{1}{2}k_\gamma^2}(1+\frac{1}{2}k_\gamma^2+\frac{1}{8}k_\gamma^4)$ \\
      2 & 4 & $\pi\left[k_\gamma^2(1+e^{-\frac{1}{2}k_\gamma^2})
                    -4(1-e^{-\frac{1}{2}k_\gamma^2})
                  \right]$ \\
      3 & 4 & $\pi^2\left[\frac{k_\gamma^4}{2}-2\left(
                 k_\gamma^2-2(1-e^{-\frac{1}{2}k_\gamma^2})
	       \right)\right]$ \\ 
      4 & 4 & $\frac{\pi^3}{6}k_\gamma^6$ \\
      1 & 5 & $\gamma=1-e^{-\frac{1}{2}k_\gamma^2}(1+\frac{1}{2}k_\gamma^2
                +\frac{1}{8}k_\gamma^4+\frac{1}{48}k_\gamma^6)$ \\
      2 & 5 & $\pi\left[(k_\gamma^2-6)
                 +(\frac{1}{4}k_\gamma^4+2k_\gamma^2+6)e^{-\frac{1}{2}k_\gamma^2}
               \right]$ \\
      3 & 5 & $\pi^2\left[
                 \frac{k_\gamma^4}{2}-4k_\gamma^2+12-(2k_\gamma^2-12)e^{-\frac{1}{2}k_\gamma^2}
               \right]$ \\
      4 & 5 & $\pi^3\left[\frac{k_\gamma^6}{6}-2\left(\frac{k_\gamma^4}{2}-2\left(
                 k_\gamma^2-2(1-e^{-\frac{1}{2}k_\gamma^2})
	       \right)\right) \right]$ \\
      5 & 5 & $\frac{\pi^4}{24}k_\gamma^8$ 
    \end{tabular}
    \label{tab:imn}
  \end{table}
 \begin{table}[!h]
    \centering
    \caption{Values of the derivatives of normalization integrals for a number of catalogue ranging from two to five.}
    \begin{tabular}{c|c|l}
      k & n & $\mathrm{d}I_{k, n}(x)$ \\ \hline
      1 & 2 & $\rchi_2(x)\mathrm{d}x$ \\
      2 & 2 & $2\pi x\mathrm{d}x$ \\
      1 & 3 & $\rchi_4(x)\mathrm{d}x$ \\
      2 & 3 & $2\pi x(1 - e^{-\frac{1}{2}x^2})\mathrm{d}x$ \\
      3 & 3 & $2\pi^2x^3\mathrm{d}x$  \\
      1 & 4 & $\rchi_6(x)\mathrm{d}x$ \\
      2 & 4 & $2\pi x\left[1-(1+\frac{1}{2}x^2)e^{-\frac{1}{2}k_\gamma^2}\right]\mathrm{d}x$\\
      3 & 4 & $2\pi^2x\left[x^2-2(1-e^{-\frac{1}{2}x^2})\right]\mathrm{d}x$ \\
      4 & 4 & $\pi^3 x^5\mathrm{d}x$ \\
      1 & 5 & $\rchi_8(x)\mathrm{d}x$ \\
      2 & 5 & $\pi x \left[2-(\frac{1}{4}x^4+x^2+2)e^{-\frac{1}{2}x^2}
	      \right]\mathrm{d}x$ \\
      3 & 5 & $2\pi^2x\left[x^2-4+(x^2+4)e^{-\frac{1}{2}x^2})\right] \mathrm{d}x$ \\
      4 & 5 & $\pi^3x\left[x^2(x^2-4)) + 8(1-e^{-\frac{1}{2}x^2}) \right]\mathrm{d}x$ \\
      5 & 5 & $\frac{\pi^4}{3} x^7\mathrm{d}x$ 
    \end{tabular}
    \label{tab:dimn}
  \end{table}

  We call $p(s|h_k)$ the marginalized probability of observing a Mahalanobis distance less than
  or equal to $k_\gamma$ in a group of $n$ sources knowing they are actually associated to $k$ real sources.
  \begin{equation}
    p(s|h_k) =
    \begin{cases}
      I_{k,n}(k_\gamma) & \text{, if } k=1; \\
      \left[\frac{\prod\limits_{i=1}^n \det \bm{V_i}}{\det \bm{V_\Sigma}}\right]^{1/2}
      \frac{1}{\Omega^{(k-1)}}I_{k,n}(k_\gamma) & \text{, if } k > 1.
    \end{cases}
    \label{eq:pshk}
  \end{equation}
  We obtain this equality by replacing $2\pi x_{g-1}$ by
  $2\pi x_{g-1}\det(V_{\Sigma_{g-1}} + V_{\Sigma_g})/\Omega$ in Eq. (\ref{eq:ikndef2})
  and then applying Eq. (\ref{eq:detprodgroups}).
  The factor $1/\Omega$ comes from the normalisation of the Poisson distribution so it
  integrates to one over the common surface area of the cross-matched catalogues. 
  For the particular case in which all sources are spurious, we logically find the
  summed terms in Eq. (\ref{eq:nfullyspur}).

  And the distribution (p.d.f) associated to the probability $p(x|h_{k})$
  of observing a given Mahalanobis distance $x$ knowing $h_{k}$
  is simply given by the derivative of $p(s|h_k)\mathrm{d}x$, so
  is proportional to $\mathrm{d}I_{k,n}(x)\mathrm{d}x$.

\section{Simple Bayesian probabilities \label{sec:1model}}
In this section, we compute Bayesian probabilities which depend
on the Mahalanobis distance only.

\subsection{General formula}

  Given a set of $n$ candidates from $n$ distinct catalogues
  satisfying the candidate selection criterion, we know
  \begin{itemize}
    \item $x$, the Mahalanobis distance (Eq. \ref{eq:dm}), or $\rchi$ value,
               which is a real value;
    \item $s$, the result of the selection criterion $x\le k_\gamma$, that is
               a boolean always equals to $true$ for the sets of associations we keep, 
	       so $p(s)=1$;
    \item $\{h_i\}$, $i \in [1, B_n]$,  the set of $B_n$ (Eq. \ref{eq:bellnumber})
                     hypotheses to be formulated for each set of association.
  \end{itemize}
  We then note $h_k$ hypotheses in which the $n$ sources are associated with
  $k$ ``real'' sources.
  
  For a given set of $n$ candidates from $n$ distinct catalogues,
  the probabilities associated with the various hypotheses are given by
  Bayes' formula
  \begin{eqnarray}
    p(h_i|x,s) & = & \frac{p(s)p(h_i|s)p(x|h_i,s)}
                          {\sum\limits_{k=1}^{B_n}p(s)p(h_k|s)p(x|h_k,s) }, \\
	       & = & \frac{p(h_i|s)p(x|h_i,s)}
	                  {\sum\limits_{k=1}^{B_n}p(h_k,s)p(x|h_k,s)}. 
    \label{eq:bayesformula}
  \end{eqnarray}
  In this formula, $p(h_i|s)$ are priors (considering only $\rchi$-matches, hence only $s=true$)
  and correspond to the number of
  associations satisfying the candidate selection ($\rchi$-matches) and hypotheses $h_i$
  over the total number of associations satisfying the candidate selection.
  We can transform the likelihood $p(x|h_i,s)$ in
  \begin{eqnarray}
    p(x|h_i,s) & = & \frac{p(x,h_i,s)}{p(h_i,s)}, \\
               & = & \frac{p(h_i)p(x|h_i)p(s|x,h_i)}{p(h_i)p(s|h_i)}, \\
	       & = & \frac{p(x|h_i)}{p(s|h_i)},
  \end{eqnarray}
  because we keep in our sample only associations satisfying the candidate
  selection criteria we have $p(s|x,h_i) = 1$.
  In other words, the likelihood we use is a classical likelihood 
  normalized so it integrates to one over the $\rchi$ test acceptance region
  (defined by $x\le k_\gamma$).\\
  We easily compute priors from the numbers estimated in \S \ref{sec:spurest}.
  And likelihoods are simply computed from \S \ref{sec:integrate}
  \begin{equation}
    p(x|h_i,s) = \frac{\mathrm{d} p(x'<x|h_k) \mathrm{d}x}{p(s|h_k)}
               = \frac{\mathrm{d}I_{k,n}(x)\mathrm{d}x}{I_{k,n}(k_\gamma)}.
    \label{eq:pxhis}
  \end{equation}
  We make explicit this result in the next section for the case of two,
  three and four catalogues.

\subsection{Likelihoods $p(x|h_i,s)$}

In this section, we compute the likelihoods $p(x|h_i,s)$, that is the
p.d.f of the Mahalanobis distance of $\rchi$-matches under hypothesis $h_i$.

 \subsubsection{Case of two catalogues \label{sec:proba2catsMD}}

\begin{figure}[!h]
    \centering
    \includegraphics[clip,width=0.4\textwidth]{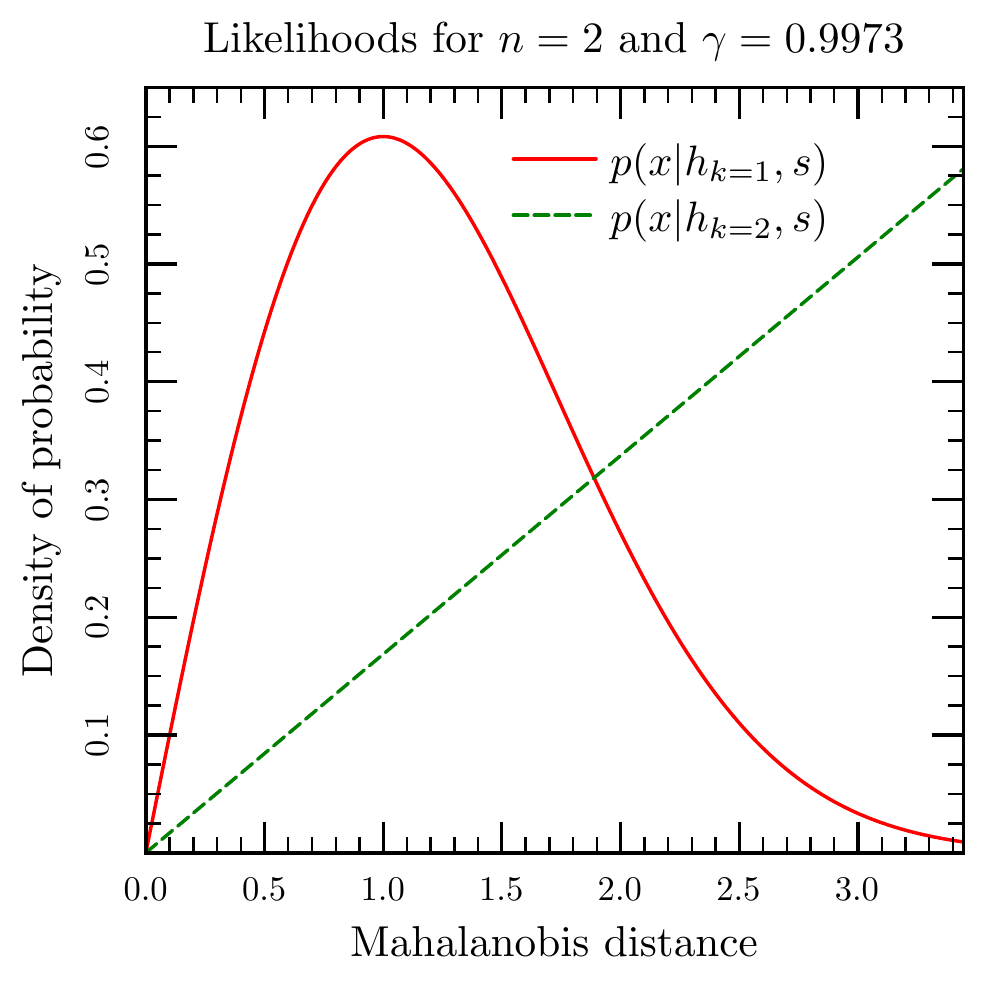}
    \caption{Two possible likelihoods for $n=2$ catalogues and $\gamma=0.9973$:
             normalized Rayleigh (red, filled curve) and Poisson (green, dashed curve) components.
	     We note that this $\gamma$ value implies the $[0, k_\gamma=3.443935]$ range for $x$.}
    \label{fig:likelihoods_n2}
  \end{figure}

    For a set of two sources from two distinct catalogues, we have only
    two hypotheses, hence the two likelihoods:
    \begin{eqnarray}
      p(x|h_{k=1}) & = & \rchi(x)\mathrm{d}x; \\
      p(x|h_{k=2}) & = & \frac{2 \pi \sqrt{\det (V_1 + V_2)} x\mathrm{d}x}{4\pi}.
    \end{eqnarray}
    Knowing that the selection criterion is satisfied, we have to normalize so the integral
    of each likelihood over the domain defined by the selection criteria equals
    one (likelihoods are p.d.f):
    \begin{eqnarray}
      p(x|h_{k=1},s) & = & \frac{\rchi(x)\mathrm{d}x}
                            {\int_0^{k_\gamma} \rchi(x)\mathrm{d}x}
                 = \frac{\rchi(x)\mathrm{d}x}{\gamma}
		 = \frac{\mathrm{d}I_{1,2}(x)\mathrm{d}x}{I_{1,2}(k_\gamma)}; \\
      p(x|h_{k=2},s) & = & \frac{x\mathrm{d}x}{\int_0^{k_\gamma} x\mathrm{d}x}
                 = \frac{2}{k_\gamma^2} x \mathrm{d}x
		 = \frac{\mathrm{d}I_{2,2}(x)\mathrm{d}x}{I_{2,2}(k_\gamma)}.
    \end{eqnarray}
    All constant terms, that is terms independent of $x$, vanish with the
    normalisation. The likelihoods are plotted in Fig. \ref{fig:likelihoods_n2}.

    Remark: $p(x|h_{k=2})$ mixes the derivative of the surface area of an ellipse
            in the Euclidean plane and the surface area of the sphere.
            It is an approximation valid as long as $\sqrt{\det (\bm{V_1} + \bm{V_2})}x$
	    is small enough so effects of the curvature of the sphere
	    are negligible.

  \subsubsection{Case of three catalogues}

    For a set of three sources A, B, and C from three distinct catalogues,
    we have five hypotheses (see \S \ref{sec:3cathyp}). In the five hypotheses,
    the number of ``real'' sources can be either one, two or three.
    We have as many likelihoods as possible distinct number of ``real'' sources.
    There are three ways of performing an iterative cross-match, leading to
    the same Mahalanobis distance $x$:
    \begin{itemize}
      \item cross-match A with B and then with C, $x^2 = x_{AB}^2 + x_{ABC}^2$;
      \item cross-match B with C and then with A, $x^2 = x_{BC}^2 + x_{BCA}^2$;
      \item cross-match A with C and then with B, $x^2 = x_{AC}^2 + x_{ACB}^2$.
    \end{itemize}
    We denote $x_{AB}$ the Mahalanobis distance between $A$ and $B$
    and we denote $x_{ABC}$ the Mahalanobis distance between $C$ and
    the weighted mean position of A and B.\\
    $x_1$ and $x_2$ are used to designate without distinction $x_{AB}$, $x_{BC}$ or $x_{AC}$
    and $x_{ABC}$, $x_{BCA}$ or $x_{ACB}$ respectively.

    Although it may be tempting to write
    \begin{eqnarray}
      p(x_{AB}, x_{ABC}|h_{ABC}) & = &
        \frac{\rchi(x_{AB})\rchi(x_{ABC})\mathrm{d}x_{AB}\mathrm{d}x_{ABC}}
	     {I_{1,3}(k_\gamma)}, \\
      p(x_{AB}, x_{ABC}|h_{AB\_C}) & = &
        \frac{\rchi(x_{AB})2\pi x_{ABC}\mathrm{d}x_{AB}\mathrm{d}x_{ABC}}
	     {I_{2,3}(k_\gamma)}, \\
      p(x_{AC}, x_{ACB}|h_{AC\_B}) & = &
        \frac{\rchi(x_{AC})2\pi x_{ACB}\mathrm{d}x_{AC}\mathrm{d}x_{ACB}}
	     {I_{2,3}(k_\gamma)}, \\
      p(x_{BC}, x_{BC\_A}|h_{A\_BC}) & = &
        \frac{\rchi(x_{BC})2\pi x_{BC\_A}\mathrm{d}x_{BC}\mathrm{d}x_{BCA}}
	     {I_{2,3}(k_\gamma)}, \\
      p(x_{AB}, x_{ABC}|h_{A\_B\_C}) & = &
        \frac{2\pi x_{AB} 2\pi x_{ABC}\mathrm{d}x_{AB}\mathrm{d}x_{ABC}}
	     {I_{3,3}(k_\gamma)}.
    \end{eqnarray}
    But we cannot directly compute probabilities $p(h_i|\vec{x})$ from those likelihoods since
    infinitesimals ($\mathrm{d}x_{AB}$, $\mathrm{d}x_{AC}$, ...) are not the same
    and so do not vanish when applying Bayes' formula.

\begin{figure}[!h]
    \includegraphics[clip,width=0.45\textwidth]{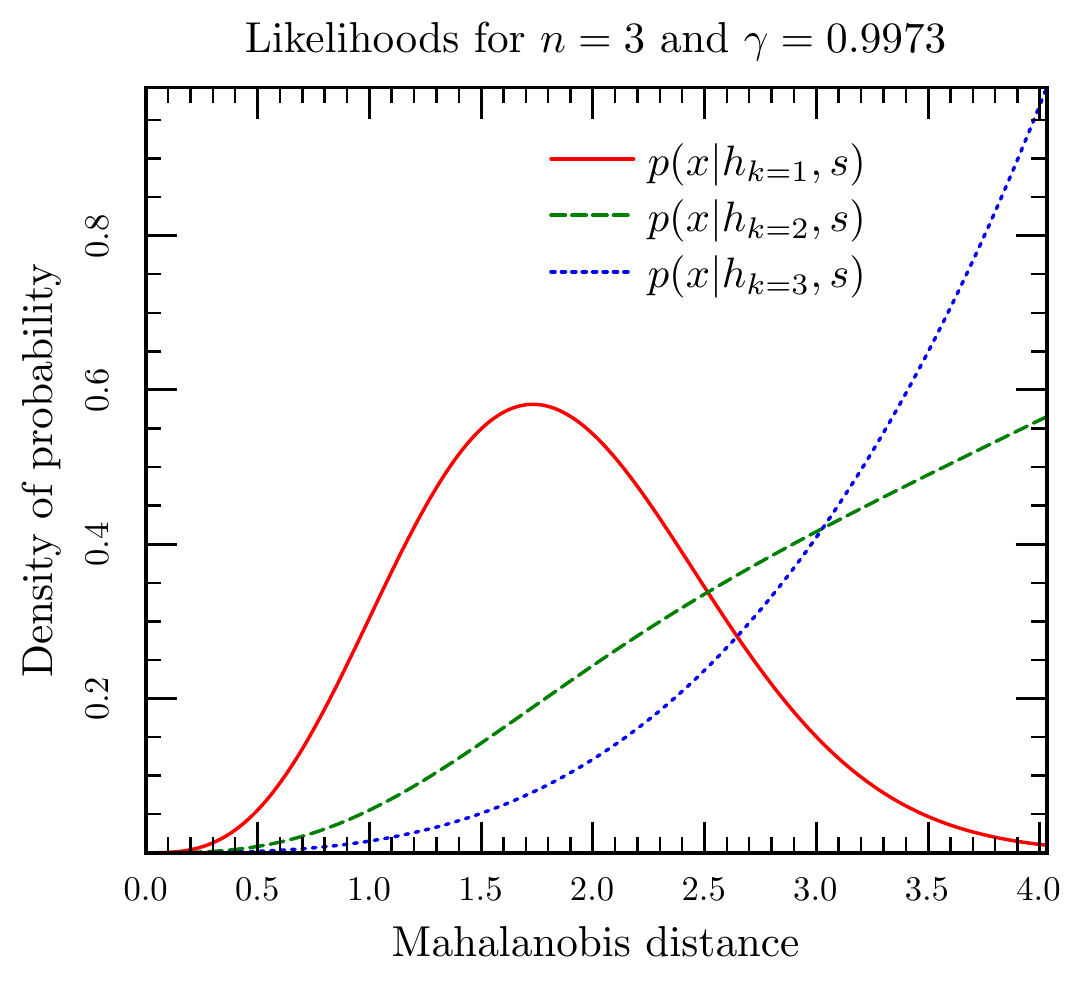}
    \caption{Three possible likelihoods $p(x|h_k,s)$ for $n=3$ catalogues and $\gamma=0.9973$:
             $\rchi$ distribution with four degrees of freedom (red, filled curve);
	     Integral of a $\rchi$ distribution with two degrees of freedom times a two-dimensional Poisson distribution (green, dashed curve);
	     Four-dimensional Poisson distribution (blue, dotted curve).}
    \label{fig:likelihoods_n3}
  \end{figure}

    It seems that the only measurement one can use to obtain coherent
    (and symmetrical) probabilities is the total Mahalanobis distance $x$.
    So we have to integrate the above probabilities over the domain defined by
    $x_1^2+x_2^2 \le x^2$ and then evaluate their derivatives for $x$. 
    We obtain the following likelihoods represented on Fig. \ref{fig:likelihoods_n3}.
    \begin{eqnarray}
      p(x|h_{k=1},s) & = & \frac{\mathrm{d}I_{1,3}(x)\mathrm{d}x}{I_{1,3}(x)} = \frac{\rchi_{dof=4}(x)\mathrm{d}x}{\gamma}, \\
      p(x|h_{k=2},s) & = & \frac{\mathrm{d}I_{2,3}(x)\mathrm{d}x}{I_{2,3}(x)} = \frac{2x(1-\exp(-x^2/2))\mathrm{x}}{k_\gamma^2-2(1-\exp(-k_\gamma^2/2))},\\
      p(x|h_{k=3},s) & = & \frac{\mathrm{d}I_{3,3}(x)\mathrm{d}x}{I_{3,3}(x)} = \frac{4x^3\mathrm{x}}{k_\gamma^4},
    \end{eqnarray}
    in which $h_{k=1}$ is the hypothesis $h_{ABC}$, $h_{k=3}$ is the hypothesis $h_{A\_B\_C}$
    and $h_{k=2}$ is either the hypothesis $h_{AB\_C}$ or $h_{AC\_B}$ or $h_{A\_BC}$.

  \subsubsection{Case of four catalogues}

\begin{figure}[!h]
      \includegraphics[clip,width=0.45\textwidth]{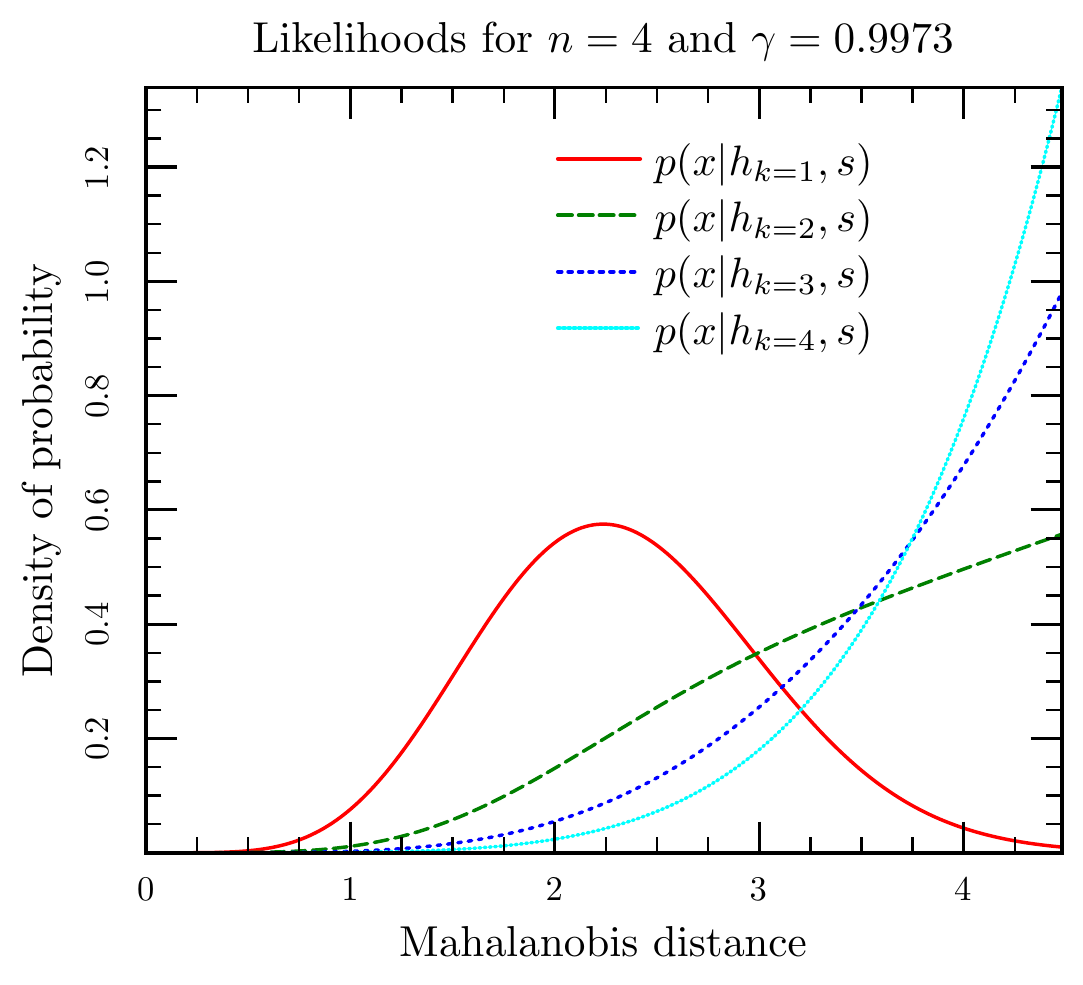}
      \caption{Four possible likelihoods $p(x|h_k,s)$ for $n=4$ catalogues and $\gamma=0.9973$:
             $\rchi$ distribution with six degrees of freedom (red, filled curve);
             Integral of a $\rchi$ distribution with two degrees of freedom times a four-dimensional Poisson distribution (green, dashed curve);
             Integral of a $\rchi$ distribution with four degrees of freedom times a two-dimensional Poisson distribution (blue, dotted curve);
             six-dimensional Poisson distribution (cyan, filled curve).}
      \label{fig:likelihoods_n4}
    \end{figure}

    For a set of four sources A, B, C and D from four distinct catalogues,
    we have fifteen hypotheses (see \S \ref{sec:4cathyp}).
    In the fifteen hypotheses, the number of ``real'' sources can be either
    one, two, three or four.
    We have as many likelihoods as possible distinct numbers of ``real'' sources.
    They are represented in Fig. \ref{fig:likelihoods_n4}:
    \begin{eqnarray}
      p(x|h_{k=1},s) & = & \frac{\mathrm{d}I_{1,4}(x)\mathrm{d}x}{I_{1,4}(x)} = \frac{\rchi_{dof=6}(x)\mathrm{d}x}{\gamma};  \\
      p(x|h_{k=2},s) & = & \frac{\mathrm{d}I_{2,4}(x)\mathrm{d}x}{I_{2,4}(x)}; \\
      p(x|h_{k=3},s) & = & \frac{\mathrm{d}I_{3,4}(x)\mathrm{d}x}{I_{3,4}(x)}; \\
      p(x|h_{k=4},s) & = & \frac{\mathrm{d}I_{4,4}(x)\mathrm{d}x}{I_{4,4}(x)} = \frac{6x^5\mathrm{d}x}{k_\gamma^6}.
    \end{eqnarray}
    In which $h_{k=1}$ is the hypothesis $h_{ABCD}$; $h_{k=4}$ is the
    hypothesis $h_{A\_B\_C\_D}$;
    $h_{k=2}$ is either the hypothesis $h_{A\_BCD}$ or $h_{B\_ACD}$ or
    $h_{C\_ABD}$ or $h_{D\_ABC}$ or $h_{AB\_CD}$ or $h_{AC\_BD}$ or $h_{AD\_BC}$;
    and $h_{k=3}$ is either the hypothesis $h_{AB\_C\_D}$ or $h_{AC\_B\_D}$ or
    $h_{AD\_B\_C}$ or $h_{BC\_A\_D}$ or $h_{BD\_A\_C}$ or $h_{CD\_A\_B}$.

\subsection{Advantage \& Limits \label{sec:advandlimit}}

  The main advantage of using $p(h_i|x,s)$ is that the likelihoods it is based on
  do not depend on the positional errors: the only input parameter
  is the Mahalanobis distance $x$.
  Although it is true that $x$ is computed from positional errors,
  once the $\rchi$-match has been performed we do not need the errors anymore:
  the distributions we use relies only on $x$.
  Changing positional errors modifies the priors, not the likelihoods.
  So we can easily add independent
  likelihoods based on magnitudes or other parameters.

  There are two main problems.
  The first problem is precisely that the likelihoods depend only
  on $x$. It means that a set of very close sources with very
  accurate positions may have the same probability than a set
  of distant sources with large positional errors, even if intuitively
  the risk the first set have to contain spurious association should be
  far lower than in the second case.
  The second limitation is due to the fact that likelihoods are the same for
  hypotheses considering the same number of ``real'' sources.
  In the three catalogues case, $p(x|h_{AB\_C},s) = p(x|h_{A\_BC},s) = p(x|h_{AC\_B},s)$.
  The priors being constants, if $p(h_{AB\_C}|s) > p(h_{A\_BC}|s) > p(h_{AC\_B}|s)$, 
  we always obtain posterior probability $p(x|h_{AB\_C},s) > p(h_{A\_BC}|s) > p(h_{AC\_B}|s)$.

\section{Bayesian probabilities with positional errors \label{sec:2model}}
In this section, we compute Bayesian probabilities which include explicitly 
positional errors.

\subsection{Warning about the non independence of positional uncertainties\label{sec:errorsnoinde}}

In surveys providing individual uncertainties, positions of unsaturated
bright sources are often more precise than positions of faint sources. 
The reason has to do with the higher photometric signal-to-noise ratio of
bright sources compared to faint sources, while the FWHM is similar.
An example of computation of positional uncertainties based on photon statistics
can be found for example in the documentation of the SExtractor software \citep{Bertin1996}.
As mentioned in the documentation, the photon statistics based error is a lower value estimate.\\
It means that we cannot blindly assume that the positional uncertainties and photometric
quantities like apparent magnitudes are independent.
Moreover, if the positional errors of sources are related to their magnitudes
and if the magnitudes of the sources in different catalogues are also
related, then positional uncertainties in the different catalogues are related
too.
It means that we cannot blindly assume that the positional uncertainties of
matching objects in different catalogues are independent from each other,
at least not for $h_{k<n}$, that is the hypothesis in which at least two sources are from a
same actual source.\\
One has to keep this in mind when using the naive independent hypothesis
to simplify Bayes probabilities.

\subsection{Probability using the Mahalanobis distance}
 
  To (at least partly) solve the first issue mentioned in \S \ref{sec:advandlimit}
  one possibility is to introduce likelihoods based for example on the volume $V$ of
  the Chi test acceptance region writing:
  \begin{equation}
    p(h_i|x,V,s) = \frac{p(h_i|s)p(x|h_i,s)p(V|x,h_i,s)}
                        {\sum\limits_{k=1}^{B_n}p(h_k|s)p(x|h_k,s)p(V|x,h_k,s)}.
    \label{eq:bayesformulaV}
  \end{equation}
  From \S \ref{sec:spurest}, it is easy -- even though it may be time consuming --
  to build the estimated histogram $n_{A\_B} p(V+\Delta V|h_{A\_B})$ (in which $\Delta V$
  is the width of the histogram's bars).
  Given this histogram and the result of a 2-catalogue cross-match,
  we can also build an estimated histogram $n_{AB} p(V+\Delta V|h_{AB})$.
  And so on for multiple catalogues, performing all possible sub-cross-matches.
  
  If $V$ and $x$ are independent for all hypotheses, and knowing
  (having estimates of) $n_{AB}$ and $n_{A\_B}$,
  we have all the ingredients to compute
  \begin{equation}
    p(h_i|x,V+\Delta V,s) = \frac{p(h_i|s)p(x|h_i,s)p(V+\Delta V|h_i,s)}
                        {\sum\limits_{k=1}^{B_n}p(h_k|s)p(x|h_k,s)p(V + \Delta V|h_k,s)},
    \label{eq:bayesformulaV2}
  \end{equation}
  even if it is not elegant to introduce a somewhat arbitrary slicing in $V$ histograms.

\subsection{Putting aside the Mahalanobis distance}

We also consider the alternative form which puts aside the Mahalanobis distance
and relies on the full sets of positions $\vec{\mu}$ and associated errors $\vec{\bm{V}}$
\begin{equation}
  p(h_i|\vec{\mu},\bm{V},s) = \frac{p(h_i|s)p(\vec{\bm{V}}|h_i,s)p(\vec{\mu}|h_i,\bm{V},s)}
                                   {\sum\limits_{k=1}^{B_n}p(h_k|s)p(\vec{\bm{V}}|h_k,s)p(\vec{\mu}|h_k,\bm{V},s)},
\end{equation}
in which the probabilities explicitly depend on the ``configuration''
of each position and on the associated errors.
It also depends on the distribution of positional errors for a given hypothesis.
Although $p(\vec{\bm{V}}|h_i,s)$ can be estimated performing all possible sub-cross-matches,
it is not trivial since it is a joint distribution in a space of dimension equal to the number
of ``actual'' sources considered in $h_i$ (using the circular error approximation).

\subsubsection{Likelihoods $p(\vec{\mu}|h_i,\vec{V},s)$}
 We make the hypothesis that $n_g$ sources are $n_g$ detections of a same
 true source having a given position $\vec{p}$.
 The probability to observe the set of positions
 $\vec{\mu_{\{g\}}} = \{\vec{\mu_1}, \vec{\mu_2}, \dots, \vec{\mu_{n_g}}\}$,
 knowing $\vec{p}$ and the set of errors
 $\bm{V_{\{g\}}}=\{\bm{V_1}, \bm{V_2}, \dots, \bm{V_{n_g}}\}$
 is
 \begin{equation}
   p( \vec{\mu_{\{g\}}} | \vec{p}, \bm{V_{\{g\}}} ) =
     \prod\limits_{i=1}^{n_g}\mathcal{N}_{\vec{\mu_i},V_i}(\vec{p})
     \mathrm{d}\vec{p}
     \mathrm{d}\vec{\mu_1}\dots\mathrm{d}\vec{\mu_{n_g}}.
 \end{equation}
 In practice we do not know the position of the real source $\vec{p}$.
 So the probability to observe the set of positions
 $\vec{\mu_{\{g\}}}$ knowing the set of errors $\bm{V_{\{g\}}}$ is obtained by
 integrating over all possible positions
 \begin{eqnarray}
   p( \vec{\mu_{\{g\}}} | \bm{V_{\{g\}}} )
   & = &
     \left( \int\int
       \prod\limits_{i=1}^{n_g}\mathcal{N}_{\vec{\mu_i},V_i}(\vec{p})
       \mathrm{d}\vec{p}
     \right) \mathrm{d}\vec{\mu_1}\dots\mathrm{d}\vec{\mu_{n_g}}, \\
   & = &
      \sqrt{
      \frac{\det \bm{V_{\Sigma_{\{g\}}}}}
           {\prod\limits_{i=1}^{n_g}\det\bm{V_i} }}
      \frac{\exp\left\{-\frac{1}{2} Q_{\rchi^2(\vec{\mu_1},\vec{\mu_2},...,\vec{\mu_{n_g}})}\right\}}
           {(2\pi)^{n_g-1}} \nonumber\\
   & & \mathrm{d}\vec{\mu_1}\dots\mathrm{d}\vec{\mu_{n_g}}.
 \end{eqnarray}
 This result is the same as Eq. (\ref{eq:gausschi2}) in \S \ref{sec:candselect}
 but applied here to the sub-set of positions
 $\{\vec{\mu_1},\vec{\mu_2},...,\vec{\mu_{n_g}}\}$.
 The difference is that in \S \ref{sec:candselect} we wanted to estimate the
 probability the sources had to be at the same location whereas here,
 knowing (making the hypothesis) they are at the same location,
 we compute the probability we had to observe this particular outcome.
 The particular case $n_g=1$ leads to
 $p( \vec{\mu_{\{g\}}} | \bm{V_{\{g\}}}) = \mathrm{d}\vec{\mu_1}$. \\ 
 We now consider $G$ groups and the selection criteria $s$ ($x\le h_\gamma)$.
 Each of the $n$ input sources is part of one, and only one group.
 Given the $G$ groups, the errors on the positions and the candidate
 selection criteria, the probability to observe positions $\vec{\mu}$ is
 \begin{eqnarray}
   p(\vec{\mu}|h_G,\bm{V},s) 
     = \frac{p(\vec{\mu}|h_G,\vec{\bm{V}})}{p(s|h_G,\vec{\bm{V}})}
     = \frac{ \prod\limits_{g=1}^G p( \vec{\mu_{\{g\}}} | \bm{V_{\{g\}}} ) }
            { \int_{x\le k_\gamma}
	      \prod\limits_{g=1}^G p( \vec{\mu_{\{g\}}} | \bm{V_{\{g\}}} ) }.
 \end{eqnarray}
 The denominator ensures that the likelihood integrates to 1 over its domain
 of definition, domain delimited by the candidate selection criteria, that is
 the region of acceptance of the $\rchi^2$ test.

 Let us compute the integral in the denominator.
 The differential of the substitution  $x=\vec{y}\bm{V^{-1}}\vec{y}$ transforms as
 $\mathrm{d}\vec{y}=\sqrt{\det \bm{V}}x\mathrm{d}x\mathrm{d}\theta$.
 $\vec{y}$ can be the difference between two positions
 (e.g. $\vec{\mu_i}-\vec{\mu_{\Sigma_{i-1}}}$) and $(x, \theta)$
 the polar coordinates of $\vec{y}$ in the basis defined by the eigenvectors of $\bm{V}$
 and reduced by its eigenvalues. $\bm{V}$ can for example be $\bm{V}_{\Sigma_{i-1}}+\bm{V}_i$.
 Using the iterative form of \S \ref{sec:iterativeform} and Eq. (\ref{eq:itdet}),
 we can rewrite $p( \vec{\mu_{\{g\}}} | \bm{V_{\{g\}}} )$
 \begin{equation}
   p( \vec{\mu_{\{g\}}} | \bm{V_{\{g\}}} ) = \frac{ \exp\left\{-\frac{1}{2}
     \sum\limits_{i=1}^{n_g-1} x_i \right\} }{(2\pi)^{n_g-1}}
     \prod\limits_{i=1}^{n_g-1}x_i\mathrm{d}x_i\mathrm{d}\theta_i.
 \end{equation}
 Integrating over all $\theta_i$ we obtain
 \begin{eqnarray}
   p( \vec{x_{\{g\}}} )
     & = &  \int_0^{2\pi}...\int_0^{2\pi}
            p( \vec{\mu_{\{g\}}} | \bm{V_{\{g\}}} ), \\
     & = & \exp\left\{-\frac{1}{2} \sum\limits_{i=1}^{n_g-1} x_i \right\}
           \prod\limits_{i=1}^{n_g-1}x_i\mathrm{d}x_i, \\
     & = & \prod\limits_{i=1}^{n_g-1}\rchi_{k=2}(x_i)\mathrm{d}x_i.
 \end{eqnarray}
 This joint p.d.f of the successive Mahalanobis distances is different
 from the p.d.f of their quadratic sum which gives the total Mahalanobis
 distance
 \begin{equation}
   p( x_{\{g\}} )
      = \int_0^{\frac{\pi}{2}}...\int_0^{\frac{\pi}{2}}
            p( \vec{x_{\{g\}}}  )
      = \rchi_{k=2(n_g-1)}(x_{\{g\}})\mathrm{d}x_{\{g\}},
 \end{equation}
 in which $x_{\{g\}}^2 = ||\vec{x_{\{g\}}}||^2 = \sum\limits_{i=1}^{n_g}x_i^2$. \\ 

Putting all together, the integral in the denominator is no other than the 
integral $I_{k=G,n}(k_\gamma)$ defined in \S \ref{sec:integrate}.
Written explicitly:
\begin{equation}
  p(\vec{\mu}|h_G,\bm{V},s) =
     \frac{ \prod\limits_{g=1}^G p( \vec{\mu_{\{g\}}} | \bm{V_{\{g\}}} ) }
          {I_{k=G,n}(k_\gamma)}.
  \label{eq:likelihodds_v2}
\end{equation}

\subsubsection{Classical two catalogues case}

  In the case of two catalogues, the probabilities are simply:
  \begin{equation}
    p(\vec{\mu_1}, \vec{\mu_2}|h_1,\bm{V_1},\bm{V_2},s) =
       \frac{\exp\left\{-\frac{1}{2} x^2\right\}
             \mathrm{d}\vec{\mu_1}\mathrm{d}\vec{\mu_2}}
            {2\pi\sqrt{\det(\bm{V_1} + \bm{V_2})} \gamma};
  \end{equation}
  \begin{equation}
    p(\vec{\mu_1}, \vec{\mu_2}|h_2,\bm{V_1},\bm{V_2},s) =
       \frac{\mathrm{d}\vec{\mu_1}\mathrm{d}\vec{\mu_2} }
            {\pi k_\gamma^2 }.
  \end{equation}
  If we compare the likelihood ratio $LR_{\vec{\mu},\bm{V}}$
  computed from those formulae with the likelihood ratio $LR_x$
  computed from the previous result (\S \ref{sec:proba2catsMD})
  we obtain:
  \begin{eqnarray}
    LR_x                  & = & \frac{k_\gamma^2 e^{-x^2/2}}{2\gamma}; \\
    LR_{\vec{\mu},\bm{V}} & = & \frac{k_\gamma^2 e^{-x^2/2}}{2\gamma\sqrt{\det(\bm{V_1} + \bm{V_2})}}.
  \end{eqnarray}
  Contrary to $LR_x$, $LR_{\vec{\mu},\bm{V}}$ accounts for the size of positional errors.
  As we will see in the next section, the drawback is that we can hardly combine the likelihoods
  $p(\vec{\mu_1}, \vec{\mu_2}|h_k,\bm{V_1},\bm{V_2},s)$ with photometry based likelihoods.

\section{Bayesian probabilities with photometric data \label{sec:phot}}
All probabilities of association discussed so far are based on the likelihood
that the positions recorded in various catalogues are consistent with that of
a unique astrophysical object. However, one may wish to make additional
assumptions on the nature of the source (e.g. star, active galactic nucleus, etc.)
that could help decrease or, conversely, increase the plausibility of a given
association of catalogue entries. This is particularly important when one seeks
to gather homogeneous samples of objects. Spectral energy distributions assembled
from photometric catalogues can be usefully compared with templates and assigned
a probability of being representative of the targeted class of objects.
This procedure has been presented in \cite{Budavari2008} and recently
used in \cite{Naylor2013} and \cite{Hsu2014}. However, following
\cite{Budavari2008} we underline that entering criteria of resemblance to a
given class of objects in the computation of association probabilities is done
at the expense of the capability to find scientifically interesting outliers.

Another possibility may consist in building colour-colour diagrams 
for random and $\rchi$-matched associations to derive
colour-colour diagrams for real associations (more precisely, we have to derive
each diagram for each possible hypothesis $h_i$).
Those normalized diagrams are p.d.f that can be interpreted as likelihoods
($p(\vec{m}|h_i,s)$ in the following equations).
Smoothing those diagrams, one can see them as the likelihoods of
the kernel density classification \citep{Richards2004},
replacing the object types by the hypothesis $h_i$, and using magnitudes
from different catalogues instead of just one.\\

We detail below how photometric data can be folded into the output
of the purely astrometric method discussed in this paper, 
without making additional assumptions on the nature of the source.

Suppose we note $\vec{a}$, a vector containing all the astrometric
information we have about a set of $n$ candidates ($\vec{a}$ may
contain the positions, the associated covariance matrices, ...).
We note $\vec{m}$ the set of photometric informations we have
about the same set of $n$ candidates ($\vec{m}$ may contain
magnitudes and/or colours, associated errors, ...).
Then we can write the Bayes formula:
\begin{equation}
  p(h_i|\vec{a},\vec{m},s) =
    \frac{p(h_i|s)p(\vec{a}|h_i,s)p(\vec{m}|h_i,s,\vec{a})}
         {\sum\limits_{k=1}^{B_n}
	  p(h_k|s)p(\vec{a}|h_k,s)p(\vec{m}|h_k,s,\vec{a})}.
  \label{eq:bg}
\end{equation}
If $\vec{a}$ and $\vec{m}$ are independent (naive hypothesis
which is not granted, see \S \ref{sec:errorsnoinde})
\begin{equation}
  p(\vec{m}|h_k,\vec{a},s) = p(\vec{m}|h_k,s),
\end{equation}
and Eq. (\ref{eq:bg}) becomes
\begin{equation}
  p(h_i|\vec{a},\vec{m}) =
    \frac{p(h_i|s)p(\vec{a}|h_i,s)p(\vec{m}|h_i,s)}
         {\sum\limits_{k=1}^{B_n}
	  p(h_k|s)p(\vec{a}|h_k,s)p(\vec{m}|h_k,s)}.
  \label{eq:bgr}
\end{equation}
Let's imagine we perform a cross-match taking into account astrometric data only.
We compute probabilities $p(h_i|\vec{a},s)$ for all possible hypotheses.
Then if $\vec{a}$ and $\vec{m}$ are independent, and if we are able to compute
likelihoods based on photometric data only $p(\vec{m}|h_i,s)$, then we can
compute the probabilities $p(h_i|\vec{a},\vec{m},s)$ in a second step from
the probabilities computed in the astrometric part:
\begin{equation}
  p(h_i|\vec{a},\vec{m},s) =
    \frac{p(h_i|\vec{a},s)p(\vec{m}|h_i,s)}
         {\sum\limits_{k=1}^{B_n}
          p(h_k|\vec{a},s)p(\vec{m}|h_k,s)
	 },
  \label{eq:bgrc}
\end{equation}
which is equivalent to Eq. (\ref{eq:bgr}).\\

Unfortunately, positional errors and magnitudes are not necessarily independent.
So one should not use $p(h_i|\vec{\mu},\vec{\bm{V}},s)$ without any due caution
in Eq. (\ref{eq:bgrc}).
However, one can use the Mahalanobis distance $x$ which is independent of the
photometry, that is probabilities $p(h_i|x,s)$  (Eq. \ref{eq:bayesformula}).

\section{Tests on synthetic catalogues \label{sec:tests}}
In the context of the ARCHES project, we developed a tool implementing
the statistical multi-catalogue cross-match described in this paper.
We added to the tool the possibility to generate synthetic catalogues
that can be cross-matched like real tables.
It has been allowing us to perform tests and to check both the software
and the theory.

We present here such a test and provide the associated script
 (see\S \ref{sec:syntscript}) so anybody can
try it independently, possibly changing the input values.
Currently the tool is accessible both via a web interface and an HTTP API\footnote{
\url{http://serendib.unistra.fr/ARCHESWebService/index.html}}.
Future plans are discussed in the conclusion \S \ref{sec:conclu}.

We generate three synthetic catalogues, setting the numbers of sources
they contain and have in common: we call $n_{ABC}$ the number of common sources in the three
catalogues $A$, $B$ and $C$;
$n_{AB}$ the number of common sources in $A$ and $B$ only;
$n_A$ the number of sources in catalogue $A$ only; and so on.
Knowing a priori common and distinct sources in the catalogues, we can
track the associations which are real and the spurious ones
in the cross-match output.
We can also check for missing associations.

The error associated to each individual position is a random value which 
follows a user define distribution. A different distribution is used for
each catalogue.
For catalogue $A$, we choose a constant value equal to 0.4"; for catalogue $B$,
the positional error distribution follows a linear function between 0.8" and 1.2";
for catalogue $C$, the positional errors follow a Gaussian distribution of mean
0.75" and standard deviation $0.1$" truncated to the $0.5$--1" range.

We set the input sky area to be a cone of radius 0.42 degrees.
Each position is randomly (uniform distribution) placed in this cone.
For each catalogue in which the source is included, we randomly pick a
positional error that we associate to the source and we blur the position using its error. 

We first compute $\overline{\bm{V_{\Sigma}}}$ for pairs AB, AC and BC.
Given the chosen error distributions, the mean errors are equal to the median errors
and the mean of the inverse of the errors is quite close to the inverse of the mean errors.
So, for this particular case, we use the inverse of the mean errors $0.4$, $1$ and $0.75$
instead of the means of the inverse.
Given this approximation and using Eq. \ref{eq:detvsig}, we obtain
\begin{eqnarray}
  \overline{\sqrt{\det \bm{V_{\Sigma_{AB}}}}} & = & 
    \frac{1}{\frac{1}{\overline{\sqrt{\det{\bm{V_A}}}}}+\frac{1}{\overline{\sqrt{\det{\bm{V_B}}}}}}, \\
    & = & \frac{0.4^2 \times 1.0^2}{0.4^2 + 1.0^2} = 0.138; \\
  \overline{\sqrt{\det \bm{V_{\Sigma_{AC}}}}} & = & 
    \frac{1}{\frac{1}{\overline{\sqrt{\det{\bm{V_A}}}}}+\frac{1}{\overline{\sqrt{\det{\bm{V_C}}}}}}, \\
    & = & \frac{0.4^2 \times 0.75^2}{0.4^2 + 0.75^2} = 0.125; \\
  \overline{\sqrt{\det \bm{V_{\Sigma_{BC}}}}} & = & 
    \frac{1}{\frac{1}{\overline{\sqrt{\det{\bm{V_B}}}}}+\frac{1}{\overline{\sqrt{\det{\bm{V_C}}}}}}, \\
    & = & \frac{1.0^2 \times 0.75^2}{1.0^2 + 0.75^2} = 0.36.
\end{eqnarray}
And, to estimate the number of ``fully'' spurious associations, we have to compute the mean
of the square root of Eq. (\ref{eq:detvsigAnotherform}) over all possible source trios, which
can be approximated in this specific case by
\begin{eqnarray}
  \overline{\left(\frac{\det\bm{V_A}\det\bm{V_B}\det\bm{V_C}}{\det\bm{V_{\Sigma_{ABC}}}}\right)^{1/2}} & \approx &
    1.0^2\times 0.75^2 + 0.4^2 \times 0.75^2 \nonumber\\
    & & + 0.4^2\times1.0^2, \\
    & \approx & 0.8125.
\end{eqnarray}
We note that in this particular case, the error distribution of sources $A$ involved in 
$AB$, $AC$ and $ABC$ associations is the same. Idem for the error distribution of $B$ and $C$
sources.

\begin{figure*}
  \centering
  \includegraphics[width=0.3\textwidth]{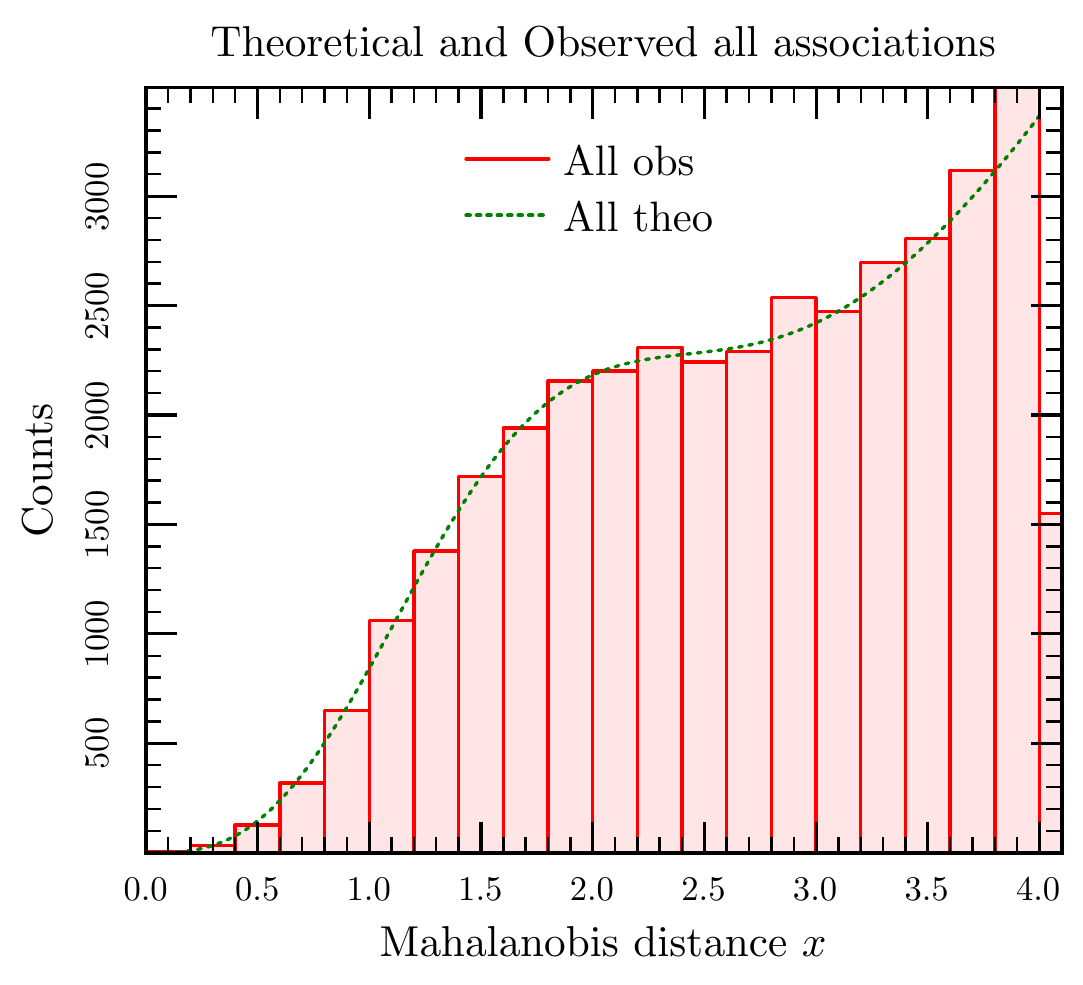}
  \includegraphics[width=0.3\textwidth]{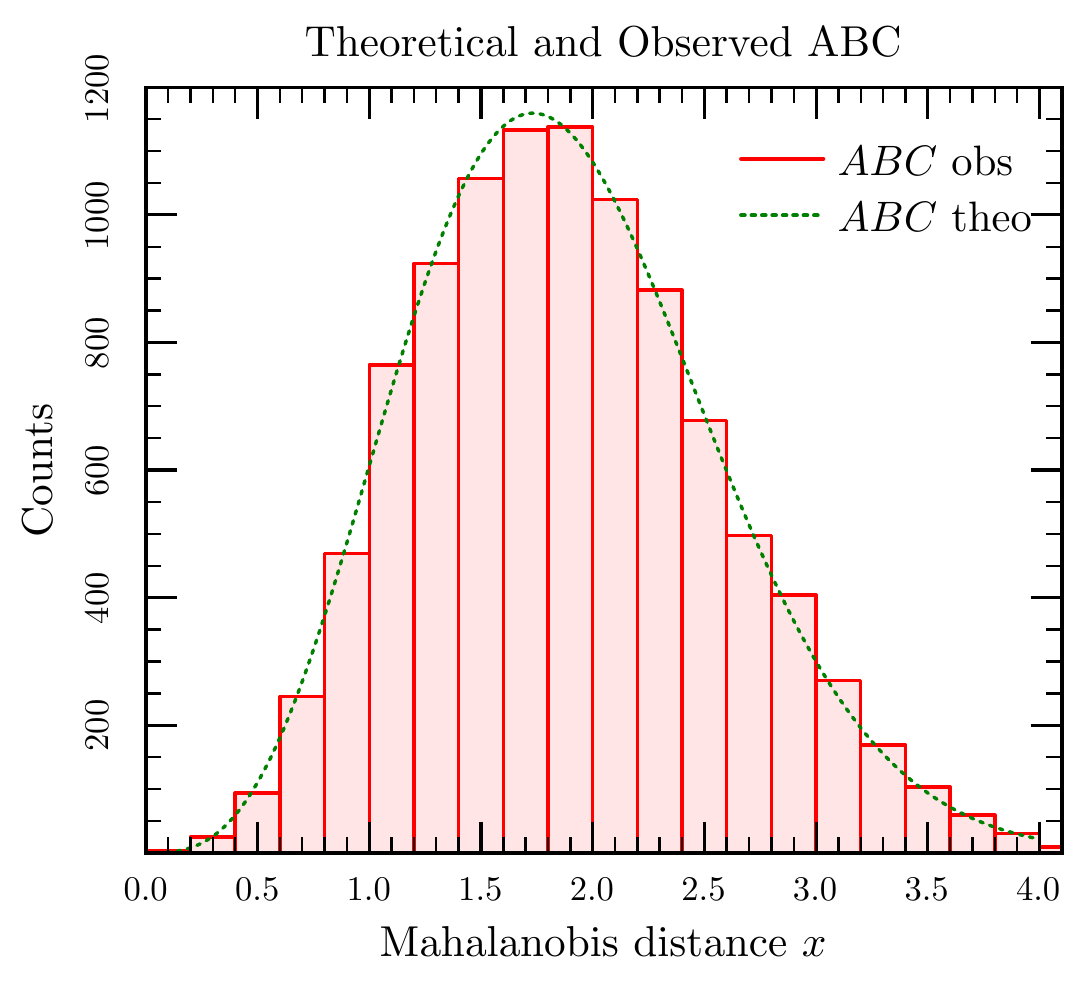}
  \includegraphics[width=0.3\textwidth]{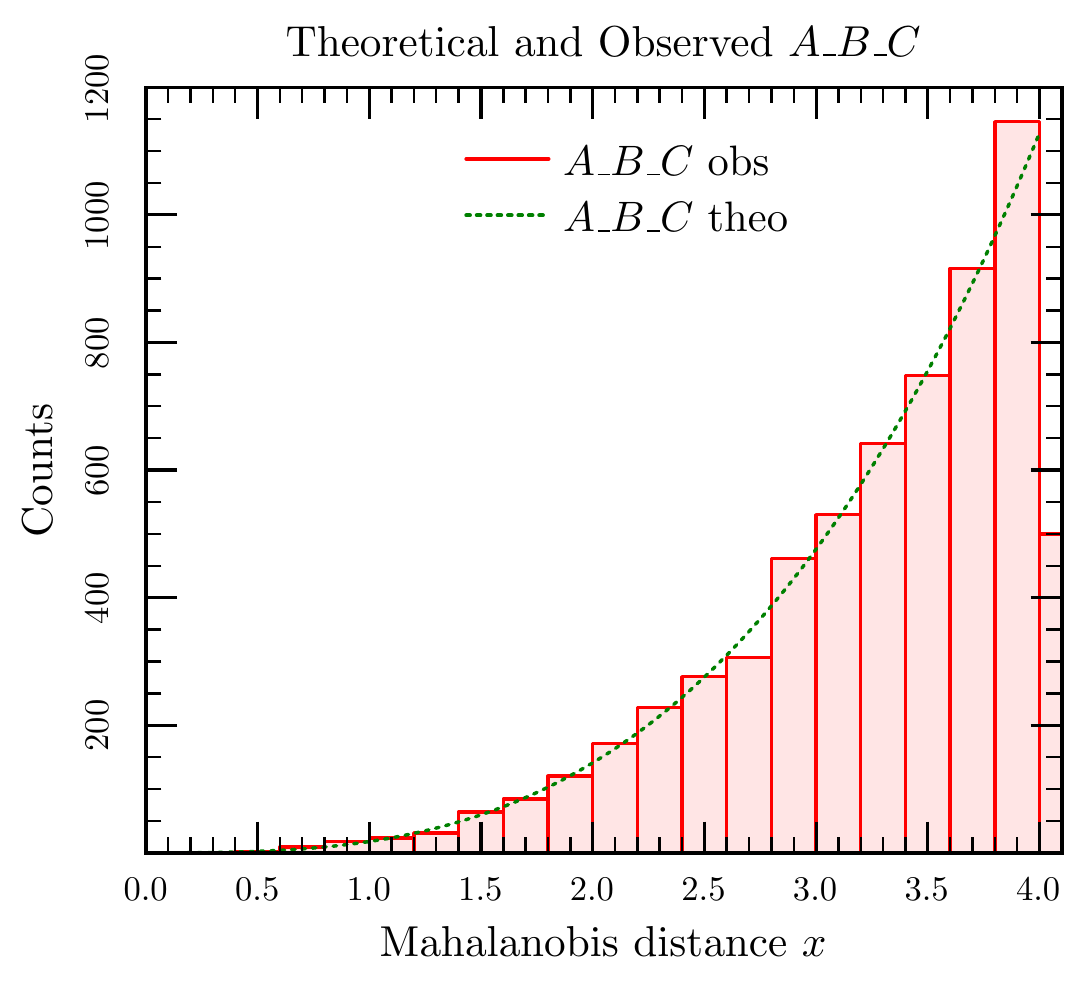} \\
  \includegraphics[width=0.3\textwidth]{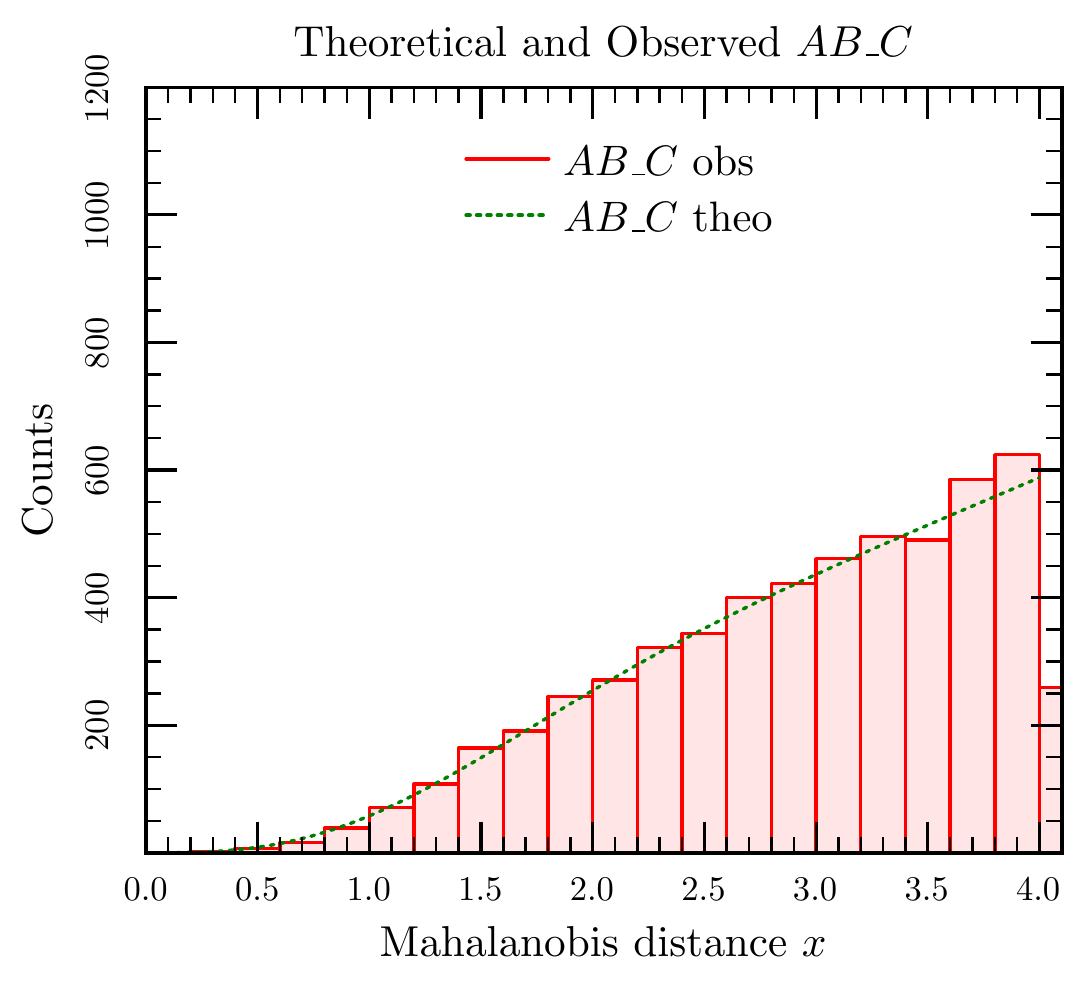}
  \includegraphics[width=0.3\textwidth]{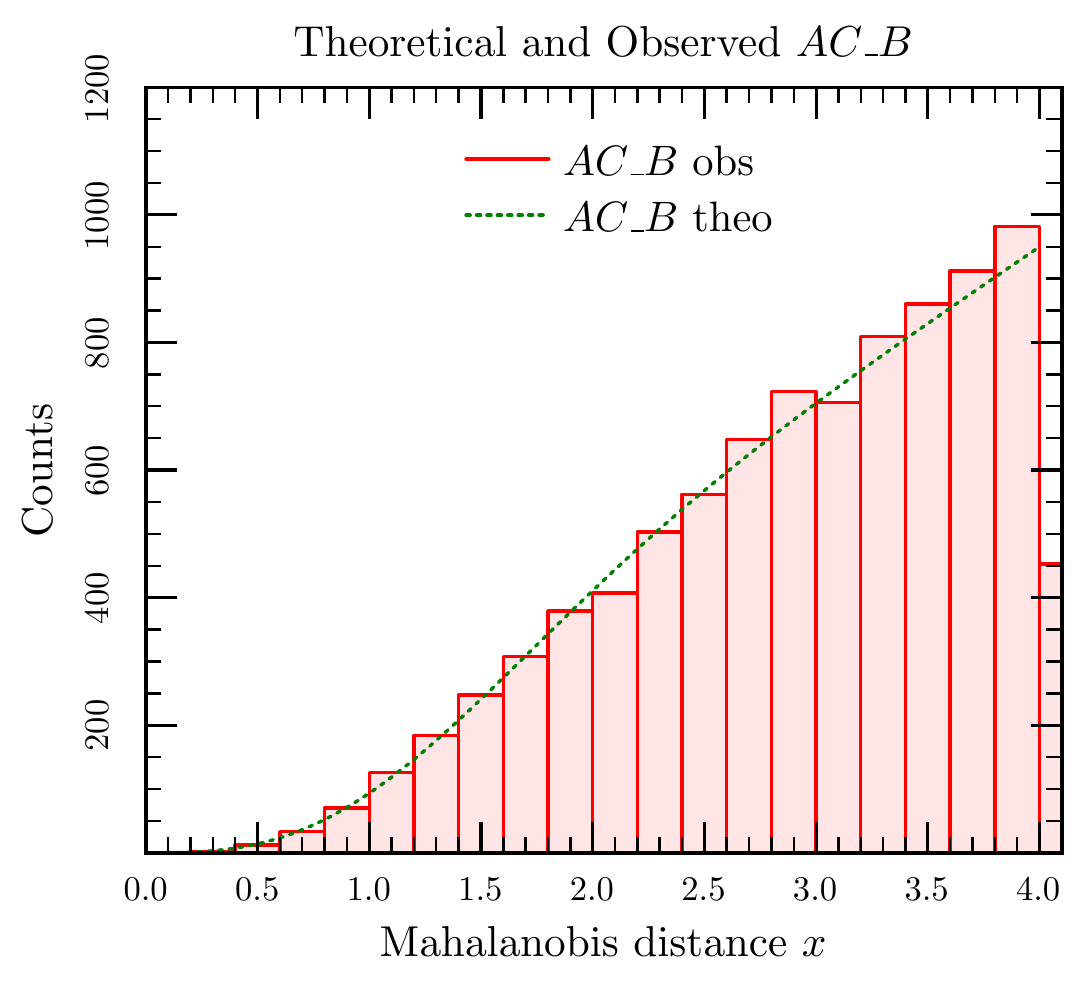}
  \includegraphics[width=0.3\textwidth]{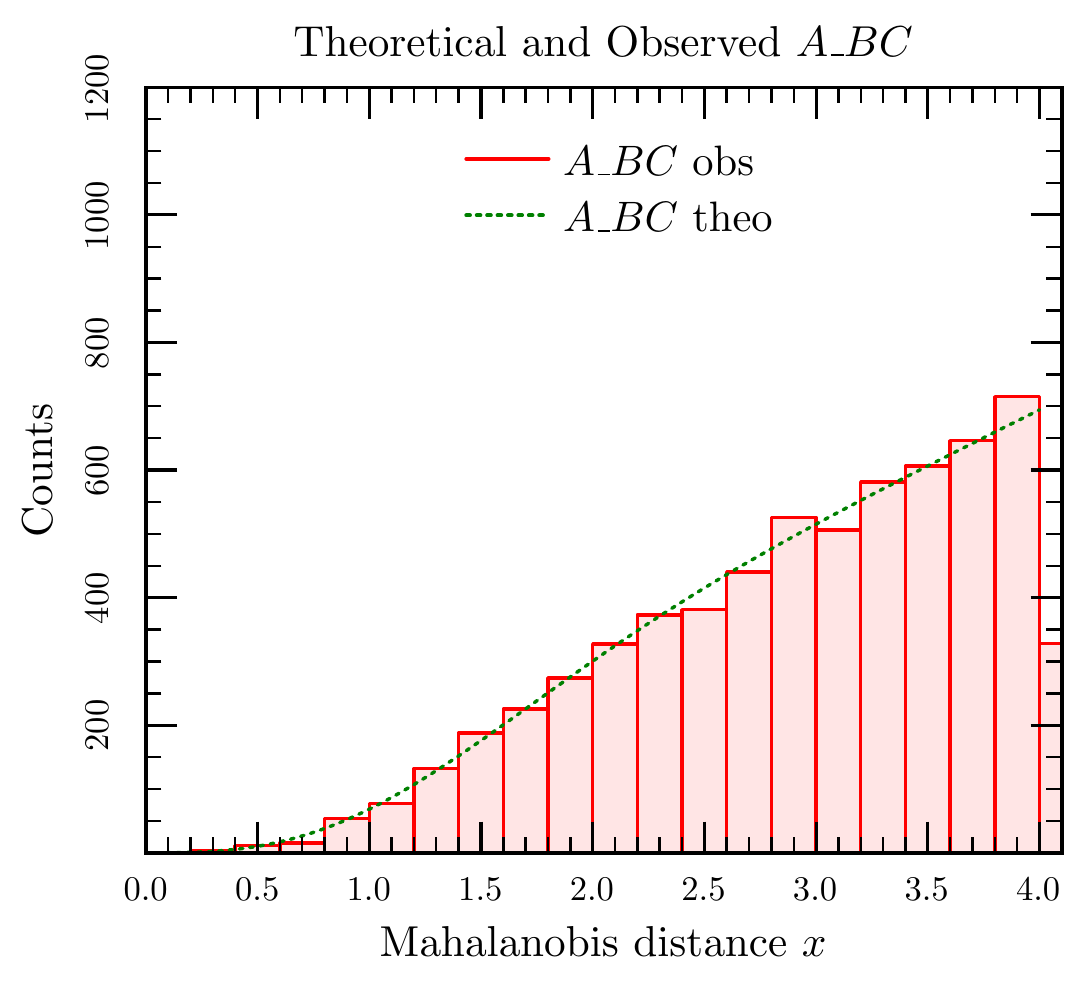}
  \caption{Result of the cross-match of three synthetic catalogues with input values
           $n_A=40\,000$, $n_B=20\,000$ $n_C=35\,000$ $n_{AB}=6\,000$, $n_{AC}=12\,000$, $n_{BC}=18\,000$
	   and $n_{ABC}=10\,000$. The error on catalogue $A$ is a constant equal to 0.4\arcsec.
	   The circular error on catalogue $B$ follows a linear distribution between 0.8 and 1.2\arcsec.
	   The circular error on catalogue $C$ follows a Gaussian distribution of mean 0.75\arcsec and standard deviation
           of 0.1\arcsec between 0.5 and 1\arcsec.
           The common surface area is a cone of radius 0.42\degr.
           Top left: histogram of all associations and theoretical curve from the input parameters.
           Top centre: histogram of real associations and theoretical curve from input parameters.
           Top right: histogram of ``fully'' spurious associations and theoretical curve from input paramameters.
           Bottom: histograms and theoretical curves of associations mixing a real association between two sources plus a spurious source.}
  \label{fig:simu3}
\end{figure*}

We are now able to compute all components, depending on the size of histograms bins ($step$).
To do this we note
\begin{eqnarray}
  n_{AB*} & = & n_{ABC} + n_{AB}, \\
  n_{AC*} & = & n_{ABC} + n_{AC}, \\
  n_{BC*} & = & n_{ABC} + n_{BC}, \\
  n_{A*}  & = & n_{ABC} + n_{AB} + n_{AC} + n_A, \\
  n_{B*}  & = & n_{ABC} + n_{AB} + n_{BC} + n_B, \\
  n_{C*}  & = & n_{ABC} + n_{AC} + n_{BC} + n_C,
\end{eqnarray}
to finally obtain
\begin{eqnarray}
  \hat{n}_{ABC}(x)     & = & step \times n_{ABC} \times \rchi_{dof=4}(x), \\
  \hat{n}_{AB\_C}(x)   & \approx & step \times n_{AB*} \times n_{C*} \frac{0.138+0.75^2}{\pi(0.42\times 3600)^2} \nonumber \\
                       & &     2\pi x (1-\exp(-\frac{1}{2}x^2)), \\
  \hat{n}_{AC\_B}(x)   & \approx & step \times n_{AC*}  \times n_{B*} \frac{0.125+1.0^2}{\pi(0.42\times 3600)^2} \nonumber \\
                       & &    2\pi x (1-\exp(-\frac{1}{2}x^2)), \\
  \hat{n}_{BC\_A}(x)   & \approx & step \times n_{BC*} \times n_{A*} \frac{0.36+0.4^2}{\pi(0.42\times 3600)^2} \nonumber\\
                       & &    2\pi x (1-\exp(-\frac{1}{2}x^2)), \\
  \hat{n}_{A\_B\_C}(x) & \approx & step \times n_{A*} n_{B*} \times n_{C*} \frac{0.8125}{\pi^2(0.42\times 3600)^4} \nonumber \\
                       & &2\pi^2x^3, \\
  \hat{n}_{Tot}(x)     & \approx & \hat{n}_{ABC}(x) + \hat{n}_{A\_B\_C}(x) \nonumber \\
                       & & + \hat{n}_{AB\_C}(x) + \hat{n}_{AC\_B}(x)  + \hat{n}_{BC\_A}(x).
\end{eqnarray}
Each component equals $step\times n \times p(s|h_i)p(x|h_i,s)$, see Eq. (\ref{eq:pshk}) and Eq. (\ref{eq:pxhis}).
We simplify the expression since $p(s|h_i)\propto I_{k,n}(k_\gamma)$ and $p(x|h_i,s) \propto 1/I_{k,n}(k_\gamma)$.
The normalized histograms associated to each component are distributed according to each likelihood $p(x|h_i,s)$.
Both histograms made from the data and theoretical curves are plotted in Fig. \ref{fig:simu3}.
The theoretical results fit very well the result of the Chi-square cross-match based on simulated data.

We also verify that the number of ``good'' ABC matches we obtain as output of the cross-match 
is coherent with $n_{ABC}$ times the input completeness $\gamma$.

When cross-matching real catalogues, the number of sources $n_{ABC}$, etc. are not known.
But the previous ``theoretical'' curves can be built after a $\rchi$-match from the number
of sources estimated to compute priors in \S \ref{sec:spurest}.

  \section{Summarized recipe \label{sec:recipe}}
  In this section, we give the main steps and equations
to perform a $\rchi$-match and to compute for each association
the probability it has to be a good match (or any other possible hypothesis).

For a small and compact sky area, project all the sources of all catalogues on
an Euclidian plane using for example the ARC projection \citep{Calabretta2002}.

To select matching candidates, for each possible set of $n$ sources from $n$ distinct catalogues:
\begin{itemize}
  \item compute their weighted mean position (Eq. \ref{eq:wmeanpos}) and the associated error (Eq. \ref{eq:wposerr});
  \item derive $x$, the Mahalanobis distance defined by the square root of Eq. (\ref{eq:qchi2form2});
  \item fix a constant threshold on all Mahalanobis distances, that is
  \begin{itemize}
    \item set the fraction $\alpha$ of real associations it is acceptable to miss -- the type I error -- and
    \item derive numerically the threshold $k_\gamma$ inverting Eq. (\ref{eq:alpha})
        based on the Chi-square distribution with $2(n-1)$ degrees of freedom
	(the result is the same computing the threshold from the Chi distribution with $2(n-1)$ degrees of freedom);
  \end{itemize}
  \item keep all sets of $n$ sources having a Mahalanobis distance less than the threshold $k_\gamma$
	(Eq. \ref{eq:candselectcri}) as possibly being $n$ observations of a same real
        source. 
\end{itemize}

To compute Bayes' probabilities, as many hypotheses as
the number of possible partitions of the set of $n$ sources
(see Eq. (\ref{eq:bellnumber}), Table \ref{tab:bell} and Fig. \ref{fig:bell5})
have to be formulated.
Depending on whether one wants to be able to account for photometry in a second step
or not, a set of likelihoods may be chosen among several such sets.\\
In the first case, the likelihood associated to each hypothesis
(knowing the selection criteria is fulfilled) depends only on
the Mahalanobis distance and on the number of real sources $k$ in the hypothesis
$h_i$ (see e.g. Fig \ref{fig:likelihoods_n4}).
The likelihoods are (Eq. \ref{eq:pxhis})
\begin{equation}
  p(x|h_i,s) = \frac{\mathrm{d}I_{k,n}(x)\mathrm{d}x}{I_{k,n}(k_\gamma)},
\end{equation}
with $I_{k,n}(k_\gamma)$ given in Eq. (\ref{eq:ikn}).
The formulae of $I_{k,n}(k_\gamma)$ and $\mathrm{d}I_{k,n}(x)$ are provided
for $n\le5$ in tables Table \ref{tab:imn} and \ref{tab:dimn} respectively.\\
In the second case (no photometry to be taken into account), one can use likelihoods
defined by Eq. (\ref{eq:likelihodds_v2}).\\
Finally, to apply Bayes' formula, priors $p(h_i|s)$ are needed.
This is more tricky and the steps detailed in section \S \ref{sec:spurest}
have to be considered.
A pre-requisite is to work on an area ($\Omega$) uniformly covered by all catalogues.
For two catalogues, the number of spurious associations can be estimated by
computing for each catalogue the mean area covered by the error ellipses
for a radius equal to the threshold $k_\gamma$ (so the mean area of the $1\sigma$ error ellipses
times $k_\gamma^2$).
The two means are summed and the result is divided by $\Omega$ to obtain the mean probability to
spuriously associate two unrelated sources.
It is then multiplied by the product of the number of sources in both catalogues
to finally obtain the mean expected number of spurious associations (Eq. \ref{eq:estimnspur1}).
Knowing the number of associations in the cross-correlation output, 
both the probability that one such association is spurious and the complementary
probability of having a real association (the two priors of the 2-catalogues cross-match)
can be estimated (see Eq. \ref{eq:estimatedPriorAB} and \ref{eq:estimatedPriorApasB}).
Similarly, performing all possible sub-$\rchi$-matches, all priors needed
for a $n$-catalogues cross-match can be derived.\\
All needed ingredients to compute the probabilities associated
with each hypothesis are thus available. 
Those probabilities can be computed applying Eq. (\ref{eq:bayesformula}).

  \section{Conclusions \label{sec:conclu}}
In this paper we developed a comprehensive framework for performing the
cross-correlation of multiple astronomical catalogues, in one pass.
The approach employs a classical $\rchi^2$-test to select candidates.
We computed two sets of likelihoods based on positions, individual elliptical
positional errors and the $\rchi^2$-test region of acceptance: 
one that can be mixed without any caution with other parameters such as photometric values; 
and one for which the naive hypothesis of independence between positional uncertainties
and magnitudes has to be tested.
We also presented a way to estimate ``priors'' from the region of acceptance
of the $\rchi^2$-test.
Probabilities for each possible hypothesis can thus be computed from those
likelihoods and ``priors''.\\
In practice the number of hypotheses, and thus the number of ``priors'',
increases dramatically with the number of catalogues.
To be able to cross-match more than six or seven catalogues, it is necessary 
to simplify the problem.
One possibility consists of merging two catalogues of similar astrometric accuracy
and similar wavelength range, considering all matches as non-spurious matches.
Doing so we would effectively reduce the number of input catalogues by one.

A large part of the statistical work carried out here depends on the
simplifying assumptions made in \S \ref{sec:assumptions}:
perfect astrometrical calibration (no systematic offsets), no proper motions,
no clustering and no blending.
In real life, the ``normalized'' distance between two detections of a same
source present in two distinct catalogues hardly follows a Rayleigh
distribution.
The ``actual distribution'' ({in practice it is not easy to build such a
distribution since it requires secure identifications) often has a
broader tail \citep[see for example][Fig. 5]{Rosen2016} and a log-normal distribution may better fit it than the Rayleigh distribution.
This is probably due to a combination of causes like small proper motions, imperfect reduction, systematics or
bias from the calibration process, under or overestimated errors, etc.\\
In practice this means that the number of associations missed by the candidate
selection criteria (based on Rayleigh) is larger that the chosen theoretical value ($\gamma$).
We could for example add larger systematics to positional errors. The risk is then
to distort (even more) the Rayleigh distribution.
We could also try to re-calibrate locally the set of catalogues we want to cross-match,
but we need secure identifications to do it properly;
for each catalogue, all sources in the local area must have been calibrated
at once (to possibly correct for a locally uniform systematic using four simple parameters
$\Delta\alpha$, $\Delta\delta$, $scale$, $\theta$).
Those two constraints (having secure identifications and at once calibration) are in practice quite hard to satisfy.\\
If we re-calibrate using a ``secure'' population (i.e. a population of objects having no proper motions like QSOs)
we introduce a bias since QSOs are fainter than most stars in the optical and thus have errors larger than the global population of objects.
And adding stars, we introduce noise due to proper motions.

For these reasons, we believe that in case of ``old'' optical surveys based on photographic plates,
a classical fixed radius cross-match may be more efficient that the $\rchi$-match to select candidates.
We are nonetheless conviced that the equations we derived in this paper can help
in building new catalogues, based for example on both multi-band and multi-epoch observations,
and can be used to assess and improve the quality of coming surveys.

We generated and processed synthetical catalogues, which meet the simplifying assumptions,
in the tool we developed for the ARCHES project.
The consistency between the theoretical results derived in this paper
-- completeness of the candidate selection criterion, likelihoods and priors --
and the outputs of the tool has allowed us to cross-validate both the method and its implementation.
The tool has also been used to generate ARCHES products which were used in the scientific work packages of the project.
Currently the CDS XMatch Service \citep{Pineau2011b, Boch2012, Pineau2015} provides a basic but very efficient facility to cross-correlate
two possibly large (> 1 billion sources) catalogues.
It is planned to include the ARCHES tool into the CDS XMatch.
This paper will be the basic reference for the extension of the latter to multi-catalogue statistical $\rchi$-match.

  \begin{acknowledgements}
A large part of this work was supported by the ARCHES project. 
ARCHES (n$^0$ 313146) was funded by the 7$^\mathrm{th}$ Framework
of the European Union and coordinated by the University of Strasbourg.\\
All figures (except Fig. 1) were made using the ctioga2 plotting program developped by Vincent Fourmond.\\ 
F. J. Carrera also acknowledges financial support through grant AYA2015-64346-C2-1-P (MINECO/FEDER).\\
Many thanks to Fran\c{c}ois Ochsenbein for pointing out an error in the $1\sigma$ conversion of FIRST positional errors (Eq. \ref{eq:first.a} and \ref{eq:first.b}):
the FIRST errors do come from an ellipse fit, not from two individual 1-dimensional fits of the semi-major and semi-minor axis.
Hence the factor is $\sqrt{2\ln 10}$ instead of $1.64485$.\\
Many thanks also to Fran\c{c}ois Mignard for pointing out an error in Eq. (\ref{eq:rhosigxsigy}) which is now correct: squares on $\cos$ and $\sin$ have been removed.
  \end{acknowledgements}
  
  \appendix
\section{Demonstrations}
\subsection{From $\bm{V_\Sigma}^{-1}$ to $\bm{V_\Sigma}$\label{sec:vsigma}}

From Eq. (\ref{eq:VSigma}), we compute for $2\times 2$ symmetric square matrices:
\begin{eqnarray}
  V_\Sigma & = & (\bm{V_\Sigma}^{-1})^{-1}, \\
           & = & \frac{1}{\text{det}\bm{V_\Sigma}^{-1}}
	         \text{adj}(\bm{V_\Sigma}^{-1}), \\
           & = & \frac{1}{\text{det}\bm{V_\Sigma}^{-1}}
	         \text{adj}\sum\limits_{i=1}^n\bm{V_i}^{-1}, \\
           & = & \frac{1}{\text{det}\bm{V_\Sigma^{-1}}}
	         \sum\limits_{i=1}^n\text{adj}\bm{V_i}^{-1}, \\
	   & = & \frac{1}{\text{det}\bm{V_\Sigma}^{-1}}
	         \sum\limits_{i=1}^n\text{adj}
		 \frac{\text{adj}\bm{V_i}}{\text{det}\bm{V_i}}, \\
	   & = & \frac{1}{\text{det}\bm{V_\Sigma}^{-1}}
	         \sum\limits_{i=1}^n\frac{\bm{V_i}}{\text{det}\bm{V_i}}, 
\end{eqnarray}
in which $\text{adj}\bm{A}$ is the adjugate matrix of $\bm{A}$,
that is the transpose of the cofactor matrix of $\bm{A}$.

\subsection{Sum of quadratics canonical form\label{sec:quad}}

First expanding and then factoring:
\begin{eqnarray}
  \sum\limits_{i=1}^nQ_i(\vec{x})
    & = & \sum\limits_{i=1}^n\transposee{(\vec{x}-\vec{\mu_i})}
          \bm{V_}i^{-1}(\vec{x}-\vec{\mu_i}), \\
    & = & \sum\limits_{i=1}^n\transposee{\vec{x}}\bm{V_i}^{-1}\vec{x}
          - 2\sum\limits_{i=1}^n\transposee{\vec{\mu_i}}\bm{V_i}^{-1}\vec{x} \nonumber \\
    &   &  + \sum\limits_{i=1}^n\transposee{\vec{\mu_i}}\bm{V_i}^{-1}\vec{\mu_i}.
\end{eqnarray}
We use
\begin{eqnarray}
  \sum\limits_{i=1}^n\transposee{\vec{\mu_\Sigma}}\bm{V_i}^{-1}
    & = & \sum\limits_{i=1}^n\transposee{\vec{\mu_i}}\bm{V_i}^{-1}, \\
  \transposee{\vec{\mu_\Sigma}}\sum\limits_{i=1}^n\bm{V_i}^{-1}
    & = & \sum\limits_{i=1}^n\transposee{\vec{\mu_i}}\bm{V_i}^{-1}, \\
  \transposee{\vec{\mu_\Sigma}}\bm{V_\Sigma}^{-1}
    & = & \sum\limits_{i=1}^n\transposee{\vec{\mu_i}}\bm{V_i}^{-1}, \\
  \transposee{\vec{\mu_\Sigma}}
    & = & \left(\sum\limits_{i=1}^n\transposee{\vec{\mu_i}}\bm{V_i}^{-1}\right)
          \bm{V_\Sigma},
\end{eqnarray}
that we introduce in the previous equations to finally find
\begin{eqnarray}
  \sum\limits_{i=1}^nQ_i(\vec{x})
    & = & \sum\limits_{i=1}^n\transposee{\vec{x}}\bm{V_i}^{-1}\vec{x}
     - 2\sum\limits_{i=1}^n\transposee{\vec{\mu_\Sigma}}\bm{V_i}^{-1}\vec{x}
     + \sum\limits_{i=1}^n\transposee{\vec{\mu_\Sigma}}
       \bm{V_i}^{-1}\vec{\mu_\Sigma} \nonumber \\
    & & 
     - \sum\limits_{i=1}^n\transposee{\vec{\mu_\Sigma}}\bm{V_i}^{-1}\vec{\mu_\Sigma}
     + \sum\limits_{i=1}^n\transposee{\vec{\mu_i}}\bm{V_i}^{-1}\vec{\mu_i}, \\
    & = & \sum\limits_{i=1}^n
     \transposee{(\vec{x}-\vec{\mu_i})}\bm{V_i}^{-1}(\vec{x}-\vec{\mu_i}) \nonumber \\
    & & 
     + \sum\limits_{i=1}^N\transposee{\vec{\mu_i}}V_i^{-1}\vec{\mu_i}
	    - \transposee{\vec{\mu_\Sigma}}V_\Sigma^{-1}\vec{\mu_\Sigma},
\end{eqnarray}
or,
\begin{eqnarray}
  \sum\limits_{i=1}^nQ_i(\vec{x})
    & = & \sum\limits_{i=1}^n
     \transposee{(\vec{x}-\vec{\mu_i})}\bm{V_i}^{-1}(\vec{x}-\vec{\mu_i}) \nonumber \\
    & &
     + \sum\limits_{i=1}^n\transposee{\vec{\mu_i}}\bm{V_i}^{-1}\vec{\mu_i} 
     - 2\sum\limits_{i=1}^n\transposee{\vec{\mu_i}}\bm{V_i}^{-1}\vec{\mu_\Sigma} \nonumber \\
    & &
     + \sum\limits_{i=1}^n\transposee{\vec{\mu_\Sigma}}\bm{V_i}^{-1}\vec{\mu_\Sigma}, \\
    & = & \sum\limits_{i=1}^n
     \transposee{(\vec{x}-\vec{\mu_i})}\bm{V_i}^{-1}(\vec{x}-\vec{\mu_i}) \nonumber \\
    & &  + \sum\limits_{i=1}^N\transposee{(\vec{\mu_i} - \vec{\mu_\Sigma})}V_i^{-1}
                                         (\vec{\mu_i} - \vec{\mu_\Sigma}).
\end{eqnarray}

\subsection{Expanding the $Q_{\rchi^2}$ term \label{sec:q2alldist}}

Noting that
\begin{equation}
    I = \bm{V_\Sigma}\bm{V_\Sigma}^{-1}
      = \bm{V_\Sigma}\sum\limits_{j=1}^n\bm{V_j}^{-1},
\end{equation}
we can write
\begin{eqnarray}
  \bm{V_i}^{-1}-\bm{V_i}^{-1}\bm{V_\Sigma}\bm{V_i}^{-1}
  & = & \bm{V_i}^{-1}(\bm{I} - \bm{V_\Sigma}\bm{V_i}^{-1}), \\
  & = & \bm{V_i}^{-1}\bm{V_\Sigma}\sum\limits_{j=1,j\ne i}^n\bm{V_j}^{-1},
\end{eqnarray}
and thus, with square symmetric matrices:
\begin{eqnarray}
  Q_{\rchi^2} & = &
    \sum\limits_{i=1}^n\transposee{\vec{\mu_i}}\bm{V_i}^{-1}\vec{\mu_i}
    - \transposee{\vec{\mu_\Sigma}}\bm{V_\Sigma}^{-1}\vec{\mu_\Sigma},\\
  & = & \sum\limits_{i=1}^n\transposee{\vec{\mu_i}}\bm{V_i}^{-1}\vec{\mu_i}
      - \sum\limits_{i=1}^n\transposee{\vec{\mu_i}}
        \bm{V_i}^{-1}\bm{V_\Sigma}\bm{V_i}^{-1}\vec{\mu_i} \nonumber \\
  & & - \sum\limits_{i=1}^n\sum\limits_{j=i+1}^n 2\transposee{\vec{\mu_i}}
        \bm{V_i}^{-1}\bm{V_\Sigma}\bm{V_j}^{-1}\vec{\mu_j}, \\
  & = & \sum\limits_{i=1}^n\transposee{\vec{\mu_i}}\bm{V_i}^{-1}\bm{V_\Sigma}
        \sum\limits_{j=1,j\ne i}^n\bm{V_j}^{-1}\vec{\mu_i} \nonumber \\
  & & - \sum\limits_{i=1}^n\sum\limits_{j=i+1}^n 2\transposee{\vec{\mu_i}}
        \bm{V_i}^{-1}\bm{V_\Sigma}\bm{V_j}^{-1}\vec{\mu_j}, \\
  & = & \sum\limits_{i=1}^n\sum\limits_{j=i+1}^n
              \transposee{(\vec{\mu_i}-\vec{\mu_j})}
              \bm{V_i^{-1}}\bm{V_\Sigma}\bm{V_j^{-1}}
	      (\vec{\mu_i}-\vec{\mu_j}).
\end{eqnarray}

 \subsection{Sum of 2 quadratics: $Q_\rchi$ term\label{sec:sum2q}}

    We first develop~:
    \begin{eqnarray}
      \transposee{\vec{\mu_{\Sigma_2}}}\bm{V_{\Sigma_2}}^{-1}\vec{\mu_{\Sigma_2}}
        & = & (\transposee{\vec{\mu_1}}\bm{V_1}^{-1} + \transposee{\vec{\mu_2}}\bm{V_2}^{-1})V_{\Sigma_2} \nonumber \\
	&   & (\bm{V_1}^{-1}\vec{\mu_1} + \bm{V_2}^{-1}\vec{\mu_2}), \\
        & = & (\transposee{\vec{\mu_1}}\bm{V_1}^{-1} + \transposee{\vec{\mu_2}}\bm{V_2}^{-1})
	      \frac{1}{|\frac{\bm{V_1}}{|\bm{V_1}|} + \frac{\bm{V_2}}{|\bm{V_2}|}|} \nonumber \\
	  & & (\frac{\bm{V_1}}{|\bm{V_1}|} + \frac{\bm{V_2}}{|\bm{V_2}|}) (\bm{V_1}^{-1}\vec{\mu_1} + \bm{V_2}^{-1}\vec{\mu_2}), \\
        & = & |\bm{V_{\Sigma_2}}|\left[\transposee{\vec{\mu_1}}\frac{\bm{V_1}^{-1}}{|\bm{V_1}|}\vec{\mu_1}
	      + \transposee{\vec{\mu_2}}\frac{\bm{V_2}^{-1}}{|\bm{V_2}|}\vec{\mu_2}\right. \nonumber \\
	  & & + \left.\transposee{\vec{\mu_1}}\frac{\bm{V_2}^{-1}}{|\bm{V_1}|}\vec{\mu_2}
	      + \transposee{\vec{\mu_2}}\frac{\bm{V_1}^{-1}}{|\bm{V_2}|}\vec{\mu_1} \right] + \nonumber \\
          & & |\bm{V_{\Sigma_2}}|\left[\transposee{\vec{\mu_1}}\frac{\bm{V_1}^{-1}}{|\bm{V_2}|}\vec{\mu_2}
              + \transposee{\vec{\mu_2}}\frac{\bm{V_2}^{-1}}{|\bm{V_1}|}\vec{\mu_1}\right. \nonumber \\
	  & & + .\transposee{\vec{\mu_1}}\bm{V_1}^{-1}\frac{\bm{V_2}}{|\bm{V_2}|}\bm{V_1}^{-1}\vec{\mu_1} \nonumber \\
	  & & + \left.\transposee{\vec{\mu_2}}\bm{V_2}^{-1}\frac{\bm{V_1}}{|\bm{V_1}|}\bm{V_2}^{-1}\vec{\mu_2}\right].
    \end{eqnarray}
    Computing $\bm{V_{\Sigma_2}}\bm{V_{\Sigma_2}}^{-1}$~:
    \begin{eqnarray}
      \bm{V_{\Sigma_2}}\bm{V_{\Sigma_2}}^{-1} & = & |\bm{V_{\Sigma_2}}|(\frac{\bm{V_1}}{|\bm{V_1}|}
                                                   + \frac{\bm{V_2}}{|\bm{V_2}|})(\bm{V_1}^{-1}+\bm{V_2}^{-1}), \\
                             \bm{I} & = & \frac{|\bm{V_{\Sigma_2}}|}{|\bm{V_1}|} \bm{I} + \frac{|\bm{V_{\Sigma_2}}|}{|\bm{V_1}|}\bm{V_1}\bm{V_2}^{-1} \nonumber \\
				    &   & + \frac{|\bm{V_{\Sigma_2}}|}{|\bm{V_2}|}\bm{V_2}\bm{V_1}^{-1} + \frac{|\bm{V_{\Sigma_2}}|}{|\bm{V_2}|}\bm{I}, \\
      (1-\frac{|\bm{V_{\Sigma_2}}|}{|\bm{V_1}|}-\frac{|\bm{V_{\Sigma_2}}|}{|\bm{V_2}|})\bm{I}
                                    & = & \frac{|\bm{V_{\Sigma_2}}|}{|\bm{V_1}|}\bm{V_1}\bm{V_2}^{-1} 
                                        +  \frac{|\bm{V_{\Sigma_2}}|}{|\bm{V_2}|}\bm{V_2}\bm{V_1}^{-1} \\
    \end{eqnarray}
    We can write~:
    \begin{eqnarray}
      \bm{V_2}^{-1}(1-\frac{|\bm{V_{\Sigma_2}}|}{|\bm{V_1}|}-\frac{|\bm{V_{\Sigma_2}}|}{|\bm{V_2}|})I
        & = & \frac{|\bm{V_{\Sigma_2}}|}{|\bm{V_1}|}\bm{V_2}^{-1}\bm{V_1}\bm{V_2}^{-1}
            + \frac{|\bm{V_{\Sigma_2}}|}{|\bm{V_2}|}\bm{V_1}^{-1}, \\
      \frac{|\bm{V_{\Sigma_2}}|}{|\bm{V_1}|}\bm{V_2}^{-1}\bm{V_1}\bm{V_2}^{-1}
        & = & (1 - \frac{|\bm{V_{\Sigma_2}}|}{|\bm{V_1}|} - \frac{|\bm{V_{\Sigma_2}}|}{|\bm{V_2}|})\bm{V_2}^{-1} \nonumber \\
        &   & - \frac{|\bm{V_{\Sigma_2}}|}{|\bm{V_2}|}\bm{V_1}^{-1},
    \end{eqnarray}
    and, similarly~:
    \begin{eqnarray}
      \bm{V_1}^{-1}(1-\frac{|\bm{V_{\Sigma_2}}|}{|\bm{V_1}|} - \frac{|\bm{V_{\Sigma_2}}|}{|\bm{V_2}|})\bm{I}
        & = & \frac{|\bm{V_{\Sigma_2}}|}{|\bm{V_1}|}\bm{V_2}^{-1} + \frac{|\bm{V_{\Sigma_2}}|}{|\bm{V_2}|}\bm{V_1}^{-1}\bm{V_2}\bm{V_1}^{-1}, \\
      \frac{|\bm{V_{\Sigma_2}}|}{|\bm{V_2}|}\bm{V_1}^{-1}\bm{V_2}\bm{V_1}^{-1}
        & = & (1 - \frac{|\bm{V_{\Sigma_2}}|}{|\bm{V_1}|} - \frac{|\bm{V_{\Sigma_2}}|}{|\bm{V_2}|})\bm{V_1}^{-1} \nonumber \\
        &   & - \frac{|\bm{V_{\Sigma_2}}|}{|\bm{V_1}|}\bm{V_2}^{-1}.
    \end{eqnarray}
    We use the above 3 relations to develop
    \begin{equation}
      \transposee{\vec{\mu_1}}\bm{V_1}^{-1}\vec{\mu_1} + \transposee{\vec{\mu_2}}\bm{V_2}^{-1}\vec{\mu_2}
      - \transposee{\vec{\mu_{\Sigma_2}}}\bm{V_{\Sigma_2}}^{-1}\vec{\mu_{\Sigma_2}},
    \end{equation}
    which leads to
    \begin{equation}
       \transposee{(\vec{\mu_1} - \vec{\mu_2})}\left( 
         \frac{|\bm{V_{\Sigma_2}}|}{|\bm{V_2}|}\bm{V_1}^{-1} + \frac{|\bm{V_{\Sigma_2}}|}{|\bm{V_1}|}\bm{V_2}^{-1}
       \right)(\vec{\mu_1} - \vec{\mu_2}).
    \end{equation}
    Using Eq. (\ref{eq:detequals}) we found that
    \begin{equation}
       \transposee{\vec{\mu_1}}\bm{V_1^{-1}}\vec{\mu_1} + \transposee{\vec{\mu_2}}\bm{V_2}^{-1}\vec{\mu_2}
      - \transposee{\vec{\mu_{\Sigma_2}}}\bm{V_{\Sigma_2}}^{-1}\vec{\mu_{\Sigma_2}} 
    \end{equation}
    is equal to
    \begin{equation}
    \transposee{(\vec{\mu_1} - \vec{\mu_2})}(\bm{V_1} + \bm{V_2})^{-1}(\vec{\mu_1} - \vec{\mu_2}).
    \end{equation}

\subsection{$\rchi$ and $\rchi^2$ distributions}

  \subsubsection{Definition}

  The $\rchi$ and $\rchi^2$ distributions with $k=2(n-1)$ degrees of freedom are
  defined as
  \begin{eqnarray}
    \rchi_k(x)   & = & \frac{2^{1-(n-1)}}
                           {\Gamma(n-1)} x^{2(n-1)-1}e^{-\frac{x^2}{2}}, \\
    \rchi_k^2(x) & = & \frac{2^{-(n-1)}}{\Gamma(n-1)}
                      x^{(n-1)-1}e^{-\frac{x}{2}},
  \end{eqnarray}
  with the gamma function $\forall l \in \mathbb{N}$, $\Gamma(l) = (l-1)!$
  it leads to
  \begin{eqnarray}
    \rchi_{k=2(n-1)}(x)   & = & \frac{2^{2-n}}{(n-2)!} x^{2n-3}e^{-\frac{x^2}{2}}, \\
    \rchi_{k=2(n-1)}^2(x) & = & \frac{2^{1-n}}{(n-2)!} x^{n-2}e^{-\frac{x}{2}}.
  \end{eqnarray}
  So for
  \begin{align}
    n & = 2 
    & \rchi_{k=2}(x)   & = xe^{-\frac{x^2}{2}}
    & \rchi_{k=2}^2(x) & = \frac{1}{2}e^{-\frac{x}{2}} \\
    n & = 3
    & \rchi_{k=4}(x)   & = \frac{1}{2}x^3e^{-\frac{x^2}{2}}
    & \rchi_{k=4}^2(x) & = \frac{1}{4}xe^{-\frac{x}{2}} \\
    n & = 4
    & \rchi_{k=6}(x)   & = \frac{1}{8}x^5e^{-\frac{x^2}{2}}
    & \rchi_{k=6}^2(x) & = \frac{1}{16}x^2e^{-\frac{x}{2}} \\
    n & = 5
    & \rchi_{k=8}(x)   & = \frac{1}{48}x^7e^{-\frac{x^2}{2}}
    & \rchi_{k=8}^2(x) & = \frac{1}{96}x^3e^{-\frac{x}{2}}
  \end{align}
  and so on.
  
  \subsubsection{Sum of two $\rchi$ functions}

  We show here that
  $\rchi_{k=2p}(x_1) + \rchi_{k=2q}(x_2) = \rchi_{k=2(p+q)}(x)$.
  The distribution we are looking for is the density function of
  $\rchi_{k=2p}(x_1)\rchi_{k=q}(x_2)\text{d}x_1\text{d}x_2$ given
  $x=\sqrt{x_1^2+x_2^2}$.
  We use polar coordinates so,
  $\text{d}x_1\text{d}x_2=x\text{d}x\text{d}\theta$,
  $x_1=x\cos\theta$ and $x_2=x\sin\theta$~:\\
  \begin{eqnarray}
    \rchi
      & = & \rchi_{k=2(p+q)}(x)\text{d}x, \\
      & = & \int_0^{\frac{\pi}{2}}
              \rchi_{k=2p}(x_1)\rchi_{k=2q}(x_2)
              x\text{d}x\text{d}\theta, \\
      & = & \int_0^{\frac{\pi}{2}}
              \frac{2^{1-p}}{\Gamma(p)} x^{2p-1}\cos^{2p-1}\theta \nonumber \\
      &   &   \frac{2^{1-q}}{\Gamma(q)} x^{2q-1}\sin^{2q-1}\theta
	      e^{-\frac{x^2}{2}}
              x\text{d}x\text{d}\theta, \\
      & = & \frac{2^{2-(p+q)}}{\Gamma(p)\Gamma(q)} x^{2(p+q)-2}e^{-\frac{x^2}{2}}
            B(p, q) x\text{d}x, \\
      & = & \frac{2^{1-(p+q)}}{\Gamma(p+q)} x^{2(p+q)-1}e^{-\frac{x^2}{2}} \text{d}x,
  \end{eqnarray}
  in which $B(p, q)$ is the beta function
  \begin{eqnarray}
    B(p,q) & = & 2 \int_0^{\frac{\pi}{2}}\cos^{2p-1}\theta\sin^{2q-1}\theta\text{d}\theta, \\
           & = & \frac{\Gamma(p)\Gamma(q)}{\Gamma(p+q)}.
  \end{eqnarray}

  \subsubsection{$\rchi_{k=2(n-1)}$ cumulative distribution function \label{sec:app:chicumul}}

    Directly integrating for $k=2$
    \begin{equation}
      F_{\rchi_{k=2}}(x)
      = \int_0^x \rchi_{k=2}(x')\mathrm{d}x'
      = \left[-e^{-\frac{1}{2}x'^2}\right]_0^x = 1 - e^{-\frac{1}{2}x^2}
    \end{equation}
    For $k=4$, we integrate by parts with
    \begin{equation}
      \begin{array}{lclclcl}
        u(x') & = & -\frac{1}{2}x'^2     & \qquad & u'(x') & = & -x' \\
        v(x') & = & e^{-\frac{1}{2}x'^2} & \qquad & v'(x') & = & -x'e^{-\frac{1}{2}x'^2}
      \end{array}
    \end{equation}
    so
    \begin{eqnarray}
      F_{\rchi_{k=4}}(x)
      & = & \int_0^x \rchi_{k=4}(x')\mathrm{d}x', \\
      & = & F_{\rchi_{k=2}}(x) + \left[-\frac{1}{2}x'^2e^{-\frac{1}{2}x'^2}\right]_0^x, \\
      & = & F_{\rchi_{k=2}}(x) - \frac{1}{2}x^2 e^{-\frac{1}{2}x^2}.
    \end{eqnarray}
    For $k=6$, also integrating by parts, we note
    \begin{equation}
      \begin{array}{lclclcl}
        u(x') & = & -\frac{1}{8}x'^4     & \qquad & u'(x') & = & -\frac{1}{2}x'^3 \\
        v(x') & = & e^{-\frac{1}{2}x'^2} & \qquad & v'(x') & = & -x'e^{-\frac{1}{2}x'^2}
      \end{array}
    \end{equation}
    and so
    \begin{eqnarray}
      F_{\rchi_{k=6}}(x)
      & = & \int_0^x \rchi_{k=6}(x')\mathrm{d}x', \\
      & = & F_{\rchi_{k=4}}(x) + \left[-\frac{1}{8}x'^4e^{-\frac{1}{2}x'^2}\right]_0^x, \\
      & = & F_{\rchi_{k=4}}(x) - \frac{1}{8}x^4 e^{-\frac{1}{2}x^2}.
    \end{eqnarray}
    We deduce the general form for $k=2(n-1)$~:
    \begin{equation}
      F_{\rchi_{k=2(n-1)}}(x) = 1 - e^{-\frac{1}{2}x^2}\sum\limits_{i=2}^{n}\frac{2^{2-i}}{(i-2)!}x^{2(i-2)}.
      \label{eq:Fchi}
    \end{equation}

\subsection{Computing the $I_{k,n}(x)$ integral \label{sec:app:ikn}}

  Let us first expand a few notations for a better readability
  \begin{eqnarray}
    \det J_F
      & = & x^{n-2}\prod_{i=1}^{n-3}\cos^{n-i-2}\theta_i, \\
      & = & x^{n-2}\cos^{n-3}\theta_1\cos^{n-4}\theta_2...\cos\theta_{n-3}.
  \end{eqnarray}
  We are supposed to use $|\det J_F|$ but $x$ is positive and all angles
  $\theta$ are $\in [0, \pi/2]$, so $\det J_F$ is always positive.
  Let us also expand
  \begin{equation}
    \prod\limits_{i=1}^{n-1}x_i
    = x^{n-1}\cos^{n-2}\theta_1\cos^{n-3}\theta_2...\cos\theta_{n-2}
          \prod\limits_{i=1}^{n-2}\sin\theta_i.
  \end{equation}
  Multiplying both expressions leads to
  \begin{eqnarray}
    \prod\limits_{i=1}^{n-1}x_i\det J_F
    & = & x^{2(n-1)-1}\cos^{2(n-2)-1}\theta_1\cos^{2(n-3)-1}\theta_2... \nonumber \\
    &   & \cos\theta_{n-2}\prod\limits_{i=1}^{n-2}\sin\theta_i
  \end{eqnarray}

  \subsubsection{Case $I_{k=1,n}(x)$}

    \begin{eqnarray}
      I_{k=1, n}(x)
        & = & \int_{x'=0}^{x' \le x} 
              \prod_{i=1}^{n-1} \rchi_2(x_i)
	      \prod_{i=1}^{n-1}\mathrm{d}x_i, \\
	& = & \int_{x'=0}^{x' \le x}
	      (\prod\limits_{i=1}^{n-1}x_i)
              e^{-\frac{1}{2}x'^2} \nonumber \\
        &   & (x'^{n-2}\prod_{i=1}^{n-3}\cos^{n-i-2}\theta_i)
	      \mathrm{d}x'\prod_{i=1}^{n-2}\mathrm{d}\theta_i, \\
	& = & \int_{x'=0}^{x' \le x} x'^{2(n-1)-1}e^{-\frac{1}{2}x'^2}
	                                  \mathrm{d}x' \nonumber \\
	&   & \int_0^\frac{\pi}{2}...\int_0^\frac{\pi}{2}
	      \prod\limits_{i=1}^{n-2}
	        \cos^{2(n-i-1)-1}\theta_i
                \sin\theta_i
		\mathrm{d}\theta_i, \\
	& = & \int_{x'=0}^{x' \le x}
	        x'^{2(n-1)-1}e^{-\frac{1}{2}x'^2}
	        \mathrm{d}x'
	      \prod\limits_{i=1}^{n-2} \frac{B(i,1)}{2}, \\
        & = & \int_{x'=0}^{x' \le x} \rchi_{2(n-1)}(x')\mathrm{d}x', \\
	& = & F_{\rchi_{k=2(n-1)}}(x).
    \end{eqnarray}
    The exact solution is given by Eq. (\ref{eq:Fchi}).\\
    We note the particular case $I_{k, n}(k_\gamma)=\gamma$
    if $k_\gamma$ computed for this particular value of $n$.

  \subsubsection{Case $I_{k=n,n}(x)$}

    \begin{eqnarray}
      I_{k=n, n}(x)
        & = & \int_{x'=0}^{x' \le x} 
              \prod_{i=1}^{n-1} 2\pi x_i
	      \prod_{i=1}^{n-1}\mathrm{d}x_i, \\
	& = & (2\pi)^{n-1}
	      \int_{x'=0}^{x' \le x} 
              (\prod\limits_{i=1}^{n-1}x_i) \nonumber \\
        &   & (x'^{n-2}\prod_{i=1}^{n-3}\cos^{n-i-2}\theta_i)
              \mathrm{d}x'\prod_{i=1}^{n-2}\mathrm{d}\theta_i, \\
	& = & (2\pi)^{n-1} \int_{x'=0}^{x' \le x}
	        x'^{2(n-1)-1}
	        \mathrm{d}x'
	      \prod\limits_{i=1}^{n-2} \frac{B(i,1)}{2}, \\
	& = & (2\pi)^{n-1}
	      \frac{1}{2(n-1)}x^{2(n-1)}
              \frac{2^{2-n}}{(n-2)!}, \\
	& = & \frac{\pi^{n-1}}{(n-1)!} x^{2(n-1)},
    \end{eqnarray}
    which is the volume of an hypersphere of dimension $2(n-1)$, also
    called $2(n-1)$-sphere.

  \subsubsection{Intermediate case $I_{k,n}(x)$}

    For $k > 1$ and $k < n$:
    \begin{eqnarray}
      I_{k, n}(X)
        & = & \int_{x=0}^{x \le X} 
              \prod_{i=1}^{n-k} \rchi_2(x_i)
              \prod_{i=n-k+1}^{n-1} 2\pi x_i
	      \prod_{i=1}^{n-1}\mathrm{d}x_i, \\
        & = & \int_{x=0}^{x \le X}
	      \int_{\theta_1=0}^{\theta_1=\frac{\pi}{2}}...
	      \int_{\theta_{n-2}=0}^{\theta_{n-2}=\frac{\pi}{2}}
              (2\pi)^{k-1}x^{2(n-1)-1} \nonumber \\
	&   & \sin\theta_1\cos^{2(n-2)-1}\theta_1
	      \sin\theta_2\cos^{2(n-3)-1}\theta_2... \nonumber \\
	&   & \sin\theta_{n-2}\cos\theta_{n-2}
              e^{-\frac{1}{2}x^2\cos^2\theta_1...\cos^2\theta_{n-1-k}} \nonumber \\
        &   & \mathrm{d}x\mathrm{d}\theta_1...\mathrm{d}\theta_{n-2}, \\
	& = & \int_{x=0}^{x \le X}
	      \int_{\theta_1=0}^{\theta_1=\frac{\pi}{2}}...
	      \int_{\theta_{n-1-k}=0}^{\theta_{n-1-k}=\frac{\pi}{2}}
              \left(\prod_{i=1}^{k-1}B(i,1)\right) \nonumber \\
        &   & \pi^{k-1}x^{2(n-1)-1} \sin\theta_1\cos^{2(n-2)-1}\theta_1... \nonumber \\
	&   & \sin\theta_{n-1-k}\cos^{2k-1}\theta_{n-1-k}
              e^{-\frac{1}{2}x^2\cos^2\theta_1...\cos^2\theta_{n-1-k}} \nonumber \\
        &   & \mathrm{d}x\mathrm{d}\theta_1...\mathrm{d}\theta_{n-1-k}.
    \end{eqnarray}
    In a first step, we integrated by parts using
    \begin{eqnarray}
        u(\theta_{n-1-k})  & = & \cos^{2(k-1)}\theta_{n-1-k}, \\
	u'(\theta_{n-1-k}) & = & -2(k-1)\sin\theta_{n-1-k}\cos^{2(k-1)-1}\theta_{n-1-k}, \\
	c^2  & = & \cos^2\theta_1...\cos^2\theta_{n-2-k}, \\
        v(\theta_{n-1-k})  & = & e^{-\frac{1}{2}x^2c^2\cos^2\theta_{n-1-k}}, \\
	v'(\theta_{n-1-k}) & = & x^2 c^2 \sin\theta_{n-1-k}\cos\theta_{n-1-k}
	            e^{-\frac{1}{2}x^2c^2\cos^2\theta_{n-1-k}}, \\
        \left[uv\right]_0^{\pi/2} & = & -e^{-\frac{1}{2}x^2c^2}, \\
	-\int_0^{\frac{\pi}{2}} u'v & = & \int 2(k-1)\sin\theta_{n-1-k}\cos^{2(k-1)-1}\theta_{n-1-k} \nonumber\\
                                    &   & e^{-\frac{1}{2}x^2c^2\cos^2\theta_{n-1-k}}\mathrm{d}\theta_{n-1-k}.
    \end{eqnarray}
    We thus find that
    \begin{eqnarray}
      I_{k, n}(X)
        & = & \int_{x=0}^{x \le X}
	      \int_{\theta_1=0}^{\theta_1=\frac{\pi}{2}}...
	      \int_{\theta_{n-(k+2)}=0}^{\theta_{n-(k+2)}=\frac{\pi}{2}}
              \frac{1}{(k-1)!}\pi^{k-1}x^{2(n-2)-1} \nonumber \\
	&   & \sin\theta_1\cos^{2(n-3)-1}\theta_1...
	      \sin\theta_{n-2-k}\cos^{2k-1}\theta_{n-2-k} \nonumber \\
        &   & \left[\int u'v \mathrm{d}\theta_{n-1-k} - e^{-\frac{1}{2}x^2c^2} \right]
	      \mathrm{d}x\mathrm{d}\theta_1...\mathrm{d}\theta_{n-2-k}.
    \end{eqnarray}
    We finally find the recurrence formula
    \begin{equation}
      I_{k, n}(X) = I_{k,n-1}(X) - 2\pi I_{k-1, n-1}(X),
    \end{equation}
    since
    \begin{eqnarray}
      I_{k, n-1}(X)
        & = & \int_{x=0}^{x \le X} 
              \prod_{i=1}^{n-1-k} \rchi_2(x_i)
              \prod_{i=n-k}^{n-2} 2\pi x_i
	      \prod_{i=1}^{n-2}\mathrm{d}x_i \\
        & = & \int_{x=0}^{x \le X}
	      \int_{\theta_1=0}^{\theta_1=\frac{\pi}{2}}...
	      \int_{\theta_{n-3}=0}^{\theta_{n-3}=\frac{\pi}{2}}
              (2\pi)^{k-1}x^{2(n-2)-1} \nonumber \\
	&   & \sin\theta_1\cos^{2(n-3)-1}\theta_1
	      \sin\theta_2\cos^{2(n-4)-1}\theta_2... \nonumber \\
	&   & \sin\theta_{n-3}\cos\theta_{n-3}
              e^{-\frac{1}{2}x^2\cos^2\theta_1...\cos^2\theta_{n-2-k}} \nonumber \\
        &   &   \mathrm{d}x\mathrm{d}\theta_1...\mathrm{d}\theta_{n-3} \\
	& = & \int_{x=0}^{x \le X}
	      \int_{\theta_1=0}^{\theta_1=\frac{\pi}{2}}...
	      \int_{\theta_{n-2-k}=0}^{\theta_{n-2-k}=\frac{\pi}{2}}
              \left(\prod_{i=1}^{k-1}B(i,1)\right) \nonumber \\
	&   & \pi^{k-1}x^{2(n-2)-1} \nonumber \\
	&   & \sin\theta_1\cos^{2(n-3)-1}\theta_1...
	      \sin\theta_{n-2-k}\cos^{2(k-1)-1}\theta_{n-2-k} \nonumber \\
        &   & e^{-\frac{1}{2}x^2\cos^2\theta_1...\cos^2\theta_{n-2-k}}
              \mathrm{d}x\mathrm{d}\theta_1...\mathrm{d}\theta_{n-2-k}
    \end{eqnarray}
    and
    \begin{eqnarray}
      I_{k-1, n-1}(X) & = & \int_{x=0}^{x \le X} 
              \prod_{i=1}^{n-k} \rchi_2(x_i)
              \prod_{i=n-k+1}^{n-2} 2\pi x_i
	      \prod_{i=1}^{n-2}\mathrm{d}x_i \\
        & = & \int_{x=0}^{x \le X}
	      \int_{\theta_1=0}^{\theta_1=\frac{\pi}{2}}...
	      \int_{\theta_{n-3}=0}^{\theta_{n-3}=\frac{\pi}{2}}
              (2\pi)^{k-2}x^{2(n-2)-1} \nonumber \\
	&   & \sin\theta_1\cos^{2(n-3)-1}\theta_1
	      \sin\theta_2\cos^{2(n-4)-1}\theta_2 \nonumber \\
	&   & ...\sin\theta_{n-3}\cos\theta_{n-3} \nonumber \\
        &   & e^{-\frac{1}{2}x^2\cos^2\theta_1...\cos^2\theta_{n-1-k}}
              \mathrm{d}x\mathrm{d}\theta_1...\mathrm{d}\theta_{n-3} \\
        & = & \int_{x=0}^{x \le X}
	      \int_{\theta_1=0}^{\theta_1=\frac{\pi}{2}}...
	      \int_{\theta_{n-1-k}=0}^{\theta_{n-1-k}=\frac{\pi}{2}}
              \left(\prod_{i=1}^{k-2}B(i,1)\right) \nonumber \\
	&   & \pi^{k-2}x^{2(n-2)-1} \nonumber \\
	&   & \sin\theta_1\cos^{2(n-3)-1}\theta_1...
	      \sin\theta_{n-(k+1)}\cos^{2(k-1)-1}\theta_{n-1-k} \nonumber \\
        &   & e^{-\frac{1}{2}x^2\cos^2\theta_1...\cos^2\theta_{n-1-k}}
              \mathrm{d}x\mathrm{d}\theta_1...\mathrm{d}\theta_{n-1-k}.
    \end{eqnarray}

    Knowing that
    \begin{equation}
      \int_{\theta=0}^{\theta=\frac{\pi}{2}}\sin\theta\cos^{2m-1}\theta\mathrm{d}\theta
      = \frac{1}{2} B(m,1) = \frac{1}{2}\frac{1}{m},
    \end{equation}
    we integrate the differents parts:
    \begin{eqnarray}
      \int_{\theta_{n-k}=0}^{\theta_{n-k}=\frac{\pi}{2}}...
	      \int_{\theta_{n-2}=0}^{\theta_{n-2}=\frac{\pi}{2}} & &
      \sin\theta_2\cos^{2(k-1)-1}\theta_2\sin\theta_2\cos^{2(k-2)-1}\theta_2 \nonumber \\
      & & ...\sin\theta_{n-2}\cos\theta_{n-2}\mathrm{d}\theta_1...\mathrm{d}\theta_{n-2} \nonumber\\
      & = & \prod_{i=1}^{k-1}\frac{B(i,1)}{2}, \\
      & = & \frac{2^{1-k}}{(k-1)!}.
    \end{eqnarray}

\section{Proper motion estimation and testing \label{sec:pmfit}}
\subsection{Estimating proper motions}

In this section, we show how it is possible to estimate the proper motion
$\vec{v}$ of a source if the simplifying assumption of null proper motion
made in \S \ref{sec:assumptions} is not met.
We neglect the parallax and the long term effect of the radial motion
of the source, but those extra parameters could also be fitted 
provided we have enough catalogue measurements.
In our simple case, the position $\vec{p}$ of a source at any time $t$
can be computed from its position $\vec{p_0}$ at a reference epoch (e.g. 2000):
\begin{equation}
  \vec{p}(t) = \vec{p_0} + \vec{v}(t-t_0)
  =
  \begin{pmatrix}
    x\\
    y
  \end{pmatrix}
  =
  \begin{pmatrix}
    x_0 + v_x(t-t_0) \\
    y_0 + v_y(t-t_0)
  \end{pmatrix}.
\end{equation}
We assume we have $n$ observations of the source at various epochs $t_i$.
We want to estimate the proper motion, so to estimate the 4 unknowns $(v_x, v_y, x_0, y_0)$.
To do so we use the maximum-likelihood estimate which consists in maximizing the 
likelihood
\begin{equation}
   L = \prod\limits_{i=1}^n \mathcal{N}_{\vec{\mu_i}, \bm{V_i}}(\vec{p}(t_i)),
\end{equation}
therefore minimizing
\begin{equation}
  \rchi^2(\vec{p}(t)) = \sum\limits_{i=1}^n\transposee{(\vec{\mu_i}-\vec{p}(t_i))}
                                              V_i^{-1}(\vec{\mu_i}-\vec{p}(t_i)),
  \label{eq:pm_khi2}
\end{equation}
by solving the system of equations
    \begin{equation}
      \left\{
        \begin{array}{rclr}
          \frac{\partial\rchi^2(\vec{p}(t))}{\partial v_x} & = & 0 & (a),\\
          \frac{\partial\rchi^2(\vec{p}(t))}{\partial v_y} & = & 0 & (b),\\
          \frac{\partial\rchi^2(\vec{p}(t))}{\partial x_0} & = & 0 & (c), \\
          \frac{\partial\rchi^2(\vec{p}(t))}{\partial y_0} & = & 0 & (d).
	\end{array}
      \right.
    \end{equation}
To do so, we compute the derivative of $\rchi^2(\vec{p}(t))$ according to a parameter $a_k$
    \begin{eqnarray}
      \frac{\partial \rchi^2(\vec{p}(t))}{\partial a_k} & = & -2\sum\limits_{i=1}^n
        \frac{1}{(1-\rho_i^2)}\left[
	   \frac{(\mu_{i_x} - p_x)}{\sigma_{i_x}^2}\frac{\partial p_x}{\partial a_k}
	 + \frac{(\mu_{i_y} - p_y)}{\sigma_{i_y}^2}\frac{\partial p_y}{\partial a_k}
	 \right. \nonumber \\
        & &
	 \left.
	 - \frac{\rho_i}{\sigma_{i_x}\sigma_{i_y}}\left(
             (\mu_{i_y} - p_y)\frac{\partial p_x}{\partial a_k}
	   + (\mu_{i_x} - p_x)\frac{\partial p_y}{\partial a_k}
	 \right)
	\right],
	\label{eq:dkhi2}
    \end{eqnarray}
    and the Jacobian matrix of $\vec{p}(t|\vec{v},\vec{p_0})$
    \begin{eqnarray}
      \bm{J_p}(\vec{v},\vec{p_0})
      =
      \begin{pmatrix}
        \nabla x \\
        \nabla y
      \end{pmatrix} 
      & =  &
      \begin{pmatrix}
        \frac{\partial x}{\partial v_x} & \frac{\partial x}{\partial v_y}
	  & \frac{\partial x}{\partial x_0} & \frac{\partial x}{\partial y_0} \\
        \frac{\partial y}{\partial v_x} & \frac{\partial y}{\partial v_y}
	  & \frac{\partial y}{\partial x_0} & \frac{\partial y}{\partial y_0} 
      \end{pmatrix}, \\
      & = &
      \begin{pmatrix}
        (t-t_0) & 0 & 1 & 0 \\
        0 & (t-t_0) & 0 & 1 
      \end{pmatrix}.
    \end{eqnarray}
    So we have to solve the following system of equations, noting $\Delta t_i = t_i - t_0$
    \begin{equation}
      \left\{
        \begin{array}{lclr}
          \sum\limits_{i=1}^n\frac{1}{(1-\rho_i^2)}\left[
	      \frac{(\mu_{i_x} - x)}{\sigma_{i_x}^2}
	    - \frac{\rho_i}{\sigma_{i_x}\sigma_{i_y}}(\mu_{i_y} - y)
	  \right] \Delta t_i & = & 0 & (a),\\
          \sum\limits_{i=1}^n\frac{1}{(1-\rho_i^2)}\left[
	      \frac{(\mu_{i_y} - y)}{\sigma_{i_y}^2}
	    - \frac{\rho_i}{\sigma_{i_x}\sigma_{i_y}}(\mu_{i_x} - x)
	  \right] \Delta t_i & = & 0 & (b),\\
          \sum\limits_{i=1}^n\frac{1}{(1-\rho_i^2)}\left[
	      \frac{(\mu_{i_x} - x)}{\sigma_{i_x}^2}
	    - \frac{\rho_i}{\sigma_{i_x}\sigma_{i_y}}(\mu_{i_y} - y)
	  \right] & = & 0 & (c), \\
          \sum\limits_{i=1}^n\frac{1}{(1-\rho_i^2)}\left[
	      \frac{(\mu_{i_y} - y)}{\sigma_{i_y}^2}
	    - \frac{\rho_i}{\sigma_{i_x}\sigma_{i_y}}(\mu_{i_x} - x)
	  \right] & = & 0 & (d).
	\end{array}
      \right.
    \end{equation}
We can for example use Cramer's rule to solve the general problem
\begin{equation}
  \bm{A}\bm{X} = \bm{\Lambda}
\end{equation}
with, in our case, and using the notation $\xi_i=1/(1-\rho_i^2)$
\begin{eqnarray}
   \bm{A} & = &
    \begin{pmatrix}
        a_1 & b_1 & c_1 & d_1 \\
        a_2 & b_2 & c_2 & d_2 \\
        a_3 & b_3 & c_3 & d_3 \\
        a_4 & b_4 & c_4 & d_4
      \end{pmatrix}, \\
    & = &
   \sum\limits_{i=1}^n\xi_i
   \begin{pmatrix}
        \frac{\Delta^2 t_i}{\sigma_{i_x}^2}
	  & -\frac{\rho_i\Delta^2 t_i}{\sigma_{i_x}\sigma_{i_y}}
	  & \frac{\Delta t_i  }{\sigma_{i_x}^2}
	  & -\frac{\rho_i\Delta t_i  }{\sigma_{i_x}\sigma_{i_y}} \\
        -\frac{\rho_i\Delta^2 t_i}{\sigma_{i_x}\sigma_{i_y}}
	  & \frac{\Delta^2 t_i}{\sigma_{i_y}^2}
	  & -\frac{\rho_i\Delta t_i  }{\sigma_{i_x}\sigma_{i_y}}
	  & \frac{\Delta t_i  }{\sigma_{i_y}^2} \\
        \frac{\Delta t_i}{\sigma_{i_x}^2}
	  & -\frac{\rho_i\Delta t_i}{\sigma_{i_x}\sigma_{i_y}}
	  & \frac{1}{\sigma_{i_x}^2}
	  & -\frac{\rho_i   }{\sigma_{i_x}\sigma_{i_y}} \\
        -\frac{\rho_i\Delta t_i}{\sigma_{i_x}\sigma_{i_y}}
	  & \frac{\Delta t_i}{\sigma_{i_y}^2}
	  & -\frac{\rho_i   }{\sigma_{i_x}\sigma_{i_y}}
	  & \frac{1  }{\sigma_{i_y}^2}
      \end{pmatrix},
\end{eqnarray}
\begin{equation}
\vec{X} = \begin{pmatrix}
        x_1 \\
        x_2 \\
        x_3 \\
        x_4
      \end{pmatrix}
      =
      \begin{pmatrix}
        v_x \\
        v_y \\
        x_0 \\
        y_0
      \end{pmatrix},
\end{equation}
and
\begin{equation}
\vec{\Lambda} = 
     \begin{pmatrix}
        e_1 \\
        e_2 \\
        e_3 \\
        e_4 
      \end{pmatrix}
      =
     \begin{pmatrix}
        \sum\limits_{i=1}^N\xi_i\left[
	    \frac{\mu_{i_x}}{\sigma_{i_x}^2} - \frac{\rho_i\mu_{i_y}}{\sigma_{i_x}\sigma_{i_y}}
	  \right] \Delta t_i \\
        \sum\limits_{i=1}^N\xi_i\left[
	    \frac{\mu_{i_y}}{\sigma_{i_y}^2} - \frac{\rho_i\mu_{i_x}}{\sigma_{i_x}\sigma_{i_y}}
	  \right] \Delta t_i \\
        \sum\limits_{i=1}^N\xi_i\left[
	    \frac{\mu_{i_x}}{\sigma_{i_x}^2} -  \frac{\rho_i\mu_{i_y}}{\sigma_{i_x}\sigma_{i_y}}
	  \right] \\
        \sum\limits_{i=1}^N\xi_i\left[
	     \frac{\mu_{i_y}}{\sigma_{i_y}^2} - \frac{\rho_i\mu_{i_x}}{\sigma_{i_x}\sigma_{i_y}}
	  \right]
      \end{pmatrix},
\end{equation}
leading to the solution
    \begin{equation}
      x_i = \frac{|\bm{A_i}|}{|\bm{A}|},
      \label{eq:cramer_res}
    \end{equation}
    where
    \begin{center}
    $ \bm{A_1} = 
    \begin{pmatrix}
        e_1 & b_1 & c_1 & d_1 \\
        e_2 & b_2 & c_2 & d_2 \\
        e_3 & b_3 & c_3 & d_3 \\
        e_4 & b_4 & c_4 & d_4
    \end{pmatrix}
    $, $ \bm{A_2} = 
    \begin{pmatrix}
        a_1 & e_1 & c_1 & d_1 \\
        a_2 & e_2 & c_2 & d_2 \\
        a_3 & e_3 & c_3 & d_3 \\
        a_4 & e_4 & c_4 & d_4
    \end{pmatrix}
    $, \\
    $ \bm{A_3} = 
    \begin{pmatrix}
        a_1 & b_1 & e_1 & d_1 \\
        a_2 & b_2 & e_2 & d_2 \\
        a_3 & b_3 & e_3 & d_3 \\
        a_4 & b_4 & e_4 & d_4
    \end{pmatrix}
    $ and $ \bm{A_4} = 
    \begin{pmatrix}
        a_1 & b_1 & c_1 & e_1 \\
        a_2 & b_2 & c_2 & e_2 \\
        a_3 & b_3 & c_3 & e_3 \\
        a_4 & b_4 & c_4 & e_4
    \end{pmatrix}.
    $
    \end{center}

\subsection{Estimating the error on the proper motion estimate}

      The covariance matrix on the estimated proper motion parameters is provided by the inverse of the Hessian
      matrix $\bm{Hf}$ of $\ln\left(\prod\limits_{i=1}^n \mathcal{N}_b(\vec{p}(t_i)) \right)$, evaluated with the estimated
      parameters, that is by the matrix
      \begin{equation}
        -\bm{H_f}^{-1} =
        \left.
        \begin{pmatrix}
            \frac{\partial^2 f}{\partial v_x^2}
	  & \frac{\partial^2 f}{\partial v_x \partial v_y}
	  & \frac{\partial^2 f}{\partial v_x \partial x_0}
	  & \frac{\partial^2 f}{\partial v_x \partial y_0} \\
            \frac{\partial^2 f}{\partial v_y \partial v_x}
	  & \frac{\partial^2 f}{\partial v_y^2}
	  & \frac{\partial^2 f}{\partial v_y \partial x_0}
	  & \frac{\partial^2 f}{\partial v_y \partial y_0} \\
	    \frac{\partial^2 f}{\partial x_0 \partial v_x}
	  & \frac{\partial^2 f}{\partial x_0 \partial v_y}
	  & \frac{\partial^2 f}{\partial x_0^2}
	  & \frac{\partial^2 f}{\partial x_0 \partial y_0} \\
            \frac{\partial^2 f}{\partial y_0 \partial v_x}
	  & \frac{\partial^2 f}{\partial y_0 \partial v_y}
	  & \frac{\partial^2 f}{\partial y_0 \partial x_0}
	  & \frac{\partial^2 f}{\partial y_0^2}
        \end{pmatrix}^{-1}
        \right|_{(\tilde{v_x},\tilde{v_y}\tilde{x_0},\tilde{y_0})}
      \end{equation}
      \begin{equation}
        \bm{-H_f^{-1}} = \sum\limits_{i=1}^n\xi_i
        \begin{pmatrix}
          \frac{\Delta^2t_i}{\sigma_{i_x}^2}
	    & -\frac{\rho_i\Delta^2t_i}{\sigma_{i_x}\sigma_{i_y}}
	    & \frac{\Delta t_i}{\sigma_{i_x}^2}
	    & -\frac{\rho_i\Delta t_i}{\sigma_{i_x}\sigma_{i_y}} \\
          -\frac{\rho_i\Delta^2t_i}{\sigma_{i_x}\sigma_{i_y}}
	    & \frac{\Delta^2 t_i}{\sigma_{i_y}^2}
	    & -\frac{\rho_i\Delta t_i}{\sigma_{i_x}\sigma_{i_y}}
	    & \frac{\Delta t_i}{\sigma_{i_y}^2} \\
          \frac{\Delta t_i}{\sigma_{i_x}^2}
	    & -\frac{\rho_i\Delta t_i}{\sigma_{i_x}\sigma_{i_y}}
	    & \frac{1}{\sigma_{i_x}^2}
	    & -\frac{\rho_i}{\sigma_{i_x}\sigma_{i_y}} \\
          -\frac{\rho_i\Delta t_i}{\sigma_{i_x}\sigma_{i_y}}
	    & \frac{\Delta t_i}{\sigma_{i_y}^2}
	    & -\frac{\rho_i}{\sigma_{i_x}\sigma_{i_y}}
	    & \frac{1}{\sigma_{i_y}^2}
        \end{pmatrix}
      \end{equation}
      \begin{equation}
        \bm{V_{f}} = -\frac{1}{|\bm{H_f}|}\transposee{\mathrm{com}(\bm{H_f})}
      \end{equation}

\subsection{Simple case~: no covariance}

      If all positional errors are circles (i.e. $\forall i \in [1,n], \rho_i=0$),
      the simplifications leads to the classical formulae in which $x$ and $y$ are computed independently
      \citep[see][p.781, \S 15.2 ``Fitting Data to a Straight Line'']{nr3}
      \begin{equation}
        \begin{array}{lcrclcr}
	      \Delta_x & = & S_xS_{t_xt_x} - (S_{t_x})^2,
	  & & \Delta_y & = & S_yS_{t_yt_y} - (S_{t_y})^2, \\
	       v_x     & = & \frac{S_{t_xt_x}S_{\mu_x}}{\Delta_x},
	  & &  v_y     & = & \frac{S_{t_yt_y}S_{\mu_y}}{\Delta_y}, \\
	          x_0  & = & \frac{S_xS_{t_x\mu_x} - S_{t_x}S_{\mu_x}}{\Delta_x},
	  & &     y_0  & = & \frac{S_yS_{t_y\mu_y} - S_{t_y}S_{\mu_y}}{\Delta_y},
	\end{array}
	\label{eq:solution_pm_nocov}
      \end{equation}
      where
      \begin{equation}
        \begin{array}{rclcrclcrcl}
            S_x         & = & \sum\limits_{i=1}^n\frac{1}{\sigma_{i_x}^2},
	  & & S_{t_x}   & = & \sum\limits_{i=1}^n\frac{\Delta t_i}{\sigma_{i_x}^2},
	  & & S_{\mu_x} & = & \sum\limits_{i=1}^n\frac{\mu_{i_x}}{\sigma_{i_x}^2},  \\
	    S_{t_xt_x}     & = & \sum\limits_{i=1}^n\frac{\Delta^2 t_i}{\sigma_{i_x}^2},
	  & & S_{t_x\mu_x} & = & \sum\limits_{i=1}^n\frac{\mu_{i_x}\Delta t_i}{\sigma_{i_x}^2},
	  & &              &   & \\
            S_y         & = & \sum\limits_{i=1}^n\frac{1}{\sigma_{i_y}^2},
	  & & S_{t_y}   & = & \sum\limits_{i=1}^n\frac{\Delta t_i}{\sigma_{i_y}^2},
	  & & S_{\mu_y} & = & \sum\limits_{i=1}^n\frac{\mu_{i_y}}{\sigma_{i_y}^2},  \\
	    S_{t_yt_y}     & = & \sum\limits_{i=1}^n\frac{\Delta^2 t_i}{\sigma_{i_y}^2},
	  & & S_{t_y\mu_y} & = & \sum\limits_{i=1}^n\frac{\mu_{i_y}\Delta t_i}{\sigma_{i_y}^2},
	  & &              &   & 
	\end{array}
      \end{equation}
      and associated errors are
      \begin{equation}
        \begin{array}{lcrclcr}
	      \sigma_{v_x}^2  & = & \frac{S_{t_xt_x}}{\Delta_x},
	  & & \sigma_{v_y}^2  & = & \frac{S_{t_yt_y}}{\Delta_y}, \\
	      \sigma_{x_0} & = & \frac{S_x}{\Delta_x},
	  & & \sigma_{y_0} & = & \frac{S_y}{\Delta_y}, \\
	      \rho\sigma_{v_x}\sigma_{x_0} & = & \frac{-S_{t_x}}{\Delta_x},
	  & & \rho\sigma_{v_y}\sigma_{y_0} & = & \frac{-S_{t_y}}{\Delta_y}.
	\end{array}
      \end{equation}

\subsection{Verifying the results}

        We have implemented and tested the result given by equation Eq. (\ref{eq:cramer_res}).
	We compare the results with a modified version of the Levenberg-Marquardt (LM) method
	\citep[see][p. 801, \S 15.5.2 ``Levenberg-Marquardt method'']{nr3} we designed to 
	handle binormal distributions.
	The algorithm is the same except that we replace the term $\frac{\partial \rchi^2}{\partial a_k}$
	in $\beta_k$ by Eq. (\ref{eq:dkhi2}) and $\alpha_{kl}$ by
	\begin{eqnarray}
          \sum\limits_{i=1}^n\frac{1}{(1-\rho_i^2)}
	   & & \left[
	      \frac{1}{\sigma_{i_x}^2} \frac{\partial p_x}{\partial a_k}\frac{\partial p_x}{\partial a_l}
	    + \frac{1}{\sigma_{i_y}^2} \frac{\partial p_y}{\partial a_k}\frac{\partial p_y}{\partial a_l}
	    \right. \nonumber \\
	   & & \left.
	    - \frac{\rho_i}{\sigma_{i_x}\sigma_{i_y}} \left(
	          \frac{\partial p_x}{\partial a_k}\frac{\partial p_y}{\partial a_l}
	- \frac{\partial p_y}{\partial a_k}\frac{\partial p_x}{\partial a_l}
              \right)
          \right].
        \end{eqnarray}
	We initialize the LM parameters with the approximate solutions given in Eq. (\ref{eq:solution_pm_nocov}).
        The results obtained using both methods (LM and Eq. \ref{eq:cramer_res}) are identical.

\subsection{Testing the unique source hypothesis}

        When estimating the proper motion, we formulated the hypothesis $H$ than our $n$ observations come
	from a single underlying source.
        The Chi-square of equation Eq. (\ref{eq:pm_khi2}) follows a Chi-square distribution with
	$2n-4 = 2(n-2)$ degrees of freedom.\\
	Therefore the criteria not to reject $H$ is
	\begin{equation}
           \rchi^2(\vec{p}(t)) = \sum\limits_{i=1}^n\transposee{(\vec{\mu_i}-\vec{p}(t_i))}
                                              \bm{V_i}^{-1}(\vec{\mu_i}-\vec{p}(t_i))
	   \le k_{\gamma\approx~0.9973}^2,
	\end{equation}
	in which $k_{\gamma\approx0.9973}^2 = F_{\rchi_ {2(n-2)}^2}^{-1}(\gamma)$.

\onecolumn
\section{Synthetic catalogues generation script \label{sec:syntscript}}
Here is the script used to generate three synthetical tables and cross-match them 
with the on-line ARCHES XMatch Tool.
The language of the script is specific to the tool.
Both the tool and its documentation are available at the following URL: \url{http://serendib.unistra.fr/ARCHESWebService/index.html}

\begin{verbatim}
synthetic seed=1 nTab=3 prefix=true \
  geometry=cone ra=22.5 dec=33.5 r=0.42 \
  nA=40000 nB=20000 nC=35000 \
  nAB=6000 nAC=12000 nBC=18000 \
  nABC=10000 \
  poserrAtype=CIRCLE poserrAmode=formula paramA1=0.4 \
  poserrBtype=CIRCLE poserrBmode=function paramB1func=x \
                     paramB1xmin=0.8 paramB1xmax=1.2 \
                     paramB1nstep=100 \
  poserrCtype=CIRCLE poserrCmode=function \
                     paramC1func=exp(-0.5*(x-0.75)*(x-0.75)/0.01)/(0.1*sqrt(2*PI)) \
                     paramC1xmin=0.5 paramC1xmax=1 \
                     paramC1nstep=100
save prefix=simu3 suffix=.fits common=simu3.fits format=fits

cleartables

get FileLoader file=simu3A.fits
set pos ra=posRA dec=posDec
set poserr type=CIRCLE param1=ePosA param2=ePosB param3=ePosPA
set cols *

get FileLoader file=simu3B.fits
set pos ra=posRA dec=posDec
set poserr type=CIRCLE param1=ePosA param2=ePosB param3=ePosPA
set cols *

xmatch chi2 completeness=0.9973 nStep=1 nMax=2 join=inner
merge pos chi2
merge dist mec

get FileLoader file=simu3C.fits
set pos ra=posRA dec=posDec
set poserr type=CIRCLE param1=ePosA param2=ePosB param3=ePosPA
set cols *

xmatch chi2 completeness=0.9973 nStep=2 nMax=2 join=inner
merge pos chi2
merge dist mec

save simu3.ABC.fits fits
\end{verbatim}
\twocolumn

  \bibliography{biblio} 

\end{document}